\documentclass[titlepage]{article}

\usepackage{nicefrac}
\usepackage{xfrac}
\usepackage{authblk}
\usepackage[T1]{fontenc}
\usepackage{ae}
\usepackage{aecompl}
\usepackage[utf8]{inputenc}
\usepackage[ngerman,english]{babel}
\usepackage{natbib}
\usepackage[export]{adjustbox} 
\usepackage{color} 
\usepackage{enumerate} 
\usepackage[left=1in,right=1in]{geometry} 
\usepackage{longtable} 
\usepackage{arydshln} 
\usepackage{rotating} 
\usepackage{pdflscape} 
\usepackage{amsmath}
\usepackage{mathtools}
\usepackage{amssymb}
\usepackage{pbox} 
\usepackage{booktabs}
\usepackage{fancyvrb} 
\usepackage{grffile} 
			
\usepackage{pifont}

\usepackage{pdflscape}
\usepackage{color}
\usepackage{todonotes}
\usepackage{float}
\usepackage{wrapfig}
\usepackage{subcaption}
\usepackage{setspace} 
\usepackage{multirow} 
\usepackage{hyperref} 
\hypersetup{
      breaklinks=true,  
      colorlinks=true,
      urlcolor=blue,
      linkcolor=blue,
      citecolor=blue,
}
\usepackage{fancyhdr}
\usepackage{abstract}

\usepackage{tikz}
\usetikzlibrary{positioning}
\tikzset{every picture/.style={font=\sffamily}}

\usepackage{lineno}
\modulolinenumbers[5]
\definecolor{darkgreen}{rgb}{0.05,0.35,0.1}
\newcommand{\revised}[1]{#1}
\newcounter{result}[section]

\title{\rule{\textwidth}{1pt}\\\LARGE{\revised{\textbf{
Brain Age Revisited: Investigating the State vs. Trait Hypotheses of EEG-derived Brain-Age Dynamics with Deep Learning
}}}\\\rule{\textwidth}{1pt}\\\
\\
}

\author[a, b]{Lukas A. W. Gemein\thanks{Corresponding author: lukas.gemein@uniklinik-freiburg.de}}
\author[a, c]{Robin T. Schirrmeister}
\author[b, e]{Joschka Boedecker}
\author[a, d, e]{Tonio Ball}
\affil[a]{
    Neuromedical AI Lab,
    Department of Neurosurgery, 
    Medical Center – University of Freiburg,
    Faculty of Medicine, University of Freiburg, 
    Engelbergerstr. 21, 79106 Freiburg, Germany}
\affil[b]{
    Neurorobotics Lab, 
    Computer Science Department – University of Freiburg,
    Faculty of Engineering, University of Freiburg, 
    Georges-Köhler-Allee 80, 79110 Freiburg, Germany}
\affil[c]{
    Machine Learning Lab, 
    Computer Science Department – University of Freiburg,
    Faculty of Engineering, University of Freiburg, 
    Georges-Köhler-Allee 74, 79110 Freiburg, Germany}
\affil[d]{
    Freiburg Epilepsy Center,
    Department of Neurosurgery, 
    Medical Center – University of Freiburg,
    Faculty of Medicine, University of Freiburg, 
    Breisacher Str. 64, 79106 Freiburg, Germany}
\affil[e]{
    BrainLinks-BrainTools, 
    Institute for Machine-Brain Interfacing Technology,
    University of Freiburg, 
    Georges-K\"ohler-Allee 201, 79110 Freiburg, Germany}

\newcommand{\beginsupplement}{%
        \setcounter{table}{0}
        \renewcommand{\thetable}{S\arabic{table}}%
        \setcounter{figure}{0}
        \renewcommand{\thefigure}{S\arabic{figure}}%
     }
\begin{document}

\maketitle

\doublespacing



\begin{abstract}
The brain’s biological age has been considered as a promising candidate for a neurologically significant biomarker. However, recent results based on longitudinal magnetic resonance imaging data have raised questions on its interpretation. A central question is whether an increased biological age of the brain is indicative of brain pathology and if changes in brain age correlate with diagnosed pathology (state hypothesis). Alternatively, could the discrepancy in brain age be a stable characteristic unique to each individual (trait hypothesis)? To address this question, we present a comprehensive study on brain aging based on clinical EEG, which is complementary to previous MRI-based investigations. We apply a state-of-the-art Temporal Convolutional Network (TCN) to the task of age regression. We train on recordings of the Temple University Hospital EEG Corpus (TUEG) explicitly labeled as non-pathological and evaluate on recordings of subjects with non-pathological as well as pathological recordings, both with examinations at a single point in time and repeated examinations over time. Therefore, we created four novel subsets of TUEG that include subjects with multiple recordings: I) all labeled non-pathological; II) all labeled pathological; III) at least one recording labeled non-pathological followed by at least one recording labeled pathological; IV) similar to III) but with opposing transition (first pathological then non-pathological). The results show that our TCN reaches state-of-the-art performance in age decoding with a mean absolute error of 6.6 years. Our extensive analyses demonstrate that the model significantly underestimates the age of non-pathological and pathological subjects (-1 and -5 years, paired t-test, $p \leq 0.18$ and $p \leq 0.0066$). Furthermore, there exist significant differences in average brain age gap between non-pathological and pathological subjects both with single examinations and repeated examinations (-4 and -7.48 years, permutation test, $p \leq 0.016$ and $p \leq 0.00001$). However, the brain age gap biomarker is not indicative of pathological EEG in different datasets (53.37\%, 51.24\%, 53.01\%, and 54.41\% balanced accuracy, permutation test, $p \leq 0.167$, $p \leq 0.239$, $p \leq 0.825$, and $p \leq 0.43$) and we could not find evidence that acquisition or recovery of pathology relates to a significant change in brain age gap in different datasets (0.46 and 1.35 years, permutation test, $p \leq 0.825$ and $p \leq 0.43$; and  Wilcoxon-Mann-Whitney and Brunner-Munzel test, $p \leq 0.13$). Our findings thus support the trait rather than the state hypothesis for brain age estimates derived from EEG. 
\end{abstract}



\section{Introduction}
\label{sec:introduction}
To describe the speed of aging of people and their individual organ systems, the biological age has been introduced. 
While the chronological age (CA) is precisely defined by the birth date of a person, the biological age is subject to variation and can diverge from the CA. The biological age can presumably be influenced by environmental factors, by harmful behavior, by acquisition and recovery of diseases and disorders, by cognitive training, and by congenital factors. The divergence of CA and biological age might have clinical importance as it might contribute to a better understanding of healthy or pathological variations, e.g. in the context of blood and heart [\cite{pavanello2020biological}], the skin [\cite{kuklinski2017skin}], or the brain [\cite{ye2020association}]. 

To estimate the biological age of the brain (BA), a decoder model needs to be fit to data of the brain to begin with. Typically, only data of healthy subjects is included in training and the model is optimized to reduce the error between prediction and CA of the subject. While taking potential model bias into account, the final predictions are then considered to be the BA.

A variety of different terms have emerged to describe the difference of BA to CA, e.g. ‘brain age gap’ [\cite{franke2010estimating}], ‘brain age delta’ [\cite{vidal2021individual}], ‘brain-predicted age difference’ [\cite{cole2018brain}], and ‘brain age index‘ [\cite{sun2019brain}]. For the rest of this manuscript, we will refer to the difference of predicted BA and CA of the subject as \textit{brain age gap}.

So far, aging of the brain has primarily been studied based on magnetic resonance imaging (MRI) scans [ \cite{franke2010estimating} \cite{franke2010estimating}, \cite{franke2012brain}, \cite{franke2013advanced}, \cite{franke2019ten}, \cite{cole2017predicting}, \cite{gaser2013brainage}, \cite{lowe2016effect}, \cite{cole2017predicting}, \cite{vidal2021individual}, \cite{smith2019estimation}, \cite{smith2020brain}]. The best results were reported by \cite{vidal2021individual}, \cite{smith2019estimation}, and \cite{smith2020brain} who were able to decode the BA with a decoding error of 5, 3.3, and 2.9 years mean absolute error (MAE), respectively. For a review of brain aging research based on MRI, please refer to \cite{franke2019ten}.

In another line of research, BA has been studied based on magnetoencephalographic (MEG) and electroencephalographic (EEG) data [\cite{xifra2021estimating}, \cite{sabbagh2019manifold}, \cite{sabbagh2019manifold}, \cite{sabbagh2020predictive}, \cite{engemann2022reusable},\cite{bonet2023sliced},  \cite{xifra2021estimating}, \cite{sun2019brain}, \cite{ye2020association}, \cite{paixao2020excess}]. The best reported scores are 4.88 years MAE for a combination of MEG and MRI data, 6.6 years MAE for MEG data, and 6.87 years MAE for EEG data by \cite{xifra2021estimating}, \cite{bonet2023sliced} and \cite{al2018predicting}, respectively. Just recently, \cite{engemann2022reusable} also published a benchmark for brain age decoding from multiple MEG and EEG datasets that can be used for thorough model comparisons. 

The review of related literature above indicates that brain age studies based on MRI data generally yield superior decoding results, i.e. lower MAE. However, the advantages of EEG over MRI justify continued interest in developing reliable brain age estimation methods using EEG data. MRI is more expensive, requires specialized facilities and is less comfortable for patients due to the loud noises and confinement. EEG is less expensive, silent during operation, suitable for individuals with claustrophobia, and potentially portable for use in home settings. The availability of EEG systems is also considerably higher than that of MRI systems, due to their lower acquisition and running cost. Therefore, the development of a reliable and precise BA decoder based on EEG signals would be valuable.

Several studies have put the decoded BA and the resulting brain age gap in relation to pathologies. \cite{franke2013brainage} found an increase in brain age gap in patients with type 2 diabetes mellitus which also increased with longer diabetes duration. 
\cite{ye2020association} found an increased brain age gap in patients with dementia.
Similarly, \cite{sun2019brain} presented excess brain age in patients with significant neurological or psychiatric diseases, hypertension, and diabetes.
\cite{liem2017predicting} found advanced brain age in patients with cognitive impairment.
Ultimately, \cite{paixao2020excess} reported an decreased life expectancy in patients with excess brain age gap.
Similarly, \cite{cole2017predicting} found early mortality in subjects with higher brain age.

In contrast to the works listed above, \cite{vidal2021individual} argue that the brain age gap is primarily influenced by congenital factors like birth weight and other polygenic factors.

From the review of related literature above, it appears that two fundamentally different views on the brain age gap exist, which we will refer to as the state and trait hypotheses:

\begin{itemize}
    \item [I)] \textbf{State Hypothesis}: The brain age gap and its trajectory are subject to variation over time and can hence indeed be influenced, e.g., by life events, environmental factors, illness, etc \ldots
    \item [II)] \textbf{Trait hypothesis}: The brain age gap is determined by early-life or genetic factors and follows a predetermined trajectory unaffected by life events, environmental factors, illness, etc \ldots
\end{itemize}

The two views are reminiscing the concept of trait and state anxiety as widely used in psychology [\cite{spielberger1971state}]). We propose that the brain age might be conceptualized in a similar way. It might have a trait component (as described in II)) that is fixed and predetermined by congenital factors. Additionally, there might be a state component (as described in I)) that changes over time, for example with acquisition of a disease. 
Furthermore, the state hypothesis implies that this component of the BA might also be reversible, i.e. by recovery of the disease or by cognitive training. 
Finally, it is important to note that the two views as described above are not mutually exclusive: There might be both, a state and a trait component in BA.

However, studies testing these two fundamentally different views of the brain age gap against each other are so far scarce and, to the best of our knowledge, there is no study so far addressing this issue based on clinical EEG. 

To address this open issue, in this study, we first aimed to improve the performance of BA estimation from EEG signals beyond the current state-of-the-art. To this aim we adapted an established decoder model from EEG pathology classification. We fit the model to EEG recordings of non-pathological subjects minimizing the error between predictions and CA of the subjects. Next, we predicted recordings of non-pathological as well as pathological subjects and computed their brain age gap.
Then, we analyzed the predictive value of the brain age gap with respect to clinically defined EEG pathology. We further investigated the concepts of state and trait brain age gap by examining both populations with recordings at a single point in time and populations with repeated recordings over time with and without change of EEG pathology status in the meantime based on novel derivatives of the largest currently available open clinical EEG dataset.

\section{Material and Methods}
\label{sec:methods}
\subsection{Temple University Hospital EEG Corpus}
The TUH EEG Corpus (TUEG) (v1.2.0) [\cite{obeid2016temple}] is the largest publicly available resource of clinical EEG recordings to date, with almost 70.000 recordings from over 15.000 subjects (7.664 female, 7.321 male, 16 undetermined). Along with recordings in European Data Format (EDF) [\cite{kemp1992simple}], the dataset included medical reports in plain text, providing additional information, such as medication, anamneses, and EEG findings. 
TUEG features multiple predefined subsets, including a Seizure Corpus [\cite{shah2018temple}], an Epilepsy Corpus [\cite{veloso2017big}], a Slowing Corpus [\cite{von2017electroencephalographic}], an Artifact Corpus [\cite{buckwalter2021recent}], and a general Abnormal Corpus [\cite{abnormalLopez}]. In this study, we used the TUH Abnormal EEG Corpus to train our age decoder and created four novel subsets of TUEG to study aging effects over time.

\subsubsection{Temple University Hospital Abnormal EEG Corpus}
\label{sec:tuab}
The TUH Abnormal EEG Corpus (TUAB) (v2.0.0) [\cite{abnormalLopez}] is a subset of TUEG [ \cite{obeid2016temple}] and comprises a total of 2993 recordings (1715 non-pathological / 1.278 pathological) of 2.329 subjects (1.255 female / 1.074 male). The dataset is divided into a predefined training set (2.717 recordings) and a final evaluation set (FE) (276 recordings), which were not aligned in terms of age distribution. To avoid resulting bias when training our decoders, we re-split the dataset. We split the data on a subject-wise basis to prevent patient leak between the training and FE sets (Figure~\ref{fig:tuab_age_pyramid}).

\subsubsection{Novel Temple University Hospital Dataset Derivatives}
\label{sec:longitudinal_datasets}
To study the effects of aging over time, we created four novel data derivatives from TUEG. 
First, we excluded recordings from the Seizure, Epilepsy, and Artifact Corpora to avoid misleading EEG activity. We then used pathology labels for the remaining recordings created by mining their medical reports and applying a rule-based text classifier [\cite{kiessner2023extended}]. We selected recordings of at least two minutes duration that featured all 21 electrodes of the international 10-20 placement [\cite{jasper1958report}]. Finally, we only included patients with at least two recordings.

These datasets include:

\begin{itemize}
    \item [I)] \textbf{Repeated Non-Pathological (RNP)}:\newline
    This dataset only includes patients with recordings exclusively labeled non-pathological, yielding a total of 4068 recordings of 956 subjects (554 female, 402 male).
    \item [II)] \textbf{Repeated Pathological (RP)}:\newline
    This dataset only includes patients with recordings labeled pathological, yielding a total of 18.338 recordings of 2.892 subjects (1.478 female, 1.414 male).
    \item [III)] \textbf{Transition Non-Pathological Pathological (TNPP)}:\newline
    This dataset includes patients with at least one recordings labeled non-pathological and at least one recording labeled pathological, where all pathological recordings chronologically follow the non-pathological recordings. This selection yielded 914 recordings (438 non-pathological, 476 pathological) of 195 subjects (104 female, 91 male). 
    \item [IV)] \textbf{Transition Pathological Non-Pathological (TPNP)}:\newline
    This dataset includes patients with at least one recorded labeled non-pathological and at least one recording labeled pathological, where all non-pathological recordings chronologically follow the pathological recordings. This selection yielded 1273 recordings (686 non-pathological, 587 pathological) of 242 subjects (123 female, 119 male). 
\end{itemize}


\subsubsection{Comparison of Sources of Age Information}
\label{sec:age_sources}
\begin{figure}[htb!]
    \centering
    \centering
    \begin{minipage}[c]{.32\linewidth}
    \centering
    \includegraphics[width=\linewidth]{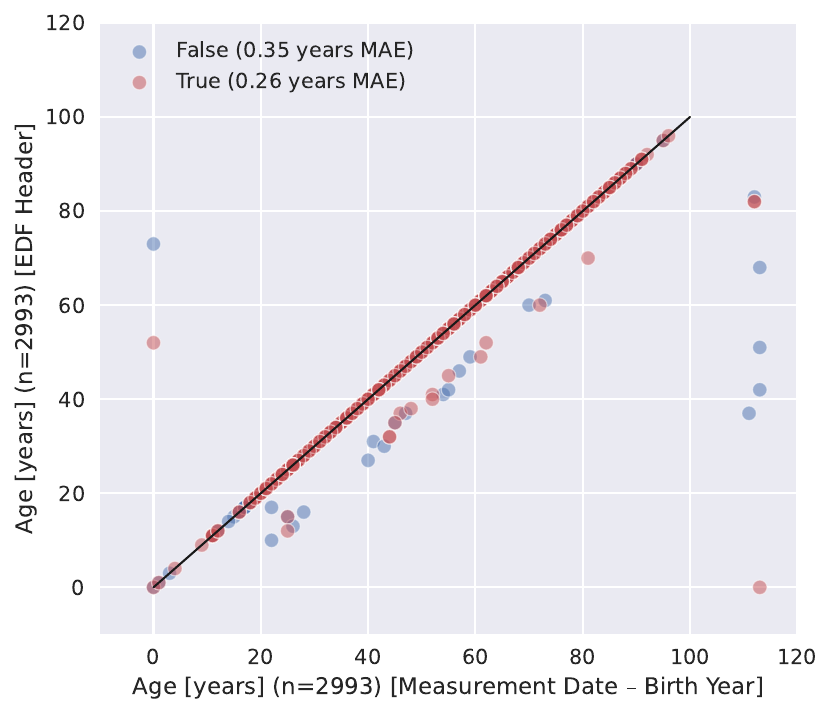}      
    a)
    \end{minipage}
    \begin{minipage}[c]{.32\linewidth}
    \centering
    \includegraphics[width=\linewidth]{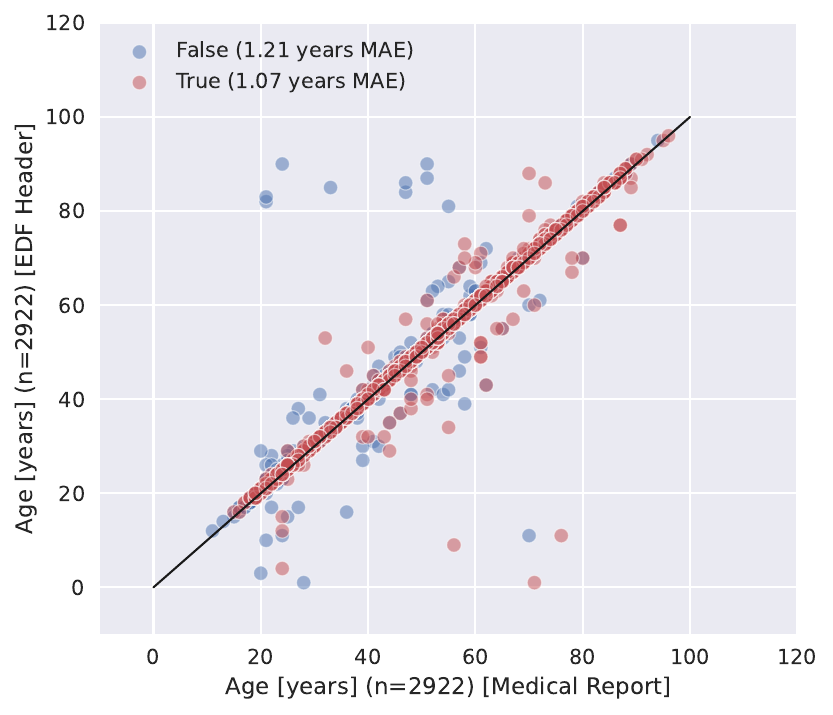}
    b)
    \end{minipage}
    \begin{minipage}[c]{.32\linewidth}
    \centering
    \includegraphics[width=\linewidth]{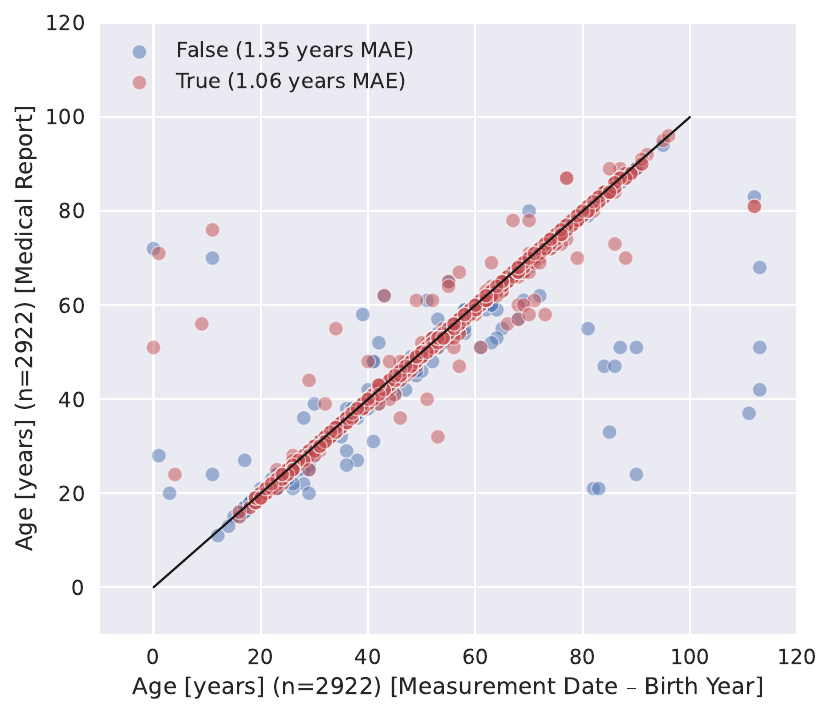}
    c)
    \end{minipage}
    \caption{Different sources of age information in TUAB: a) date vs. header, b) report vs. header, c) date vs. report. Markers represent individual recordings. Orientation along the diagonal signifies agreement of age in both sources. Ages of the different sources do not match in all cases. The highest consensus can be found in date and header age (a)), although there is a systematic error of 10 years for some recordings in date age. Matching report age with the other sources reveals a diffuse pattern of higher/lower age in one of the sources.
    }
    \label{fig:tuab_age_sources}
\end{figure}

\begin{figure}
    \centering
    \begin{minipage}[c]{.32\linewidth}
    \centering
    \includegraphics[width=\linewidth]{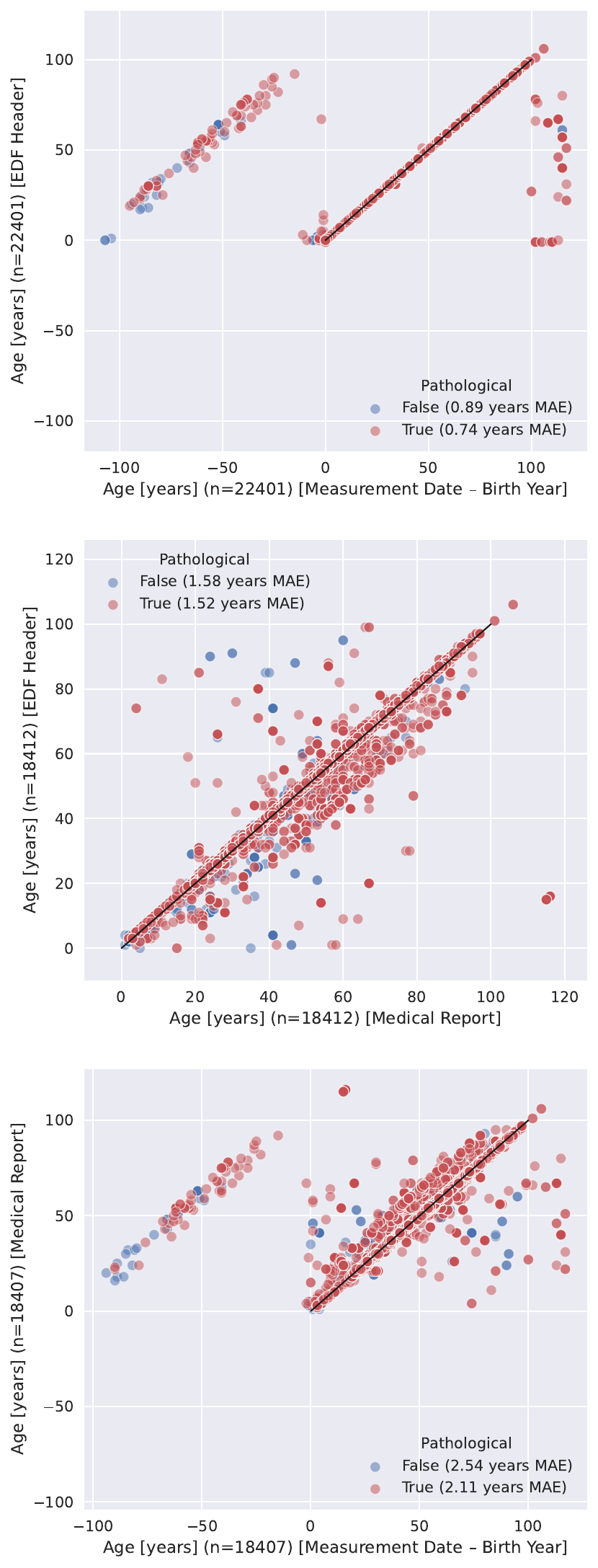}      
    a)
    \end{minipage}
    \begin{minipage}[c]{.32\linewidth}
    \centering
    \includegraphics[width=\linewidth]{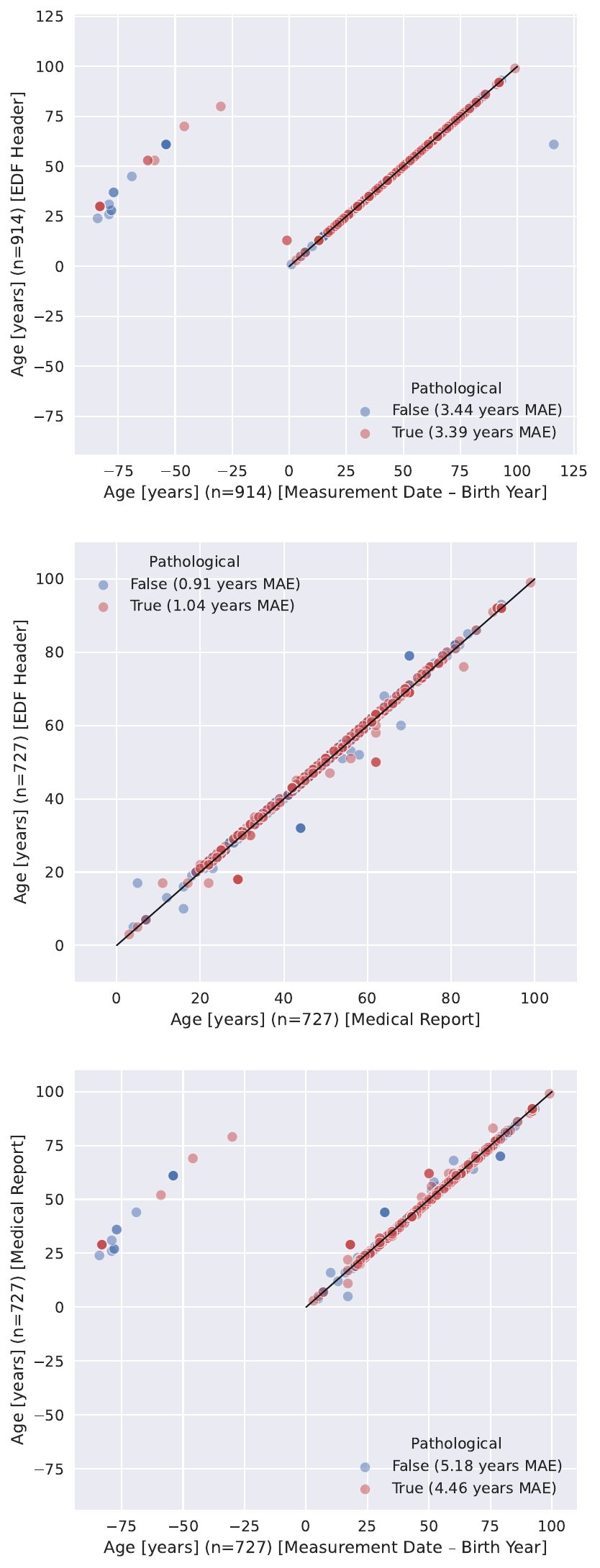}
    b)
    \end{minipage}
    \begin{minipage}[c]{.32\linewidth}
    \centering
    \includegraphics[width=\linewidth]{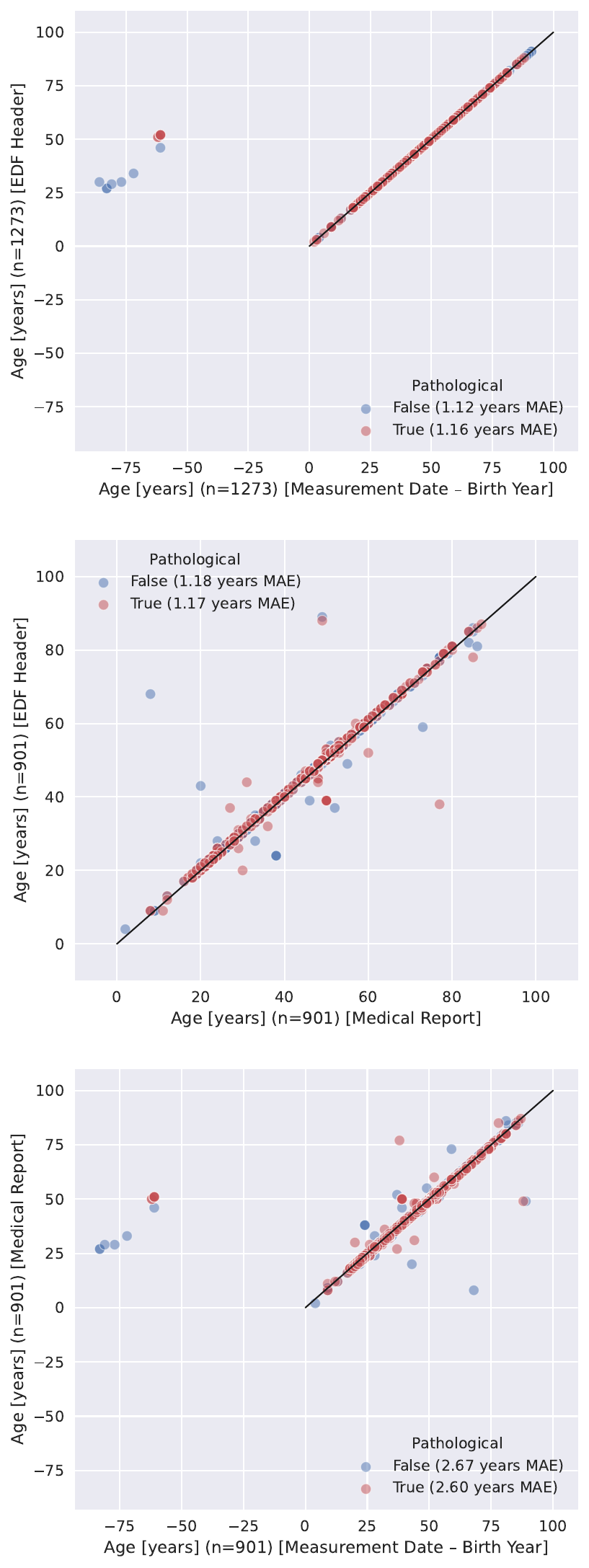}
    c)
    \end{minipage}
    \caption{Different sources of age information (from top to bottom: date vs. header, report vs. header, date vs. report) in a) RNP vs. RP,  b) TNPP, and c) TPNP. Markers represent individual recordings. Orientation along the diagonal signifies agreement of age in both sources. Ages of the different sources do not match in all cases. There is a systematic error of -100 years for some recordings in date age.
        Matching report age with other sources reveals a diffuse pattern of higher/lower age in one of the sources.
    }
    \label{fig:longitudinal_age_sources}
\end{figure}

To extract age information from TUEG, we found multiple possible methods.
\begin{itemize}
    \item[I)] \textbf{Header age:} Age can be parsen from the EDF recording file header where bytes 8 to 88 of the file are dedicated to patient information [\cite{kemp1992simple}] and age is usually presented as “Age: XX”. 
    \item[II)] \textbf{Report age:} Age can be parsed from the medical text reports, which typically start with phrases like “This is a XX-year old / XX y.o.” or variations thereof, however, some medical reports did not follow this convention.
    \item [III)] \textbf{Date age:} Age can be computed based on the anonymized patient birth date and the date of the recording. The birth date is also included in the patient information section of the EDF header, while the recording date can be obtained from measurement information contained in the EDF file.
\end{itemize}
We compared all three extraction methods in Figures~\ref{fig:tuab_age_sources} and \ref{fig:longitudinal_age_sources}. 

From the comparison plots of age sources, we observed some interesting patterns. For example, in the comparison of date and header age in TUAB (Figure~\ref{fig:tuab_age_sources}a)) we noticed a systematically higher age of ten years for some recordings. When comparing report age (Figures~\ref{fig:tuab_age_sources}b) and c)) we observed a diffuse pattern of over- and underestimation for some recordings. The reason for these differences is unclear, but one possible explanation could be a misalignment of EDF files and medical reports within TUEG. According to the mean absolute error (MAE) metric (Section~\ref{sec:metrics}), the agreement of age is best between date and header age.

We made similar observations when comparing age sources in our datasets with repeated examinations (Figure~\ref{fig:longitudinal_age_sources}). Matching date age revealed a systematic age difference of -100 years for some recordings, e.g. Figure~\ref{fig:longitudinal_age_sources}a) top, while matching report age revealed a diffuse pattern as in TUAB. According to the MAE metric, the agreement of age is best between date and header age for RNP, RP, and TPNP and between report and header age in TNPP.

Since these comparisons revealed a substantial amount of label noise, we use the presented age sources to reject recordings (Section~\ref{sec:inclusion_exclusion}). We considered age information extracted from the EDF header to be most reliable, as it does not require matching files (medical report and EDF) or computation (date from EDF measurement info and birth year extracted from EDF header), and chose it as our decoding target. However, the header age might suffer from anonymization imprecision of one year due to the omission of day and month of birth date. For the rest of this study, when we refer to (chronological) subject age or age labels, we refer to the age extracted from EDF headers. 

\subsubsection{Inclusion and Exclusion Criteria}
\label{sec:inclusion_exclusion}
To assess the trade-off between a larger quantity of less reliable data and a smaller quantity of  more reliable data, we generated two derivatives of each of the four datasets introduced above (Section~\ref{sec:longitudinal_datasets}). For the first derivative, we selected only recordings with a minimum duration of 15 minutes to better align with TUAB. For the second derivative, we additionally excluded recordings with a large age deviation (i.e. more than one year) from any of the age sources mentioned earlier (Section~\ref{sec:age_sources}). Note that rejecting individual recordings can trigger the dropping of additional recordings. For example, if a subject was included in the datasets with two recordings and one was rejected due to length or large age deviation, it would cause a drop of the second recording as well, since the subject no longer qualifies for analysis over time. 
For the rest of this study, when we refer to the datasets RNP, RP, TNPP, and TPNP, we are referring to the derivatives with a recording duration of at least 15 minutes where we additionally rejected recordings with a large age deviation. The effect of these rejection parameters on subject and recording numbers can be seen in Table~\ref{tab:longitudinal_subjects_recs}.

\begin{table}[htb!]
    \centering
    \begin{tabular}{|c|c|c|c|}
        \hline
        \textbf{Dataset} & \textbf{$\geq$ 2 min} & \textbf{$\geq$ 15 min} & \textbf{$\geq$ 15 min \& clean age} \\ 
        \hline
        \textbf{TUAB} & 2.993 / 2.329  & 2993 / 2.329  & \textbf{2647 / 2.159} \\
        \textbf{RNP}  & 4.068 / 956    & 933 / 372     & \textbf{621 / 245} \\
        \textbf{RP}   & 18.338 / 2.892 & 5.112 / 1.672 & \textbf{3.963 / 1.321} \\
        \textbf{TNPP} & 914 / 195      & 431 / 170     & \textbf{330 / 133} \\
        \textbf{TPNP} & 1.273 / 242    & 563 / 216     & \textbf{347 / 136} \\
        \hline
    \end{tabular}
    \caption{Number of recordings / subjects of datasets used for this study when changing inclusion criteria (i.e. longer recordings and clean ages across the different sources).}
    \label{tab:longitudinal_subjects_recs}
\end{table}

\subsection{Decoding Pipeline}
\subsubsection{Preprocessing}
\label{sec:preprocessing}
We applied simple preprocessing steps established in our previous studies [\cite{schirrmeister2017patho}, \cite{gemein2020machine}, \cite{kiessner2023extended}]. We picked 21 channels according to the international 10-20 system and clipped extreme technical outlier values to $\pm$800$\mu$V. We re-referenced all recordings to common average referencing and resampled the signals to 100Hz. As we have observed a higher number of outliers in the first minute of recordings, we dropped it (Section~\ref{sec:tuab_outliers}). These preprocessing steps were applied to all datasets used in this study, including TUAB, RNP, RP, TNPP, and TPNP. Preprocessing was performed through braindecode\footnote{Available for download at \url{https://braindecode.org}} [\cite{schirrmeister2017hbm}, \cite{10.3389/fnins.2013.00267}] and relies on the MNE\footnote{Available for download at \url{https://mne.tools}} library [\cite{gramfort2014mne}].

\subsubsection{Model}
We used a temporal convolutional network (TCN) [\cite{bai2018empirical}] as implemented in the braindecode (BD) library which we refer to as BD-TCN. This model has previously been successfully applied in our pathology decoding study [\cite{gemein2020machine}], where it achieved superior results. However, in this study, we used the BD-TCN in a regression setting, as opposed to the classification setting used previously. To accomplish this, we replaced the final softmax layer by a sigmoid layer. The hyperparameters of the model (Table~\ref{tab:model_hyperparameters}) were optimized to decode age based on TUAB in a previous Masters Thesis [\cite{chrab2018architecture}]. For more information about the model architecture, please refer to \cite{bai2018empirical}.

\subsubsection{Training}
We trained our model to decode age of non-pathological recordings of  TUAB only. This yielded superior results compared to including recordings that contain pathological activity during training, as observed in preliminary experiments. To facilitate model training, we standardized the recordings channel-wise to zero mean and unit variance by computing the mean and standard deviation of all channels across all recordings in the training split. Additionally, we min-max scaled targets to [0, 1] with respect to ages in the training split of the data. 

We employed a training technique knows as "cropped decoding" [\cite{schirrmeister2017hbm}] which produces maximally overlapping data crops through a sliding window approach by shifting one sample at a time. To artificially increase the amount of training data and enhance generalizability, we applied channel dropout data augmentation [\cite{saeed2021learning}], which ermerged as the most promising technique from preliminary experiments. 

As in previous studies [\cite{schirrmeister2017patho}, \cite{gemein2020machine}, \cite{kiessner2023extended}], we used the AdamW optimizer [\cite{loshchilov2017fixing}] with a cosine annealing learning rate scheduling [\cite{loshchilov2016sgdr}] to optimize the parameters of the network. We trained our network for 35 epochs with a batch size of 128, which was was sufficient to reach convergence of the L1 loss. 

The model and its training were implemented using the braindecode library [\cite{schirrmeister2017hbm}, \cite{10.3389/fnins.2013.00267}] which relies on Skorch\footnote{Available for download at \url{https://github.com/skorch-dev/skorch}} [\cite{skorch}], that implements the well-known scikit-learn\footnote{Available for download at \url{https://scikit-learn.org/stable/}} [\cite{pedregosa2011scikit}] API for aritifical neural networks, and PyTorch\footnote{Available for download at \url{https://pytorch.org/get-started/locally/}} [\cite{paszke2017automatic}].

\subsubsection{Evaluation}
We followed the same approach as in our previous study [\cite{gemein2020machine}] and performed 5-fold cross-validation (CV) on the training data of TUAB with the same fixed seed. This ensured that every recording was predicted as part of the validation set exactly once (Figure~\ref{fig:tuab_cv_age_histograms}). In addition to this, we also predicted the pathological recordings of TUAB (which were not included in the training data) for comparison purposes.

To avoid leakage of recordings of one subject into different subsets, we performed subject-wise splitting of the data and shuffled beforehand. Through CV, we were able to optimize our choices of data augmentation technique and loss function. 

After completing CV, we conducted five repetitions of final evaluation (FE) using varying seeds to average out effects of random weight initialization. During FE, we trained our model on all available training data and predicted the held-out final evaluation data.

\subsection{Metrics}
\label{sec:metrics}
We report all age regression results as the average over five CV runs or as the average over five repetitions of FE using the mean absolute error $(MAE)$: 
$\nicefrac{\sum_{i=1}^{n}|\hat{y}_{i} - y_{i}|}{n}$

Additionally, we also report the $R^{2}$ score, which is frequently used in the literature: 
$\nicefrac{\sum_{i=1}^{n}( \hat{y}_{i} - \Bar{y})^{2}}{\sum_{i=1}^{n}(y_{i} - \Bar{y})^{2}}$.\\
In both cases, $n$ is the total number of examples, $y$ is the decoding target, $\hat{y}$ is the prediction, and $\Bar{y} = \nicefrac{\sum_{i=1}^{n}y_{i}}{n}$.

Furthermore, we report all classification results as the balanced accuracy $(BACC)$ score: 
$\nicefrac{(\frac{TP}{TP + FN} + \frac{TN}{TN + FP})}{2}$, where \\
$\frac{1}{2}(\nicefrac{TP}{TP + FN} + \nicefrac{TN}{TN + FP})$, where \\
$TP$ is the number of examples that were correctly classified as the positive class, \\ $TN$ is the number of examples that were correctly classified as the negative class, \\ $FP$ is the number of examples that was incorrectly classified as the positive class, \\  $FN$ is the number of examples that was incorrectly classified as the negative class.

\subsection{Post-hoc Analyses}
\subsubsection{Brain age prediction bias}
\label{sec:bias}
We computed the model bias with respect to the original age distribution by fitting a quadratic regression model to decoding targets and brain age gap (BA-CA) during CV (Figure~\ref{fig:cv_model_bias}). Prior to conducting further analyses, we applied this model to remove bias from predictions of all datasets. As a result, all of the results presented in our study were calculated after this correcting step. 

This step is necessary as there are direct interactions between the decoding target (chronological age) and the EEG brain age gap (which represents both the decoding error of the model and the expected difference between the brain age and chronological age). This is often neglected in the literature, but it is important to address in order to obtain accurate results. For more information about this issue, please refer to \cite{smith2019estimation}. 

\subsubsection{EEG Brain Age Gap}
\label{sec:brain_age_gap}
We calculated the EEG brain age gap as the difference between the decoded brain age and the chronological age. A positive gap indicates an overestimation of the chronological age, while a negative gap indicates an underestimation of the chronological age. To account for multiple recordings per subject (see recording and patient numbers in Section~\ref{sec:inclusion_exclusion}), we averaged the brain age gaps for each subject. To analyze the difference between the average gaps, we conducted a permutation test. 

\subsubsection{EEG Brain Age Gap Pathology Biomarker}
\label{sec:brain_age_gap_proxy}
We analyzed the predictive value of the EEG brain age gap with respect to EEG pathology. Under the assumption that large deviations from the chronological age may indicate pathological brain activity, we computed two brain age gap thresholds. Gaps that fell within these thresholds were assigned a non-pathological label, while gaps outside of this range were assigned a pathological label.

We selected this combination of thresholds that achieved the highest BACC between the assigned labels and the real pathology labels (as provided by TUAB or as retrieved from medical reports for the datasets with repeated examinations, see \cite{kiessner2023extended}). The BACC scores were averaged over the five CV folds.

After completing FE, we applied the thresholds computed during CV and averaged the resulting BACC scores over the five runs. To assess whether the EEG brain age gap was a reliable predictor of EEG pathology, we conducted a randomized test.

\subsubsection{Aging Effects Over Time}
\label{sec:longitudinal_aging}
We computed the change of brain age gap over time. Therefore, we grouped the predictions of our novel datasets by subject and pathology status. If multiple recordings of a subject were performed on the same day, we averaged them. We computed the gap difference of subsequent chronological recordings and divided the brain age gap differences by the time passed in between (so, in other words, the slope of the brain age gap change over time). We averaged the gap change rates subject-wise and compared the resulting distributions of RNP vs. RP with a KS test, a Wilcoxon–Mann–Whitney test, and a Brunner-Munzel test.

For TNPP and TPNP we also computed the brain age gap rates but instead of averaging we selected the "moment of transition" for each subject, that is, when the pathology label changes between two subsequent recordings. Due to the design of the datasets, this happens exactly once per subject. We then compared the resulting distributions with a KS test, Wilcoxon-Mann-Whitney test, and a Brunner-Munzel test.

\subsubsection{Statistical Testing}
We used a paired t-test [\cite{student1908probable}] to assess potential differences in the means of the distributions of BA versus CA for both non-pathological and pathological subjects of TUAB.

We used a permutation test with 100.000 samples to assess potential differences in the means of the distributions of brain age gaps of non-pathological versus pathological subjects in TUAB.
Furthermore, we assessed whether pathology classification based on the brain age gap biomarker is significantly better than random label assignment in all datasets. 
We rejected the null hypotheses at $p < 0.05$.

Furthermore, we used a KS test to check for equality of underlying distributions of brain age change rates and a Wilcoxon–Mann–Whitney (WMW) test [\cite{Wilcoxon1945IndividualCB, mann1947test}] and a Brunner-Munzel test [\cite{brunner2000nonparametric}] to assess whether brain age gaps in subjects of RNP change at an equal rate over time compared to subjects of RP.
Again, we rejected the null hypotheses at $p < 0.05$. 

\subsubsection{Amplitude Gradient Analysis}
Following FE, we calculated the gradients of the five models with respect to the non-pathological and pathological recordings of the final evaluation set of TUAB. The gradients were then grouped according to pathology status and subject, and subsequently averaged over all five runs. To visually present which brain regions were most informative for the models in making brain age predictions, we plotted the resulting values on a diagram of the head. The resulting patterns show areas most informative or interesting regarding the brain age decoding task. The bigger the absolute value of the gradients, the more sensitive the network to changes in the input signal with respect to the output (age prediction).

\subsubsection{M/EEG Brain Age Benchmark}
\label{meeg_differences}
To evaluate the robustness of our decoding model and to compare it to other established methods, we participated in the recent brain age decoding benchmark \cite{engemann2022reusable}. This benchmark includes several well-known decoding strategies, ranging from traditional feature extraction with a random forest classifier and computation of covariance matrices with classification based on Riemannian geometry to shallow (BD-Shallow) and deep (BD-Deep) Convolutional Neural Network (ConvNet or CNN) architectures [\cite{schirrmeister2017hbm}]. The benchmark comprises four datasets that vary in ethnicity, demographics, and data type: 

\begin{itemize}
    \item [I)] TUAB [\cite{abnormalLopez}], recorded from a predominantly North American population,
    \item [II)] the Leipzig Mind-Brain-Body dataset (LEMON) [\cite{babayan2019mind}], recorded from a European population with a bimodal age distribution,
    \item [III)] the Cuban Human Brain Project (CHBP) [\cite{valdes2021cuban}, \cite{hernandez2011multimodal}, \cite{bosch2020resting}], recorded from a Latin American population,
    \item [IV)] the Cambridge Center for Aging and Neuroscience (Cam-CAN) [\cite{taylor2017cambridge, shafto2014cambridge}], the only magnetoencephalographic (MEG) dataset.
\end{itemize}

Further details about the datasets and their demographics can be found in \cite{engemann2022reusable}. Our decoding pipeline differs from that presented in the M/EEG brain age benchmark in several respects; an overview and rationale for these differences are provided in Table~\ref{tab:benchmark_differences}.

\begin{table}[htb!]
    \centering
    \begin{tabular}{| p{0.33\textwidth} | p{0.33\textwidth} | p{0.33\textwidth} |}
        \hline
        \textbf{Current study} & \textbf{Study by \citeauthor{engemann2022reusable}} & \textbf{Reason} \\ 
        \hline
        No conversion of raw EDF files & Conversion of raw EDF files to BIDS [\cite{pernet2019eeg}] format & Reduced overhead and sparing resources \\ 
        \hline
        Reject recordings with high (> 1 year) age derivation (Figure~\ref{fig:tuab_age_sources}) & No recording rejection based on age & Assumed to be labeling errors \\ 
        \hline
        Discard first 60s of each recording & No cropping of recordings & More outliers in first 60s of recordings (Figure~\ref{fig:tuab_outliers}) \\ 
        \hline
        Resampling to 100Hz & Resampling to 200Hz (49Hz lowpass afterwards) & Clinical EEG unlike to contain informative brain activity > 50Hz \\ 
        \hline
        Simple preprocessing rules & Application of Autoreject [\cite{jas2017autoreject}] & ConvNets are an end-to-end method which should be able to learn brain signals as well as to ignore artifacts \\ 
        \hline
        5-fold cross-validation & 10-fold cross-validation & Twice as many recordings in each validation set, hence better representation of the joint underlying distribution; additionally, strong reduction of runtime \\ 
        \hline
        Final evaluation & No final evaluation & Possibility to tune hyperparameters in CV and compute generalization error afterwards \\
        \hline
    \end{tabular}
    \caption{Differences in design of decoding pipeline of this study compared to the M/EEG Brain Age Benchmark study [\cite{engemann2022reusable}].}
    \label{tab:benchmark_differences}
\end{table}

\section{Results}
\subsection*{Descriptive Statistics}
\subsubsection*{Age Distributions}
\label{sec:age_distributions}
\begin{figure}[htb!]
    \centering
    \begin{minipage}[c]{.38\linewidth} 
    \centering
    \includegraphics[width=\linewidth]{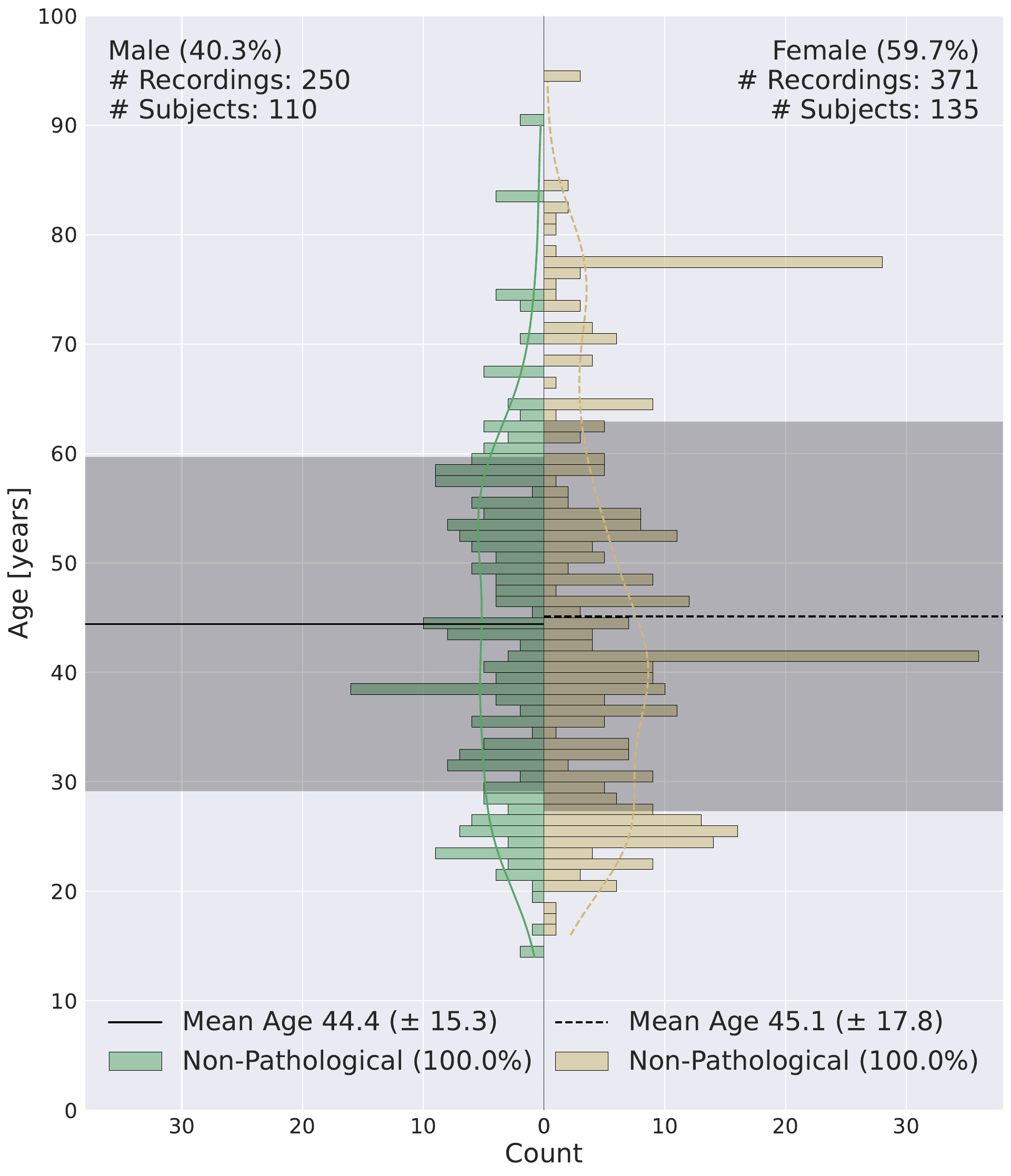}  
    a)
    \end{minipage}
    \begin{minipage}[c]{.38\linewidth}
    \centering
    \includegraphics[width=\linewidth]{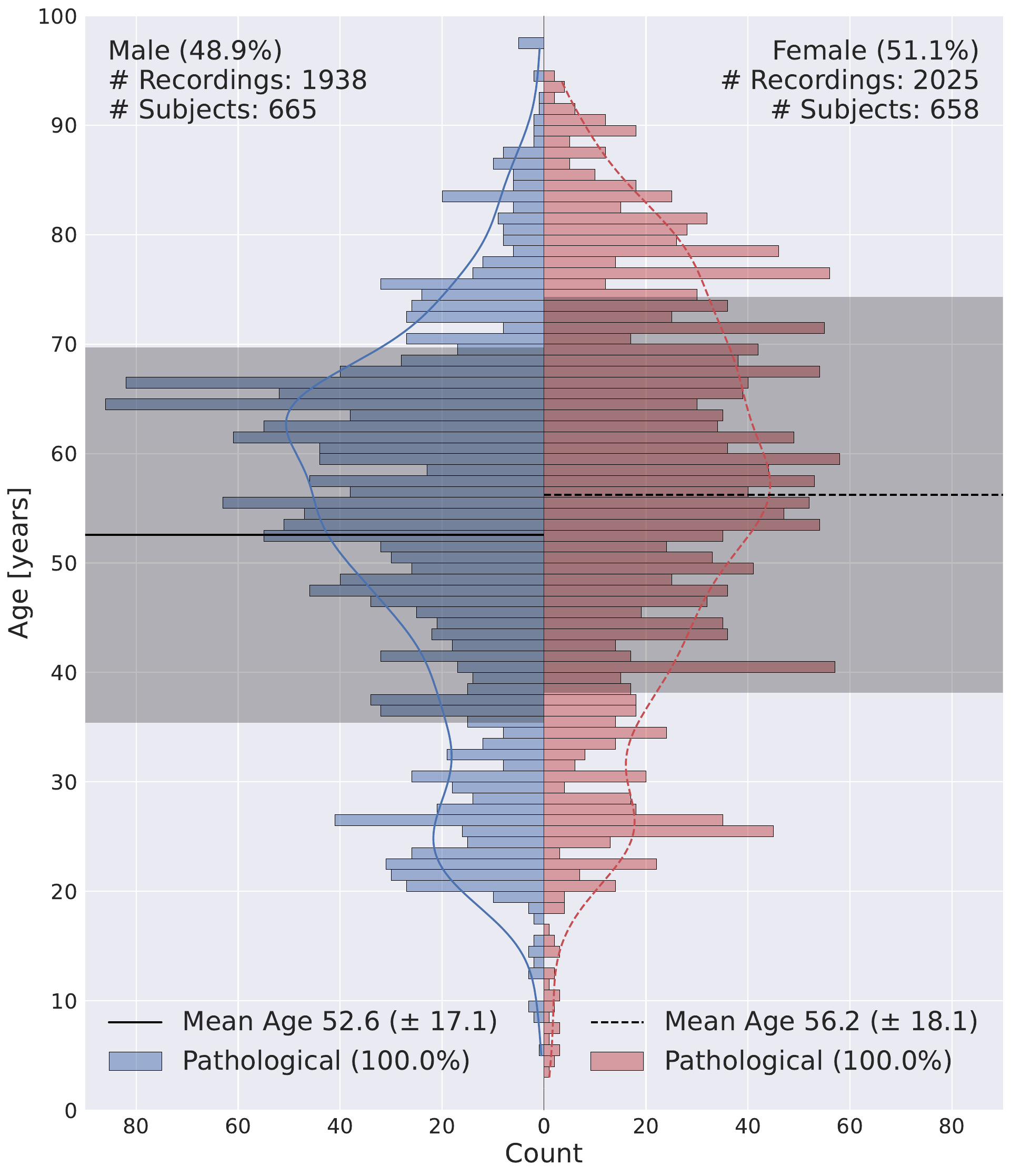} 
    b)
    \end{minipage}
    \begin{minipage}[c]{.38\linewidth}  
    \centering
    \includegraphics[width=\linewidth]{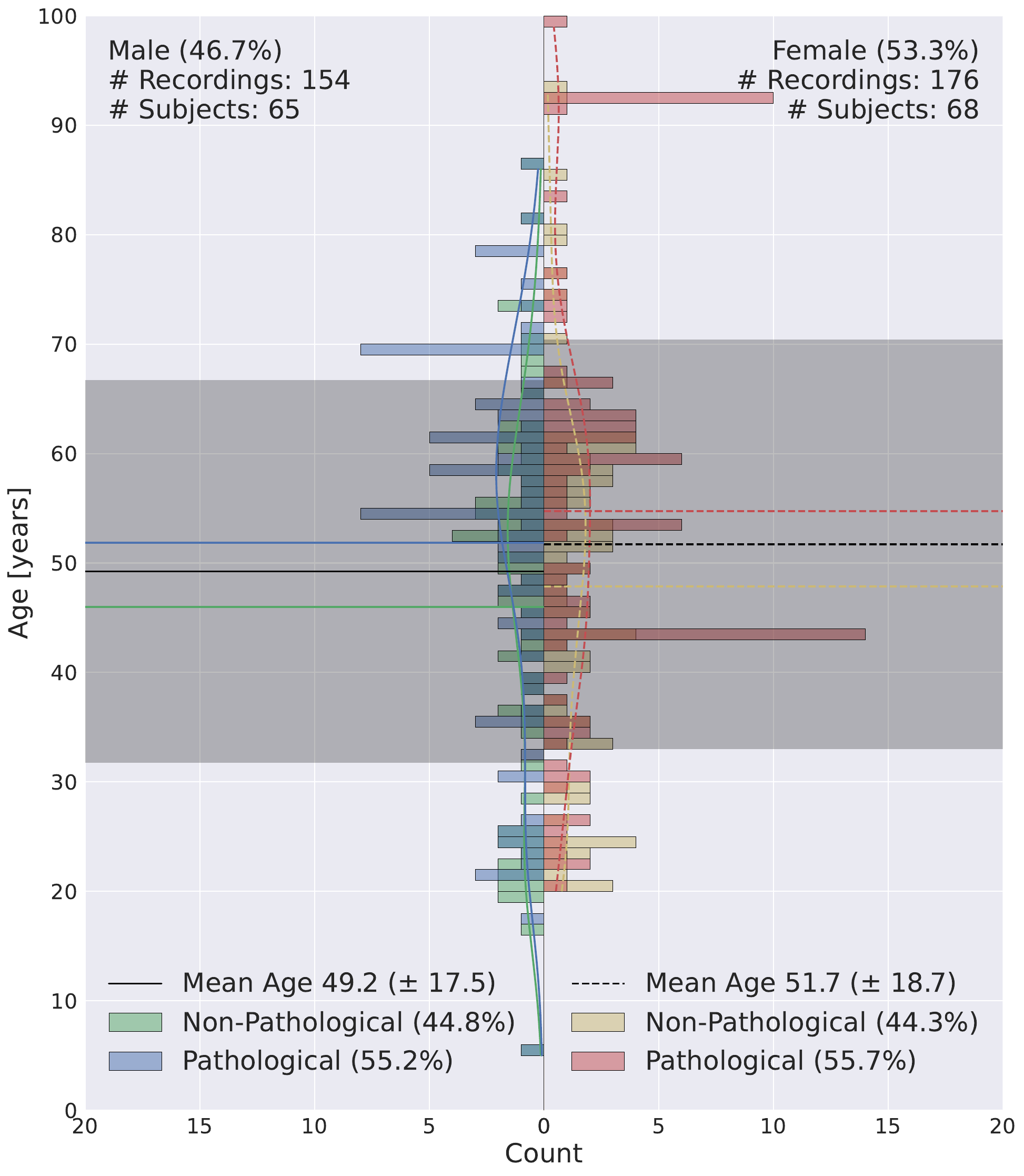}
    c)
    \end{minipage}
    \begin{minipage}[c]{.38\linewidth}  
    \centering
    \includegraphics[width=\linewidth]{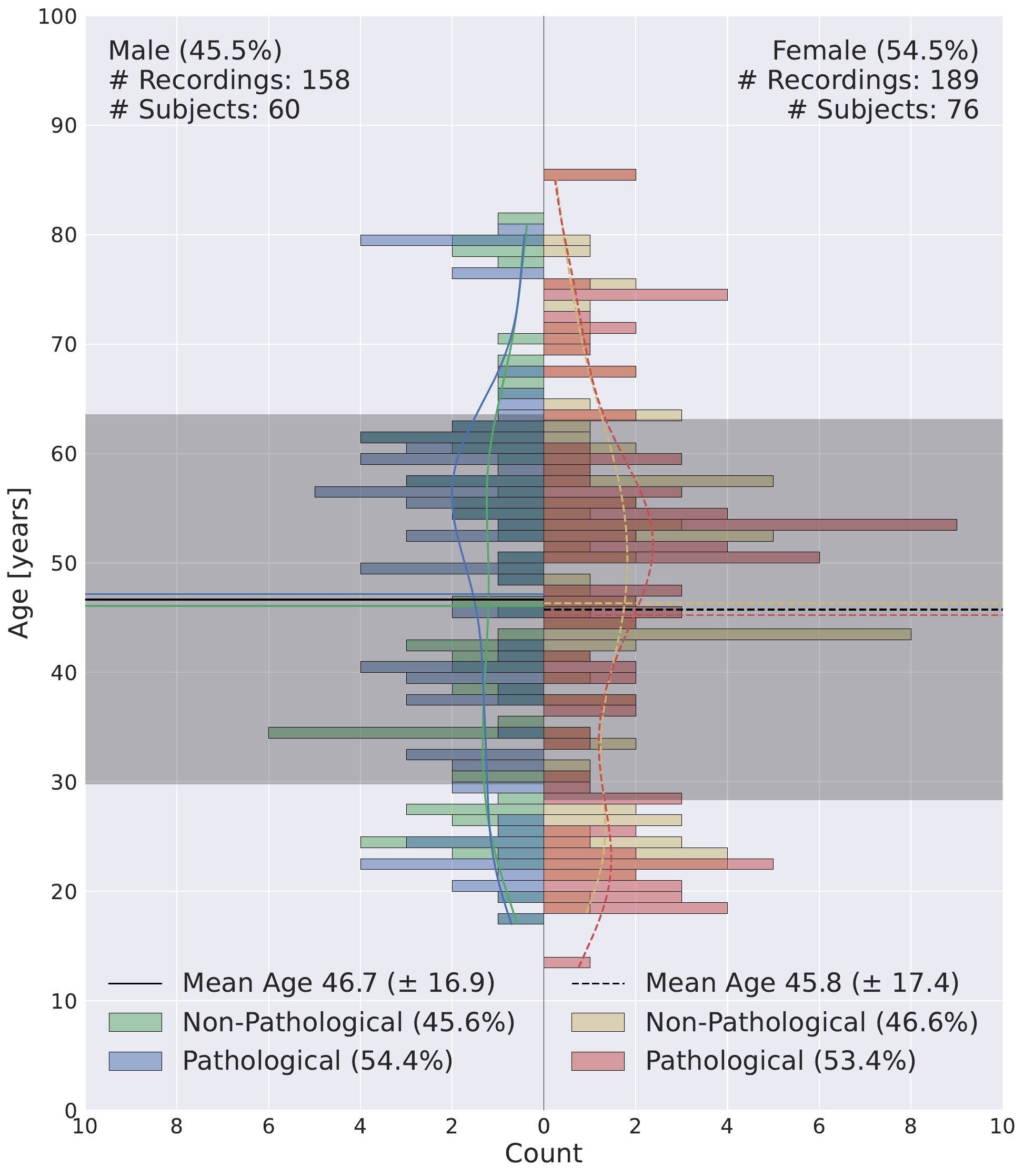}
    d)
    \end{minipage}
    \caption{Age distributions of recordings of a) RNP, b) RP, c) TNPP, and d) TPNP. Differences in age distribution can be observed independent of dataset and gender. Whereas the distribution of male subjects in RP is bimodal, it is rather uniform in RNP. Average ages of recordings in RP are considerably higher compared to the other datasets. Where average female age is larger than average male age in RNP, RP, and TNPP, this trend is inverted in TPNP. There is a considerably higher percentage of female recordings in RNP compared to RP. }
    \label{fig:longitudinal_age_pyramids}    
\end{figure}

In Figure~\ref{fig:longitudinal_age_pyramids}, we present age distributions for RNP, RP, TNPP, and TPNP. The average age is higher in RP, which can be attributed to the increased prevalence of degenerative neurological diseases with advancing age [\cite{hou2019ageing}]. Whereas there is a balanced ratio of male to female subjects in RP, there is a moderately higher ratio if female subjects in TNPP and TPNP and a considerably higher ratio of female subjects (60\%) in RNP. The variance of ages is lowest for RNP and highest for TNPP. In both TNPP and TPNP, there is a higher number of pathological subjects independent of gender. Age pyramids of the derivatives that resulted from the inclusion and exclusion criteria presented in Section~\ref{sec:inclusion_exclusion} can be found in Figures~\ref{fig:lnp_pyramids} – \ref{fig:lpnp_pyramids}. 

\subsubsection*{Recordings per Subject and Intervals}
\label{sec:durations_and_intervals}
\begin{figure}[ht!]
    \centering
    \begin{minipage}[c]{.48\linewidth}
    \centering
    \includegraphics[width=\linewidth]{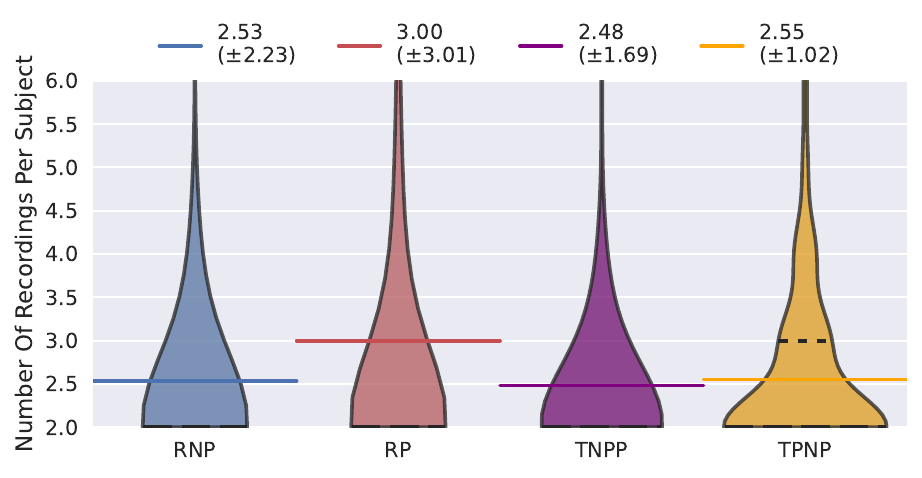}
    a)
    \end{minipage}
    \begin{minipage}[c]{.48\linewidth}
    \centering
    \includegraphics[width=\linewidth]{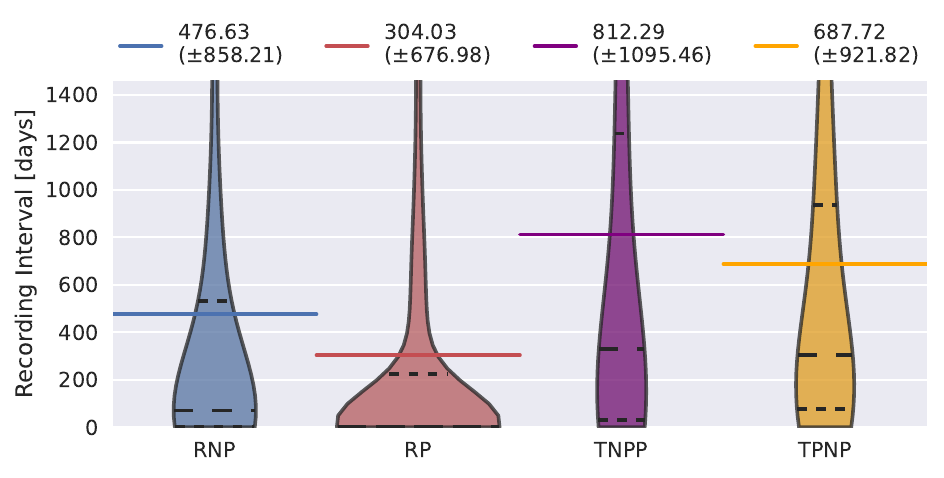}
    b)
    \end{minipage}
    \caption{a) number of recordings per subject and b) recording intervals in RNP, RP, TNPP, and TPNP. Horizontal lines represent the mean. In RP there is a higher number of recordings per subject on average compared to all other datasets.
   There are substantial time gaps of at least 10 months on average between individual recordings in all datasets.}
    \label{fig:recording_intervals}
\end{figure}

We present the number of recordings per subject in the datasets with repeated examinations in Figure~\ref{fig:recording_intervals}a).
Whereas subjects of RNP, TNPP, and TPNP have a similar number of EEG assessments (approximately 2.5) on average, subjects of RP have 3 EEG assessments on average. Variance is lowest for TPNP and highest for RP. 

We present the intervals between subsequent recordings of subjects in the novel datasets in Figure~\ref{fig:recording_intervals}b). On average, there are more than 15 months between recordings of RNP, roughly 10 months between recordings of RP, more than 26 months between recordings of TNPP, and more than 22 months between recordings of TPNP. Variance is lowest in RP and highest in TNPP.
Similar analyses of the derivatives that resulted from the inclusion and exclusion criteria presented in Section~\ref{sec:inclusion_exclusion} can be found in Figure~\ref{fig:longitudinal_rec_intervals_short_unclean}.

In RNP, more than one recording per subject suggests that these individuals may not be entirely healthy. It is possible that they presented with a condition for which they sought medical attention and underwent EEG analysis, but no signs of EEG pathology were detected despite being ill. 
In contrast, in RP, TNPP, and TPNP, multiple recordings per subject are reasonable as follow-up analyses may be performed to monitor disease progression or to prevent relapse. 

As there are several month between subsequent recordings in all datasets, we assume that observed label transitions correspond to real changes in EEG health status and that non-pathological subjects were not excessively misclassified. 

\subsection*{Finding 1: TCN Reaches State-of-the-Art Performance on Non-pathological Subjects of TUAB}
\label{sec:sota_age_decoder}
\begin{figure}[ht!]
    \centering
    \begin{minipage}[c]{.66\textwidth}
    \includegraphics[width=\textwidth]{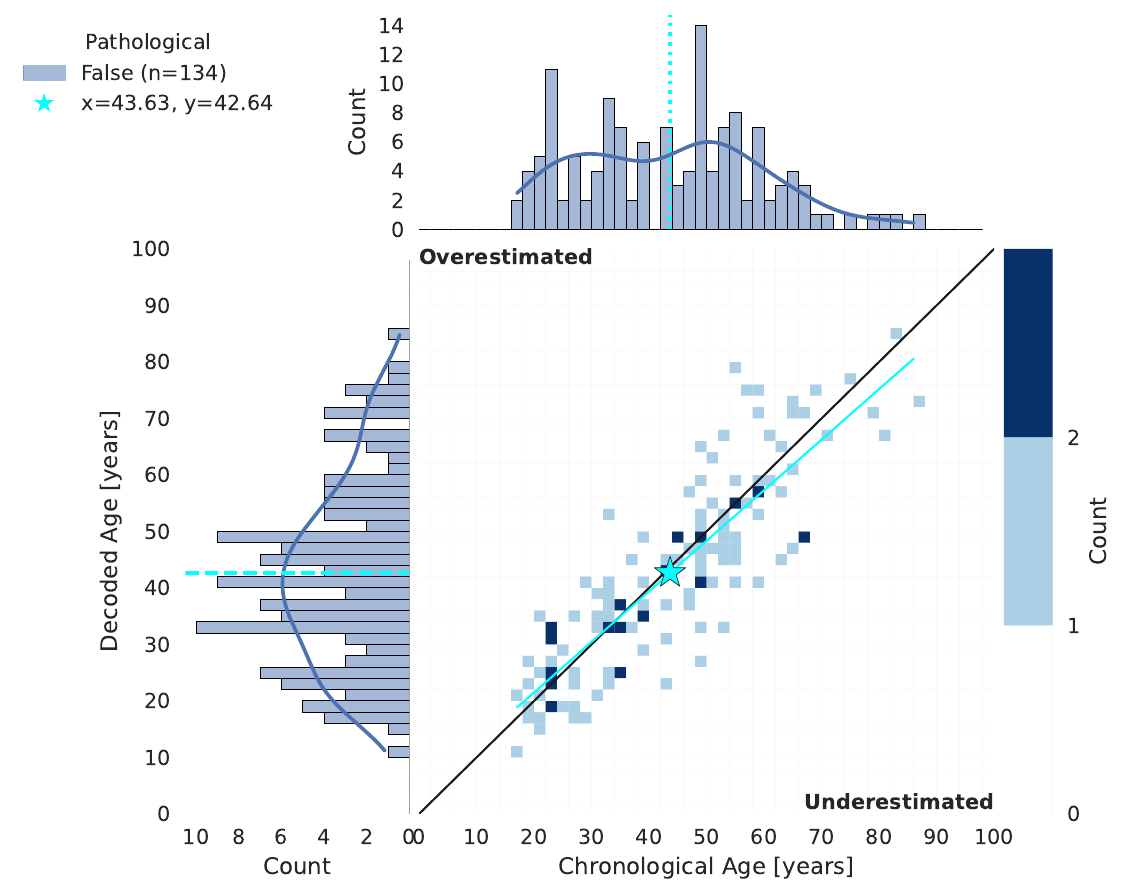}
    \end{minipage}
    \begin{minipage}[c]{.33\textwidth}
    \caption{2D histogram of chronological vs. decoded age in FE. Individual points represent individual subjects. The decoding models tend to overestimate younger subjects and to underestimate older subjects (cyan trend line). The chronological age is underestimated by one year on average (cyan star).}
    \label{fig:fe_np_age_heatmap}    
    \end{minipage}
\end{figure}

Our brain age decoders achieved an average MAE of 6.60 years on non-pathological subjects of TUAB during FE. Figure~\ref{fig:fe_np_age_heatmap} presents the decoded BA in relation to CA. Despite removal of model bias, we observe an overestimation of CA of younger subjects and an underestimation of CA older subjects. On average CA is underestimated by one year, which is not statistically significant (paired t-test, $p = 0.18$). Learning curves of the FE runs can be found in Figure~\ref{fig:fe_curves}.

To the best of our knowledge, our model achieved state-of-the-art performance. The previous best reported score for EEG brain age decoding was 6.87 years MAE by \cite{al2018predicting}, while the best reported score on TUAB was 7.75 years MAE by \cite{engemann2022reusable}. Our score even matches those presented on MEG data in \cite{bonet2023sliced} (6.6 years MAE). 
 

Despite their good performance, our models are unable to match performance of approaches based on MRI data, which achieve results around or below 3 years MAE [\cite{vidal2021individual}, \cite{smith2019estimation}, \cite{smith2020brain}]. Based on our observations and existing literature, it can be inferred that MRI facilitates a more accurate estimation of brain age as there is no clear explanation for a systematic overestimation of brain age in EEG decoding studies compared to MRI studies.
If one assumes that MRI accurately reflects the “true” age of the brain, then EEG approaches may exhibit lower precision, leading to larger errors and an increased discrepancy in brain age estimates. Further research is required to substantiate this hypothesis. One potential methodology involves the development of a combined MRI and EEG dataset to train and optimize state-of-the-art models from both domains. The resulting discrepancies in brain age estimates could then be compared to determine if they are indeed smaller for MRI compared to EEG. Additionally, the presence of individual-level correlations can be investigated. While EEG may not provide as precise an estimation of brain age as MRI, relative differences may still be preserved, such that its predictive value as a proxy measure for pathology persists.

\subsubsection*{TCN is Biased to Underestimate the CA}
\label{sec:model_bias}
\begin{figure}[ht!]
    \centering
    \begin{minipage}[c]{\linewidth}
    \centering
    \includegraphics[width=\linewidth]{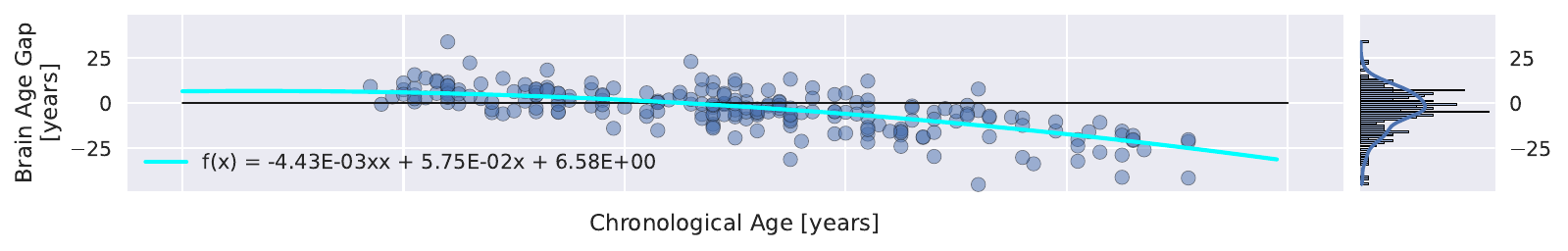}
    a)
    \end{minipage}
    \begin{minipage}[c]{\linewidth}
    \centering
    \includegraphics[width=\linewidth]{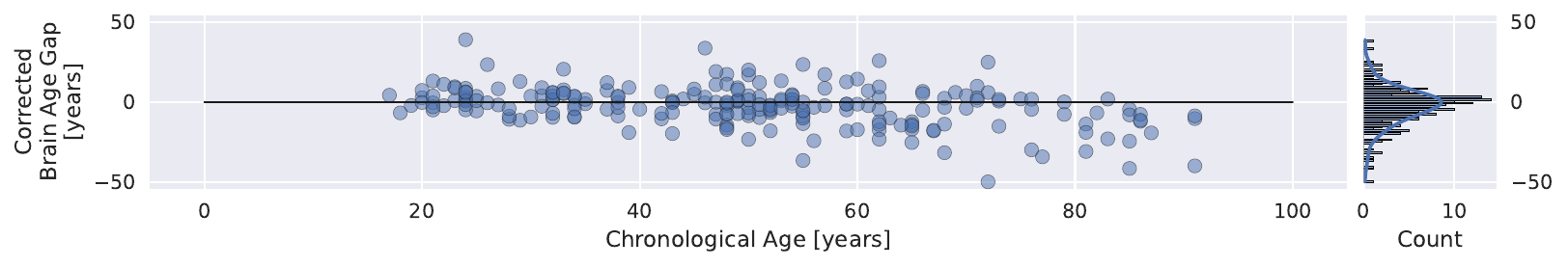}
    b)
    \end{minipage}
    \caption{Model bias in FE. a) Chronological age vs. biased brain age gap and its distribution. b) Chronological age vs. corrected brain age gap and its distribution.
    Individual points represent individual subjects.
    The cyan line represents a quadratic bias model fit to CV data. The original distribution of brain age gaps (a)) is positively skewed and not centered around zero. Application of the bias model model yields the corrected brain age gaps (b)) whose distribution is symmetrically centered around zero.}
    \label{fig:fe_model_bias}
\end{figure}

In Figure~\ref{fig:fe_model_bias}, we present the model bias on FE data. As explained in Section~\ref{sec:bias}, removing the age association bias is essential for robust follow-up analyses. The brain age gap distribution in Figure~\ref{fig:fe_model_bias}a) is skewed and not centered around zero. By applying the bias model fit during CV, we were able to account for this bias and obtain a symmetric distribution centered around zero (Figure~\ref{fig:fe_model_bias}b)). In CV we observed a similar and stronger effect (Figure~\ref{fig:cv_model_bias}). 

The removal of the model bias resulted in a minor decrease in raw decoding performance, from 6.48 years MAE (6.51 in CV) to 6.60 years MAE (6.65 in CV). Since this correction was not performed in related works (or presumably not performed, due to lack of detail detail), the resulting reduction in performance could place our work at a disadvantage when compared to other studies.

\subsubsection*{TCN Outperforms Several Competitors on Multiple Datasets}
\label{sec:benchmark_results}
\begin{figure}[ht!]
    \centering
    \includegraphics[width=\textwidth]{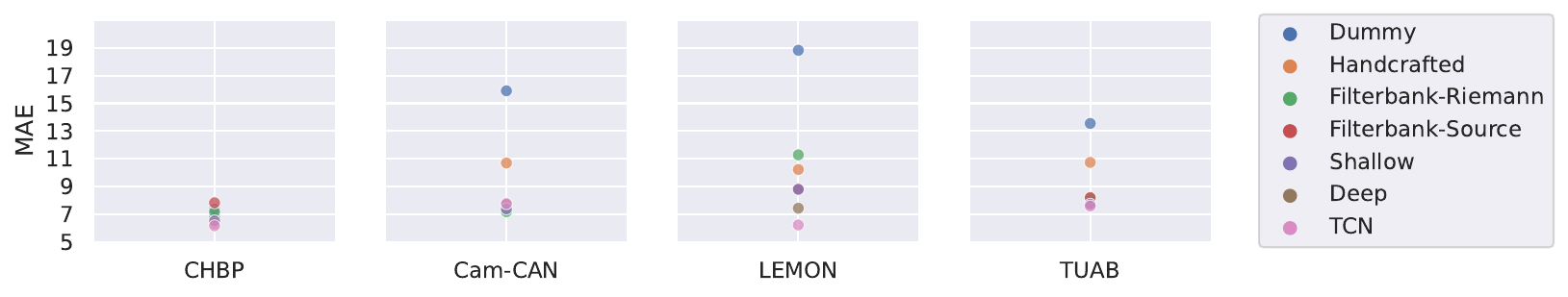}
    \caption{Performance overview and comparison on different datasets through the M/EEG brain age benchmark. The TCN presented in this work reaches top performance in CHBP, LEMON, and TUAB. On Cam-CAN it reaches a slightly lower performance compared to the Filterbank-Riemann approach.}
    \label{fig:meeg_benchmark}
\end{figure}

To demonstrate the robustness of our model, we present it in comparison to other established models and on different datasets through the M/EEG brain age benchmark [\cite{engemann2022reusable}] in Figure~\ref{fig:meeg_benchmark}. It can be seen that our TCN achieves the best score on CHBP, LEMON, and TUAB. On Cam-CAN it achieves results competitive to the other models, out of which Filterbank-Riemann performs best. 

Since our TCN was specifically designed and optimized to decode age from the EEG signals in TUAB, this could be a reason for slightly worse performance on the Cam-CAN dataset. Cam-CAN is the only magnetoencephalographic (MEG) dataset under investigation and also has a larger sensor space compared to the other datasets (Cam-CAN: 102, LEMON: 61, CHBP: 53, TUAB: 21). It is possible that our architecture would require a little update or fine tuning to improve performance on this dataset. Our model still shows very solid and robust performance across all four datasets and partially outperforms established decoding methods which underlines its general applicability to age decoding tasks.

\subsection*{Finding 2: TCN Underestimates the CA of Pathological Subjects of TUAB}
\label{sec:pathological_underestimated}
\begin{figure}[ht!]
    \centering
    \begin{minipage}[c]{.66\textwidth}
    \includegraphics[width=\textwidth]{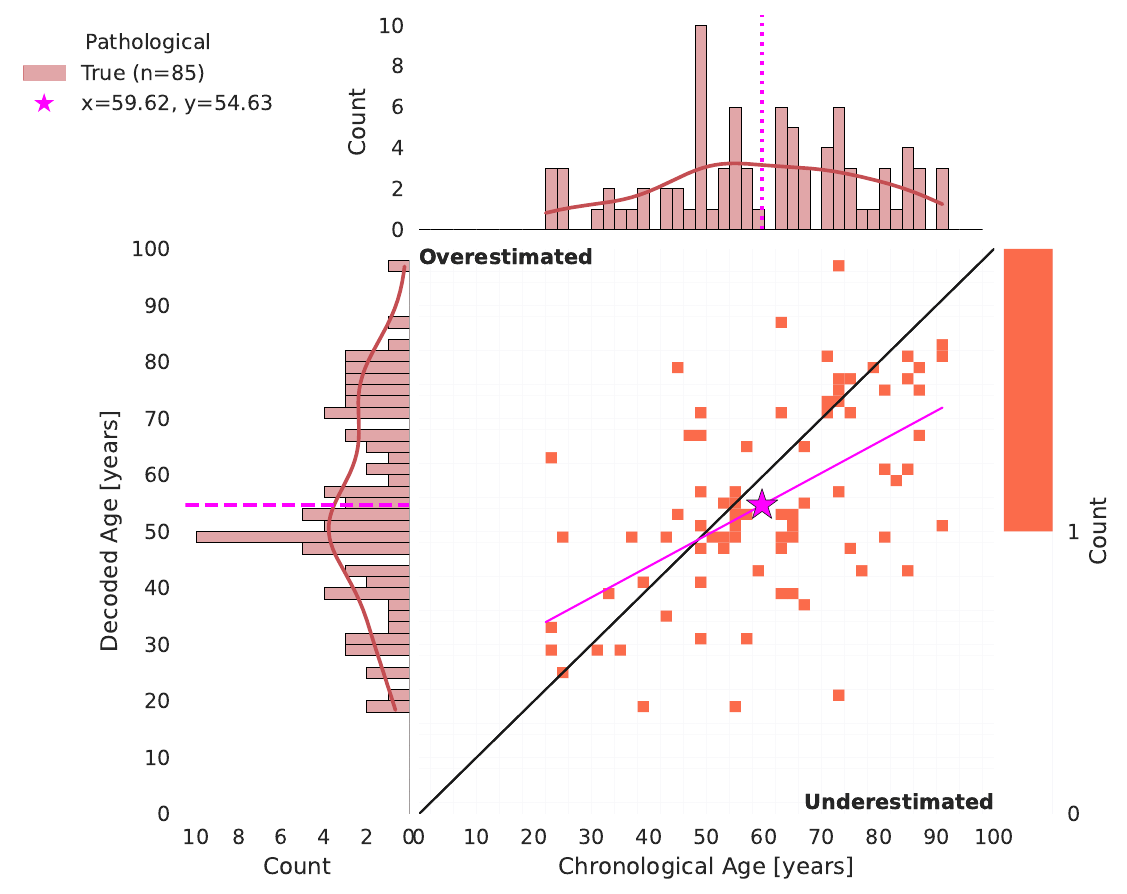}
    \end{minipage}
    \begin{minipage}[c]{.33\textwidth}
    \caption{2D histogram of chronological vs. decoded age in FE. Individual points represent individual subjects. The decoding models tend to overestimate younger subjects and to underestimate older subjects (magenta trend line). The chronological age is underestimated by five years on average (magenta star).}
    \label{fig:fe_p_age_heatmap}    
    \end{minipage}
\end{figure}

Figure~\ref{fig:fe_p_age_heatmap} shows decoded BA in relation to CA as well as their distributions of pathological subjects of TUAB. Despite removal of model bias beforehand, we again observe an overestimation of younger subjects and an underestimation of older subjects. Overall, ages are underestimated by five years on average, which is statistically significant (paired t-test, $p = 0.0066$). Our models reached 12.85 years MAE. 

Several aspects become apparent, when comparing results on non-pathological subjects (Figure~\ref{fig:fe_np_age_heatmap}) to results on pathological subjects (Figure~\ref{fig:fe_p_age_heatmap}):

\begin{itemize}
    \item [I)] The score on non-pathological subjects is substantially better compared to the score on pathological subjects (6.6 years MAE vs. 12.85 years MAE).
    \item [II)] The brain age predictions on non-pathological subjects experience substantially less variance compared to pathological subjects (compare cyan and magenta trend lines and spread of red and blue points).
    \item [III)] The brain age predictions on non-pathological subjects experience substantially less bias compared to pathological subjects (compare cyan and magenta star markers).
\end{itemize}

The presented results are in disagreement with the state hypothesis since we do not observe an increased BA in pathological subjects. On the contrary, we observe an underestimation of the pathological population overall that is even stronger compared to the non-pathological population. Regarding the trait hypothesis, no clear statement can be made as it does not specify the direction of the brain age gap. There might also be a small preference for increased brain age in pathological subjects due to neurological diseases and disorders that are induced by genetic factors.

\subsection*{Finding 3: TCN Predicts Lower BA for Pathological Than for Non-Pathological Subjects in TUAB}
\label{sec:patho_smaller_non_patho}

\begin{figure}[ht!]
    \centering
    \begin{minipage}[c]{\linewidth}
    \centering
    \includegraphics[width=\linewidth]{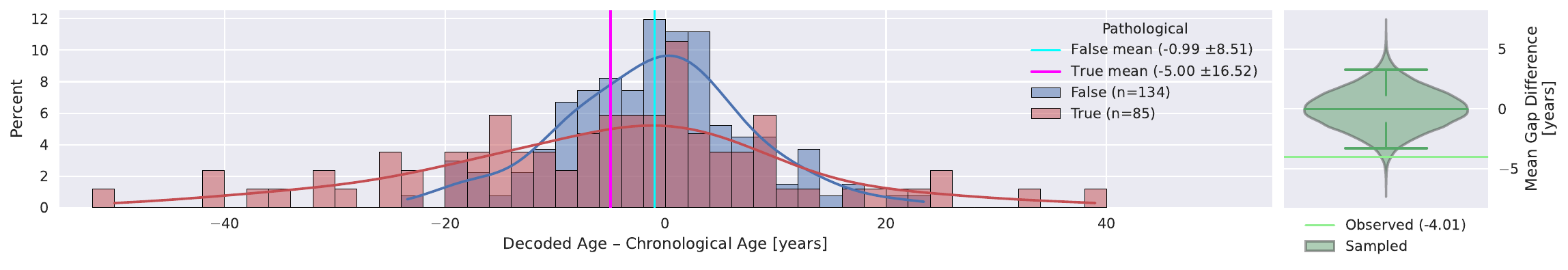}
    \end{minipage}
    \caption{Comparison of brain age gaps of non-pathological and pathological subjects in TUAB through a histogram (left) and a permutation test of average gap differences (right). The cyan and magenta lines represent the average brain age gap of non-pathological and pathological subjects, respectively. The light green line represents the actual observed difference in brain age gap between non-pathological and pathological subjects whereas the green violin represents the distribution sampled as part of a permutation test. Indicators of distribution median, 2.5th and 97.5th percentile are shown in darker green. 
    Average brain age gaps are negative while the one of pathological subjects is smaller. The variance of brain age gaps in pathological subjects is about twice as high compared to non-pathological subjects. The difference of average gaps (-4) between pathological and non-pathological subjects is statistically significant (permutation test, $p \leq 0.016$).
    }
    \label{fig:fe_age_gap_histogram}
\end{figure}

Figure~\ref{fig:fe_age_gap_histogram} shows a comparison of brain age gaps of non-pathological vs. pathological subjects of TUAB, so basically a comparison of Findings~1 and 2. We find statistically significant differences (permutation test, $p \leq 0.016$) in average brain age gaps of non-pathological compared to pathological subjects (Figure~\ref{fig:fe_age_gap_histogram} to the right). 

This finding supports and confirms the previous one: BA of pathological subjects is not only underestimated regarding the CA, the underestimation is significantly stronger than the one of non-pathological subjects. Again, this is not in agreement with the state hypothesis while it can be consistent with the trait hypothesis.

\subsection*{Finding 4: Brain Age Gap Biomarker is Not Indicative of Pathology in TUAB}
\label{sec:tuab_proxy}
\begin{figure}[ht!]
    \centering
    \begin{minipage}[c]{\linewidth}
    \centering
    \includegraphics[width=\linewidth]{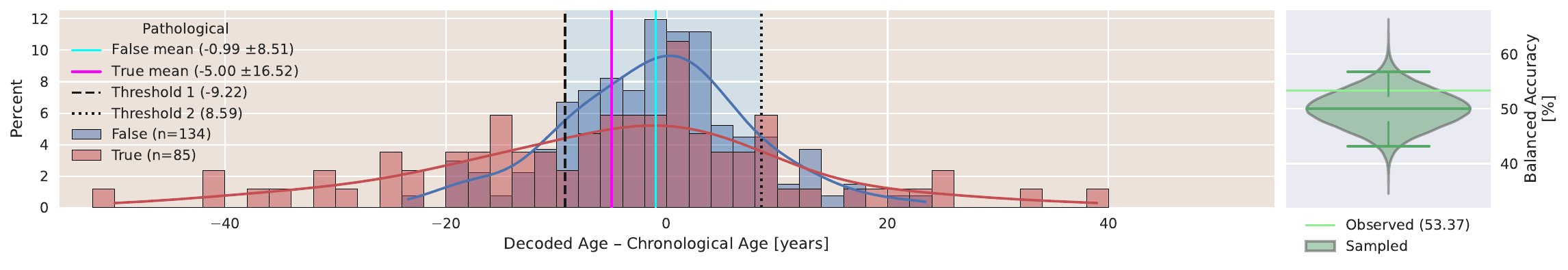}
    a) 
    \end{minipage}
    \begin{minipage}[c]{\linewidth}
    \centering
    \includegraphics[width=\linewidth]{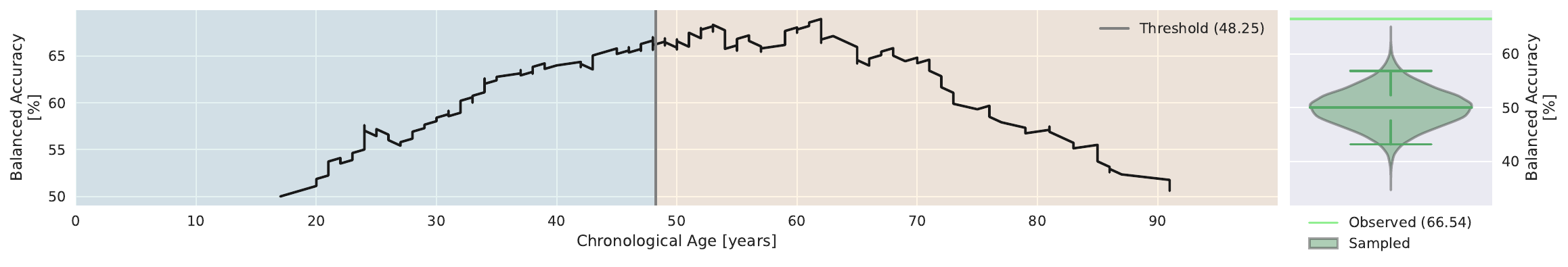}
    b) 
    \end{minipage}
    \caption{a) Histogram of brain age gaps of non-pathological and pathological subjects in TUAB (left) with permutation test of pathology proxy analysis (right). The cyan and magenta lines represent the average brain age gap of non-pathological and pathological subjects, respectively. Blue background signifies non-pathological classification and orange background signifies pathological classification based on thresholds found in CV presented as dashed lines. Light green lines represent actual observations whereas violins represent sampled distributions with indicators of median, 2.5th and 97.5th percentile in darker green. The variance of brain age gaps in pathological subjects is about twice as high compared to non-pathological subjects. The biomarker classification yielded a BACC which is not significantly better than chance level (permutation test, $p \leq 0.167$).
    b) Age threshold pathology proxy in TUAB (left) with permutation test of proxy score (right). Again, blue background signifies non-pathological classification and orange background signifies pathological classification with respect to the age threshold found in CV. The age classification yielded a BACC which is significantly better than chance (permutation test $p \leq 0.01$).
    }
\label{fig:fe_age_gap_proxy}
\end{figure}

Figure~\ref{fig:fe_age_gap_proxy}a) presents the brain age gap as a biomarker of EEG pathology which resulted in a BACC not significantly better than chance level (permutation test, $p \leq 0.167$).
Figure~\ref{fig:cv_age_thresh}b), on the contrary, shows an analysis of subject age as indicator of EEG pathology as a simple baseline for comparison which yielded a BACC significantly better than chance level (permutation test, $p \leq 0.01$).

Again, the presented finding in not in agreement with the state hypothesis but can be consistent with the trait hypothesis. If one observed a substantially higher brain age in pathological subjects, then, conversely, this implies that the brain age gap should also be indicative of EEG pathology. Again, the observations can be consistent with the trait hypothesis.
As one can achieve a better EEG pathology proxy score by simply exploiting the differences in age distributions between non-pathological and pathological subjects, we conclude that a classification of EEG pathology based on the brain age gap in its presented form is not promising.

\subsection*{Finding 5: Brain Age Gap Biomarker is Not Indicative of Pathology in RNP versus RP}
\label{sec:longitudinal_proxy}
\begin{figure}[htb!]
    \centering\includegraphics[width=\linewidth]{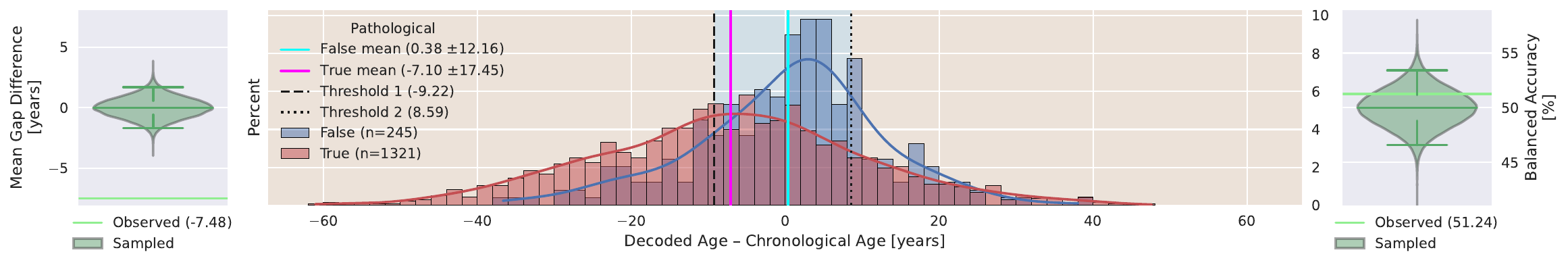}
    \caption{Histogram of brain age gaps of non-pathological and pathological subjects in RNP vs. RP (center) with permutation tests of average gap differences (left) and pathology proxy analysis (right). The cyan and magenta lines represent the average brain age gap of non-pathological and pathological subjects, respectively. Blue background signifies non-pathological classification and orange background signifies pathological classification based on thresholds found in CV presented as dashed lines. Light green lines represent actual observations whereas violins represent sampled distributions with indicators of median, 2.5th and 97.5th percentile in darker green. While there is a statistically significant difference (permutation test $p \leq 0.00001$) in average brain age gaps of subjects in RNP vs. RP, the brain age gap biomarker pathology classification does not reach a BACC statistically significantly better than chance level (permutation test, $p \leq 0.239$).
    }
    \label{fig:lnp_lp_proxy}
\end{figure}

Figure~\ref{fig:lnp_lp_proxy} presents brain age gaps of non-pathological versus pathological subjects in novel datasets RNP versus RP, whose average gaps are (as in TUAB) statistically significantly different (permutation test, $p \leq 0.00001$), while the brain age gap biomarker is (as in TUAB) not indicative of pathology (permutation test, $p \leq 0.239$).

Again, the presented finding is not in agreement with the state hypothesis but can be consistent with the trait hypothesis. The finding demonstrates reproducibility of the previous finding on TUAB on novel datasets with a larger population.

\subsection*{Finding 6: Brain Age Gap is Unaffected by Change in Pathology Status in TNPP and TPNP}
\label{sec:gap_unaffected}
\begin{figure}[htb!]
    \centering
    \begin{minipage}[c]{\linewidth}
    \centering
    \includegraphics[width=\linewidth]{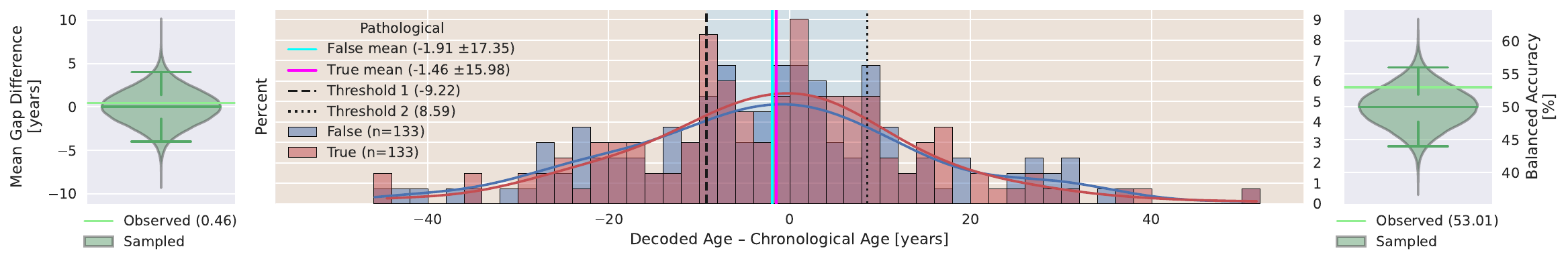}
    a)
    \end{minipage}
    \begin{minipage}[c]{\linewidth}
    \centering
    \includegraphics[width=\linewidth]{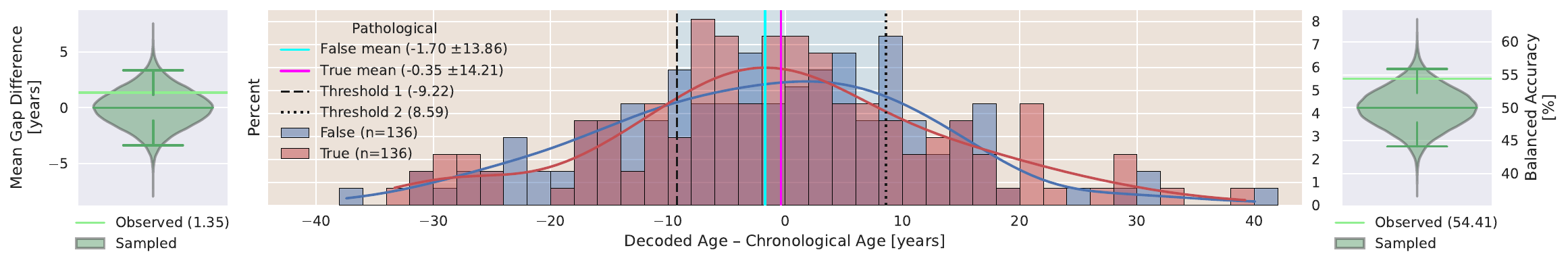}
    b)
    \end{minipage}
    \caption{Histogram of brain age gaps of non-pathological and pathological subjects in TNPP (top) and TPNP (bottom) with permutation tests of average gap differences (left) and pathology proxy analysis (right). The cyan and magenta lines represent the average brain age gap of non-pathological and pathological subjects, respectively. Blue background signifies non-pathological classification and orange background signifies pathological classification based on thresholds found in CV presented as dashed lines. Light green lines represent actual observations whereas violins represent sampled distributions with indicators of median, 2.5th and 97.5th percentile in darker green. The difference of average non-pathological and pathological gaps are not significant (permutation test, TNPP: $p \leq 0.825$, TPNP: $p \leq 0.43$). The application of the pathology proxy thresholds found in CV does not yield BACC scores significantly better than chance level (permutation test, TNPP: $p \leq 0.18$, TPNP: $p \leq 0.0788$).
    }
    \label{fig:longitudinal_age_gap_proxies}
\end{figure}

Figure~\ref{fig:longitudinal_age_gap_proxies}a) presents brain age gaps of non-pathological versus pathological subjects (center) in the novel dataset TNPP, whose average gaps (left) are not statistically significantly different (permutation test, $p \leq 0.825$) and where the brain age gap biomarker (right) is not indicative of pathology (permutation test, $p \leq 0.18$).
Figure~\ref{fig:longitudinal_age_gap_proxies}b) presents brain age gaps of non-pathological versus pathological subjects (center) in the novel dataset TPNP, whose average gaps (left) are not statistically significantly different (permutation test, $p \leq 0.43$) and where the brain age gap biomarker (right) is not indicative of pathology (permutation test, $p \leq 0.0788$).

The presented finding is not in agreement with the state hypothesis as we do not find a lager brain age gap in pathological subjects and as the brain age gap is not indicative of pathology neither in TNPP nor in TPNP. However, the finding supports the trait hypothesis as we do not find significant effects on the brain age gap by acquisition or recovery of EEG pathology.


\subsection*{Finding 7: Brain Age Change Rate in The "Moment of Transition" is Identical in TNPP and TPNP}
\label{sec:change_identical}
\begin{figure}[htb!]
    \centering
    \begin{minipage}[c]{.48\linewidth}
    \centering
    \includegraphics[width=\linewidth]{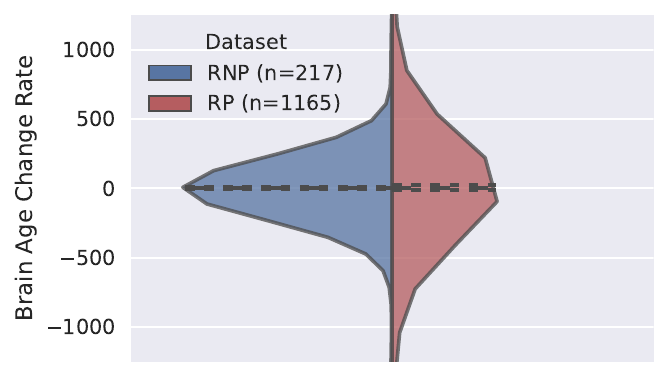}
    a)
    \end{minipage}
    \begin{minipage}[c]{.48\linewidth}
    \centering
    \includegraphics[width=\linewidth]{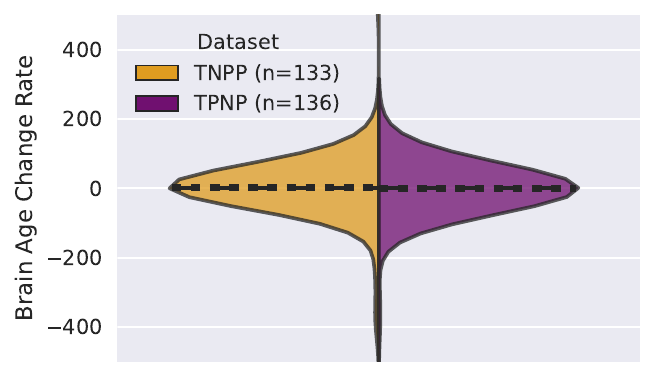}
    b)
    \end{minipage}
    \caption{a) Brain age change rate in RNP vs. RP. b) Brain age change rate in the "moment of transition" of EEG pathology label in TNPP vs. TPNP. While the distributions of RNP and RP are significantly different (KS test, $p \leq 0.003$), the distributions of TNPP and TPNP are actually the same (KS test, $p \leq 0.495$). There is neither evidence for a higher or lower brain age change rate in RNP vs. RP (WMW test, $p \leq 0.34$; Brunner-Munzel test, $p \leq 0.27$) nor in TNPP vs. TPNP (WMW test, $p \leq 0.13$; Brunner-Munzel test, $p \leq 0.13$).
    }
    \label{fig:running_age_gaps}
\end{figure}

Figure~\ref{fig:running_age_gaps}a) presents the brain age change rate in non-pathological versus pathological subjects in RNP versus RP. Whereas the distributions of brain age change rates are significantly different (KS  test, $p \leq 0.003$), there is no evidence that the brain age change rates in RNP are significantly higher or lower compared to RP (WMW test, $p \leq 0.34$; Brunner-Munzel test, $p \leq 0.27$). 
Figure~\ref{fig:running_age_gaps}b) presents the brain age change rate of subjects in TNPP versus TPNP in the "moment of transition". The distributions of TNPP and TPNP are actually the same (KS test, $p \leq 0.40$) and there is no evidence that the brain age change rates in the "moment of transition" in TNPP are significantly higher or lower compared to TPNP (WMW test, $p \leq 0.13$; Brunner-Munzel test, $p \leq 0.13$).

The presented finding is not in alignment with the state hypothesis and supports the trait hypothesis, since there is no evidence of increased or decreased brain age change rate induced by EEG pathology acquisition or recovery.

\subsection*{Finding 8: Increased Occipital Beta Activity Relates to Younger Brain Age in TUAB}
\label{sec:gradients}
\begin{figure}[htb!]
    \centering
    \includegraphics[width=\textwidth]{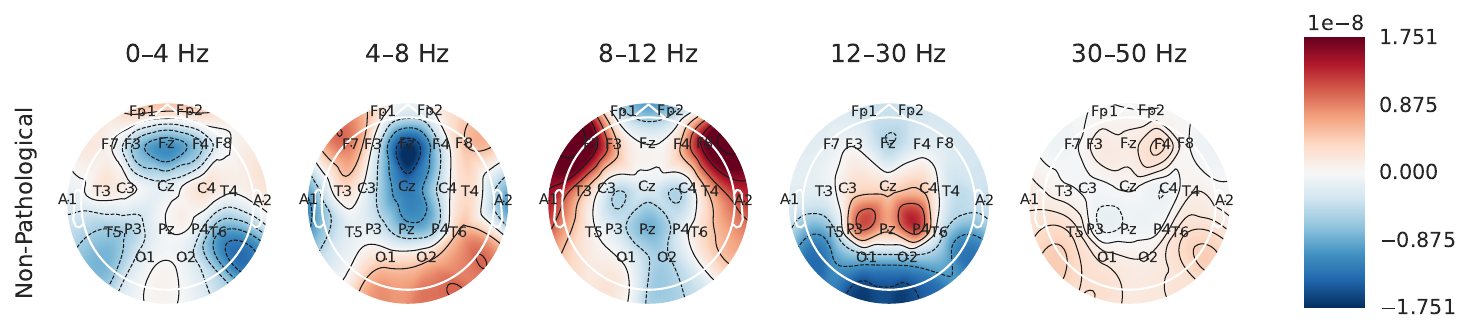}
    a)
    \includegraphics[width=\textwidth]{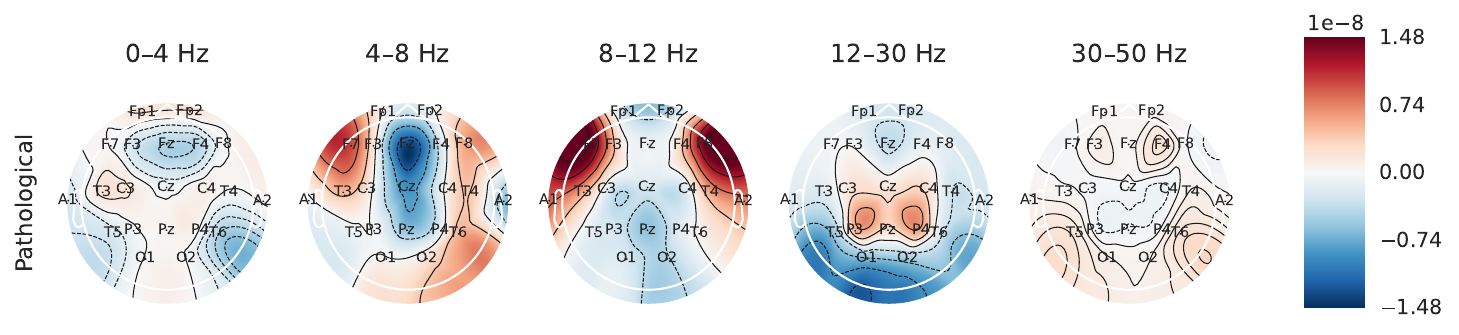}
    b)
    \includegraphics[width=\textwidth]{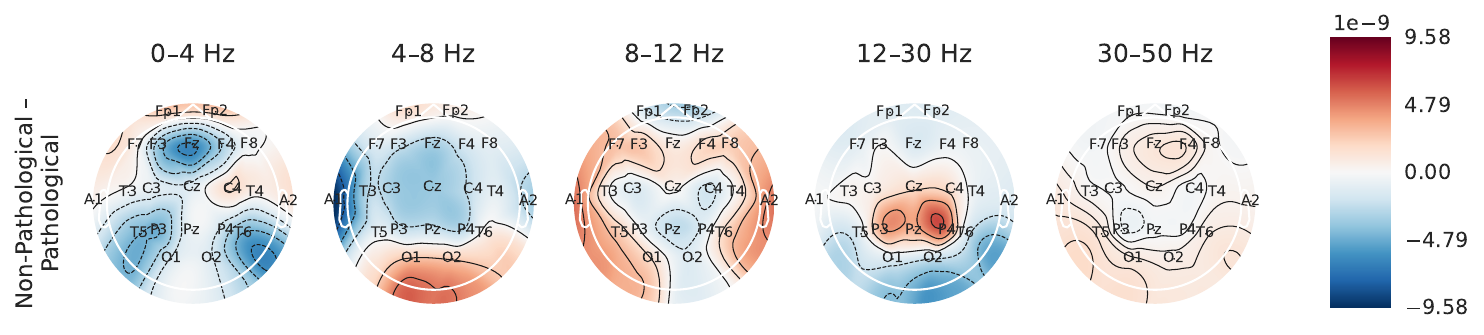}
    c)
    \caption{Amplitude gradient analysis with respect to different frequency bands on a) non-pathological, b) pathological, and c) difference of non-pathological to pathological subjects of TUAB after FE. Negative values imply lower brain age prediction while positive values imply higher brain age prediction. Prominent patterns across all frequency bands can be observed. In the delta frequency band, one can see three hotspots for negative gradients at frontocentral (F3, Fz, F4) and temporoparietal (T5, T6) electrode locations. In the theta frequency band, there are negative gradients at frontocentral (Fz) and vertex (Cz) electrode locations, whereas there are positive gradients at occipital (O1, O2) and frontotemporal (F7, F8) electrode locations. The latter are considerably bigger in alpha frequency range where there are also negative gradients at frontocentral (Fp1, Fp2), parietal (T5, T6) and occipital (O1, O2) electrode locations. In the beta frequency band, one can observe positive gradients at central (C3, Cz, C4) and parietal (P3, Pz, P4) electrode locations and surrounding negative gradients highest at occipital (O1, O2) and temporoparietal (T5, T6) electrode locations. In the gamma band, the lowest magnitude of gradients can be observed with an inverted delta pattern (positive gradients at frontocentral (F3, Fz, F4) and temporoparietal (T5, T6) electrode locations).}
    \label{fig:amplitude_perturbation}
\end{figure}

Figure~\ref{fig:amplitude_perturbation} presents topographical maps of gradients with the respect to the input amplitudes and shows strong and clear patterns across all frequency ranges. The largest absolute values can be observed in the theta, alpha, and beta frequency bands. Consequently, the network focuses less on delta and gamma bands to decode the age of subjects. A focus on the occipital brain region in the alpha band, which is one of the most prominent EEG features in healthy subjects, cannot be observed. This area is rather relevant in the beta frequency band.

\section{Discussion}
Our EEG brain age decoder achieved state-of-the-art performance of 6.6 years MAE on a subset of the largest resource of clinical EEG [\nameref{sec:sota_age_decoder}]. Although our model compared well to other established models on multiple datasets via the M/EEG brain age benchmark [\nameref{sec:benchmark_results}], we observed differences in decoding performance compared to the results listed in the original paper [\cite{engemann2022reusable}]. We hypothesize that these differences may arise from factors beyond our control, such as data releases, software and library updates, or varying seeds. Despite these differences, our training is stable, the model reaches loss convergence in CV (Figure~\ref{fig:learning_curves}) and FE (Figure~\ref{fig:fe_curves}) and outperforms the other contenders in three out of four cases, demonstrating its robustness. In general, we found it difficult to compare our results to those presented in the literature, even though similar works using EEG data of non-pathological subjects exist that additionally report the MAE score. While this metric may work well for in-dataset comparisons, we find it inadequate for cross-dataset comparisons as it does not consider the true age of subjects. The same absolute error may have a greater impact on a younger patient compared to an older patient (consider 10 years MAE for a six year old compared to a 60 year old). Furthermore, comparisons through the MAE metric seem absurd if a brain age gap is expected, since it automatically increases the MAE even if the BA is correctly decoded. Obvious alternatives like a absolute percentage error (APE) suffer from other limiting factors, e.g. the encouragement of underestimation if optimizing for low percentage error [\cite{mckenzie2011mean}], e.g. target 1, prediction 2, MAE 1, APE 100 vs. target 3, prediction 2, MAE 1, APE 33. Finding a reliable way to compare results, especially in settings where a brain age gap is expected, remains an open challenge for the brain age decoding community. For more insight on this topic please refer to [\cite{de2022mind}].

We consider the presented model together with the novel datasets [\nameref{sec:age_distributions}] sampled from TUEG the best possible foundation to date to investigate the EEG state and trait brain age hypotheses. However, there are some limiting factors, such as an unreliability both of subject age as well as of labels indicating pathology. As discussed in our previous work [\cite{gemein2020machine}], pathology labels in EEG diagnostics suffer from low inter-rate agreement [\cite{gaspard2014interrater}]. Furthermore, pathology labels were only provided for the TUAB subset of TUEG by TUH. The pathology labels of the remaining data were generated through an automatic routine [\cite{kiessner2023extended}]. While this routine yielded a conformity of 99\% when applied to TUAB as a sanity check, the labels generated for RNP, RP, TNPP, and TPNP for this study were not manually reviewed. Subject age, on the contrary, should be a more reliable target as birth date and age of subjects are typically well documented. However, in our age source analysis we found unresolved age mismatches [\nameref{sec:age_sources}] that could indicate severe mapping issues between EEG files and medical text reports within TUEG. Furthermore, the application of exclusion criteria, which we consider essential to better align with the TUAB dataset, resulted in a substantial decrease in recording and patient numbers [\nameref{sec:inclusion_exclusion}]. Despite this outstanding resource of clinical EEG, small sample sizes are a limiting factor. 

Despite removal of model bias [\nameref{sec:model_bias}], we found an underestimation of CA of the non-pathological population [\nameref{sec:sota_age_decoder}] as well as of the pathological population [\nameref{sec:pathological_underestimated}], where the latter is more extreme [\nameref{sec:patho_smaller_non_patho}]. These results were unexpected given the review of related literature. There, an increased brain age in pathological subjects is commonly reported and the state brain age hypothesis is predominant. Our findings are in disagreement with the state hypothesis.

Furthermore, we found that an application of the brain age gap as a biomarker of EEG pathology does not yield classification results significantly better than chance level [\nameref{sec:tuab_proxy}, \nameref{sec:longitudinal_proxy}, \nameref{sec:gap_unaffected}]. However, we argue that the non-pathological population might not serve well as a control group in this setting. This is because people came to the hospital for a reason, presumably suffering from a condition that prompted an EEG examination. In the RNP dataset, people even came to the hospital repeatedly, indicating the presence of a medical condition. Pathological subjects were examined only slightly (0.47 times) more often on average compared to non-pathological subjects [\nameref{sec:durations_and_intervals}]. Still, since on average there are several months between follow up EEG examinations in all datasets [\nameref{sec:durations_and_intervals}], we assume that the RNP population was not excessively misclassified. By all means, these results are again in disagreement with the state hypothesis, since, as before, it implies higher brain age in pathological subjects. Conversely, a larger brain age gap should be indicative of pathology. 


Last but not least, we did not find a significant difference in brain age change rate in the "moment of transition" in TNPP compared to TPNP [\nameref{sec:change_identical}].
(Note that the "moment of transition" actually refers to the time interval between subsequent EEG assessments where a label transition occurred and not to the actual moment of change of non-pathological to pathological brain activity, as this moment is unknown and likely fluent.) It is possible that the brain age increases or decreases significantly compared to prior label transition after some unknown time interval. An interesting direction for future research would be to investigate this time interval. The hypothesis could be investigated by curating additional data of subjects with repeated assessments and a selection of recordings based on short to long intervals between label transition and follow-up examinations. Our finding is not only in disagreement with the state hypothesis, but it additionally supports the trait hypothesis which implies no change through environmental factors or diseases acquisition and recovery.

Our analysis of amplitude gradients [\nameref{sec:gradients}] revealed symmetric patterns across all frequency ranges under investigation. The human EEG has a history of close to 100 years [\cite{haas2003hans}]. During this time, many researchers have investigated its properties regarding non-pathological and pathological as well as elderly and young subjects. For example, \cite{purdon2015ageing} have reported decreasing power in elderly compared to young subjects. One interesting line of further research would involve the application of EEG quantification methods to the presented datasets [\nameref{sec:longitudinal_datasets}] and a comparison to the presented gradient maps [\nameref{sec:gradients}]. Main interests include the differences between non-pathological and pathological EEG and age-related signal alteration. For a review of literature that analysed EEG frequencies in healthy elderly that could serve as a starting point for future analyses, please refer to [\cite{mizukami2017eeg}].

\section{Conclusion and Outlook}
The fundamental interpretation of biological vs. chronological brain age has recently been disputed by \cite{vidal2021individual}. Based on an analysis of MRI data, the authors concluded that the difference between biological brain age and chronological age is a rather stable individual trait across a lifetime, and not subject to substantial changes in response to events such as the occurrence of severe neurological disorders. In this study, we tested these state versus trait hypotheses of brain age dynamics using a different, important, and widely used modality for non-invasive assessment of the human brain: clinical EEG recordings. We found no evidence supporting the state brain age hypothesis and made observations to the contrary. In other words, our findings lend support to the trait hypothesis of brain-age dynamics. Thus, our study provides findings complementary to previous MRI-based observations. However, neither our findings nor these previous findings rule out the possibility that in addition to predominant trait-like brain age features, more subtle brain-age-related signal components, and/or putative signal components which might not be picked up by the machine learning methods used, might follow a variable, brain-state-like pattern. Therefore, further research on brain age in different imaging modalities is required. In this study, we leveraged the TUEG Corpus as the largest available clinical EEG dataset. However, this dataset was not curated specifically for research into the nature of brain age and consequently, limitations exist in terms of provided labels and criteria for subject inclusion and exclusion. For more nuanced investigations of the nature of brain age, an expanded dataset featuring a larger number of subjects with repeated EEG examinations and a larger number of EEG sessions per subject would be highly beneficial. This should ideally include non-binary pathology scores compiled by multiple reliable experts, along with accurate age information that maintains anonymity. Such a dataset could offer deeper insights into brain aging and shed more light on the state versus trait components (and also possibly their interplay) of brain age changes.

\section*{Ethical Approval}
No subjects were recruited or tested in experiments for this study. Data acquisition was performed by the Temple University Hospital, Philadelphia, Pennsylvania, United States of America.

\section*{Data and Code Availability}
The TUH Abnormal EEG Corpus [\cite{abnormalLopez}] used to train the models for our study is a subset of the TUH EEG Corpus [\cite{obeid2016temple}, DOI: 10.3389/fnins.2016.00196].
Both data sets are publicly available for download upon registration at \url{isip.piconepress.com/projects/tuh\_eeg/html/downloads.shtml}. 
Additionally, we created several subsets of the TUH EEG Corpus. We provide all required detail to recreate these datasets.
Code specific to the experiments performed for the presented study as well as results were uploaded to \url{github.com/gemeinl/eeg-brain-age}.

\section*{CRediT Author Contributions}
\begin{itemize}
    \item[] \textbf{Lukas A. W. Gemein} \textit{Conceptualization, Formal analysis, Investigation, Methodology, Software, Validation, Visualization, Writing – original draft, Writing – review \& editing}
    \item[] \textbf{Robin T. Schirrmeister} \textit{Conceptualization, Methodology, Supervision, Validation, Writing – review \& editing}
    \item[] \textbf{Joschka Boedecker} \textit{Project administration, Supervision}
    \item[] \textbf{Tonio Ball} \textit{Conceptualization, Methodology, Funding acquisition, Project administration, Resources, Supervision, Writing – review \& editing}
\end{itemize}

\section*{Funding}
This work was partly supported by AI-Trust (Baden-W\"urttemberg Foundation), Renormalized Flows (BMBF project 01IS19077C), and AI-Cog (DFG project BA 4695/4-1).

\section*{Conflict of Interest}
The authors declare no competing financial interests.


\bibliographystyle{apalike}
\bibliography{bibliography}

\appendix
\beginsupplement

\section{Supplement}
\subsection{Recording Artifacts in TUAB}
\label{sec:tuab_outliers}
\begin{figure}[htb!]
    \centering
    \includegraphics[width=\textwidth]{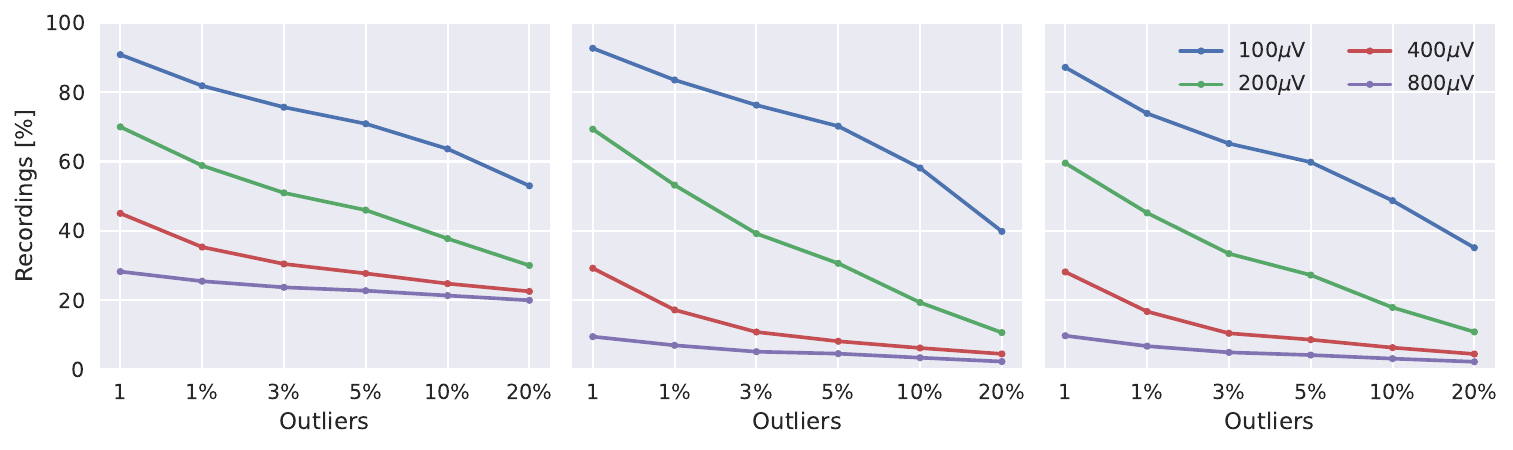}
    \caption{Unphysiologically high values at the start (first three minutes) of recordings in TUAB. From left to right: Outliers in the first, second, and third minute of recordings. In the first minute, 20\% of values are outliers in 20\% of the recordings. There is a decrease in outliers above all voltage thresholds (100, 200, 400, and 800$\mu V$) from first over second to third minute. The decrease is most prominent in high voltages (400 and 800$\mu V$) from first to second minute.}
    \label{fig:tuab_outliers}
\end{figure}

As in our previous work [\cite{gemein2020machine}], we removed outliers at the start of EEG recordings during preprocessing (Section~\ref{sec:preprocessing}). To substantiate this decision, Figure~\ref{fig:tuab_outliers} shows an analysis of the first three minutes of recordings in TUAB. One can see that the number of outliers $\pm800\mu$V is highest in the first minute and decreases in the following minutes. We assume that some readjustment of the electrode cap or finding a comfortable seating position takes place in the first minute and hence causes these outliers.

\subsection{Descriptive Statistics}
\subsubsection{Recording Durations}
\begin{figure}[htb!]
    \centering
    \includegraphics[width=\textwidth]{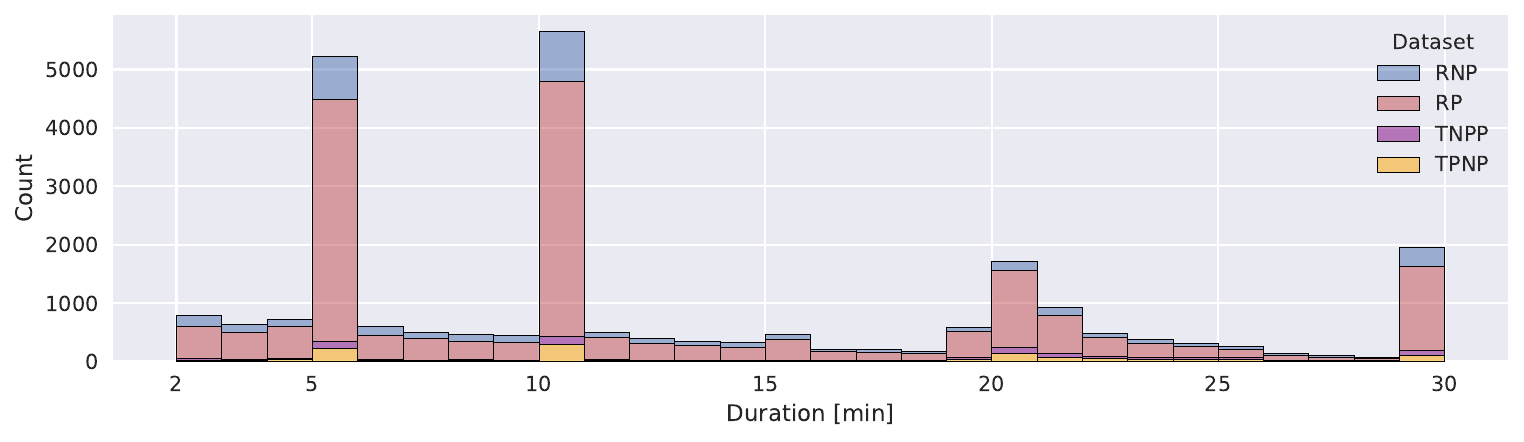}
    \caption{Durations of recordings in RNP, RP, TNPP, and TPNP. The last bin contains all recordings with a duration of 30 minutes or longer. There are prominent peaks at 5 and 10 minutes and there is a less prominent peak at 20 minutes.}
    \label{fig:longitudinal_durations}
\end{figure}

Whereas recordings of TUAB were selected to have a duration of at least 15 minutes [\cite{abnormalLopez}], this does not apply to the entirety of recordings contained in TUEG and hence our novel datasets.
Therefore, we show the distribution of recording duration in RNP, RP, TNPP, and TPNP in Figure~\ref{fig:longitudinal_durations}. We introduce a minimal recording duration of two minutes as we drop the first minute of recordings during preprocessing, due to higher amount of outliers (Section~\ref{sec:tuab_outliers}). To better align with TUAB, we made selections of recordings based on duration in our decoding pipeline (Section~\ref{sec:longitudinal_datasets}).

\subsubsection{Non-pathological Subset of TUAB}
\begin{figure}[htb!]
    \centering
    \begin{minipage}[c]{.49\linewidth}
    \centering
    \includegraphics[width=\linewidth]{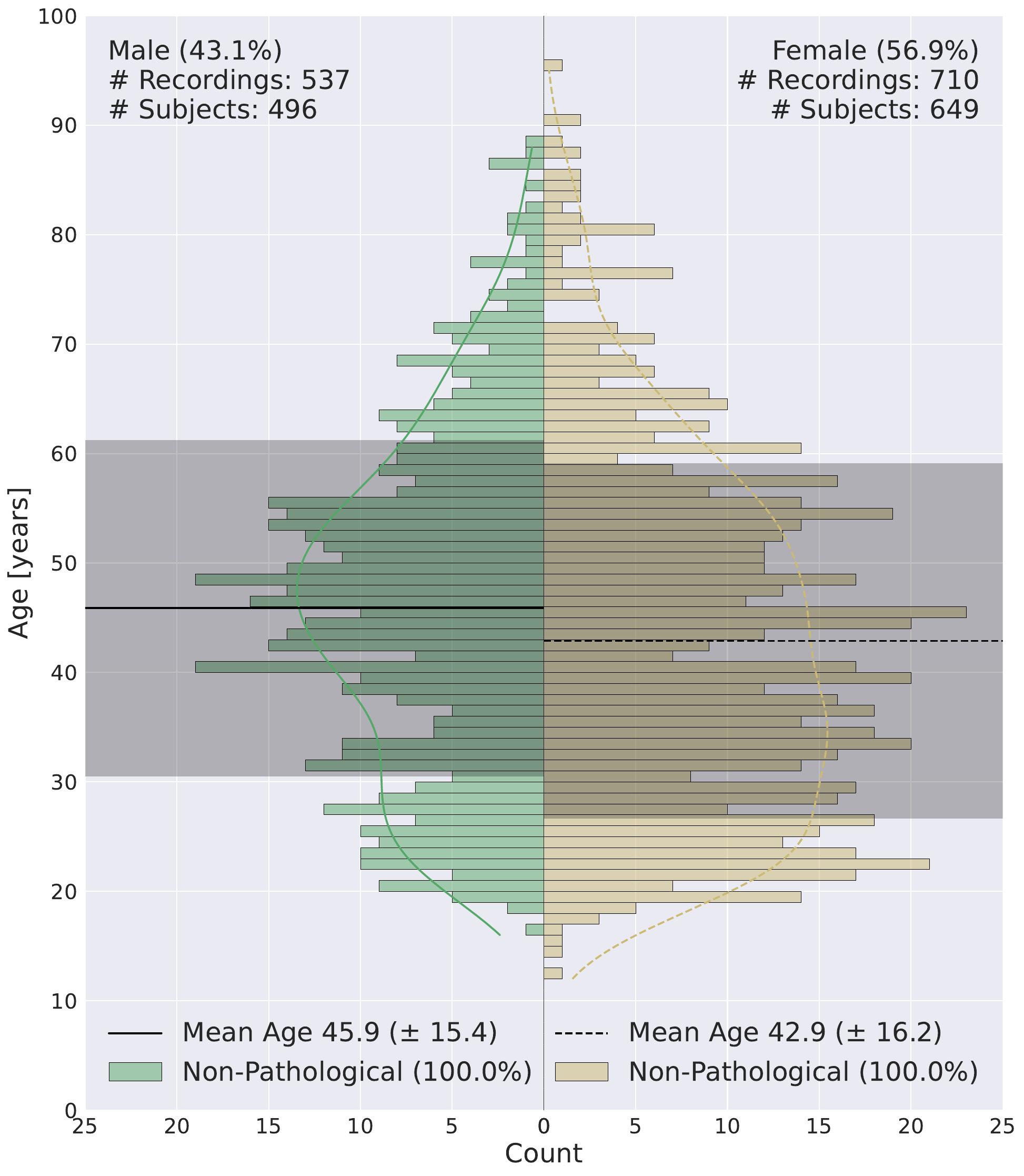}
    a) Training
    \end{minipage}
    \begin{minipage}[c]{.49\linewidth}
    \centering
    \includegraphics[width=\linewidth]{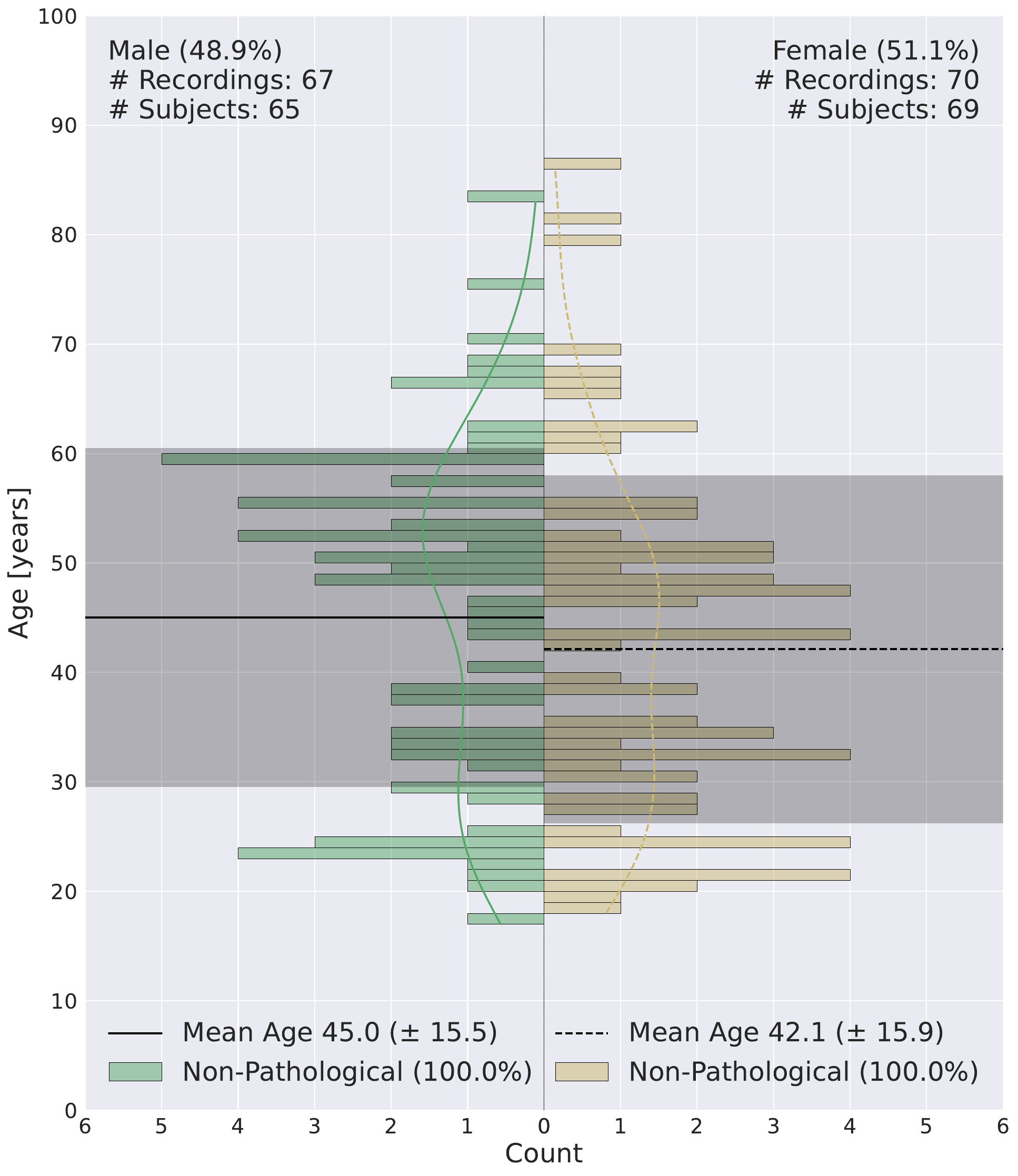}
    b) Final evaluation
    \end{minipage}
    \caption{Age pyramid of a) training and b) final evaluation split of non-pathological recordings in TUAB. There is a higher amount of female recordings and subjects (710/649) in the training split compared to the final evaluation split (70/69). Average ages and standard deviation are similar between the data splits.}
    \label{fig:tuab_age_pyramid}
\end{figure}

We present an age pyramid of the non-pathological recordings in TUAB in Figure~\ref{fig:tuab_age_pyramid}. The histograms show that the distribution of ages is similar in the training and final evaluation sets, which was the main reason we re-split the dataset. For a visualization of the entire dataset including pathological subjects, please refer to \cite{gemein2020machine}. 

\subsubsection{CV Splits}
\begin{figure}[htb!]
    \centering
    \includegraphics[width=\textwidth]{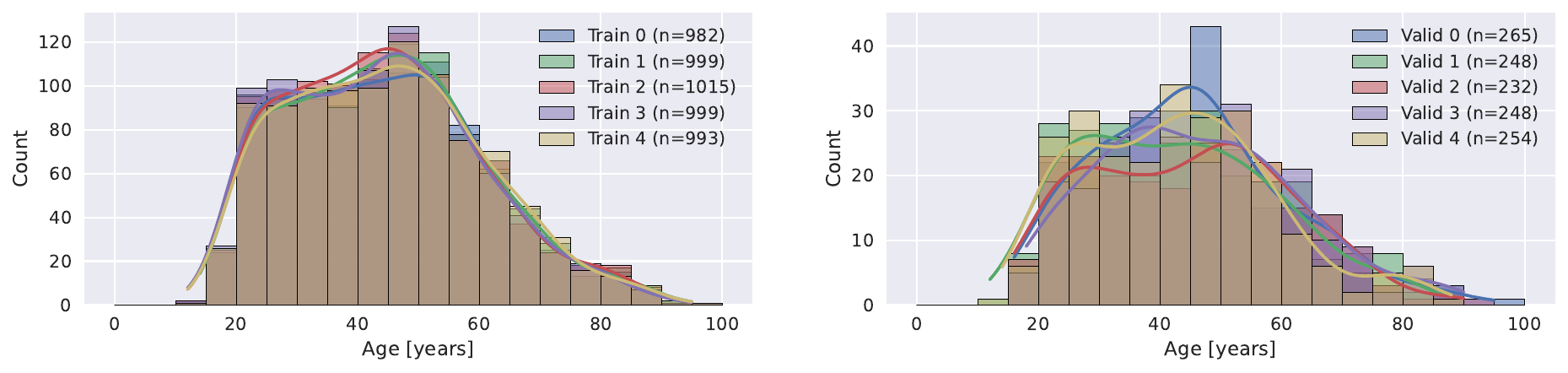}
    \caption{Left: Age distributions of non-pathological recordings in training splits of 5-fold CV. Right: Age distributions of non-pathological recordings in validation splits of 5-fold CV.}
    \label{fig:tuab_cv_age_histograms}
\end{figure}

We present age histograms of our CV splits in Figure~\ref{fig:tuab_cv_age_histograms}. For CV we can observe slight variations in age distribution, especially in the validation set. The relatively high variance regarding number of recordings is caused by the subject-wise data splitting procedure.

\subsubsection{Age Pyramids}
\begin{figure}[htb!]
    \centering
    \includegraphics[width=.48\textwidth]{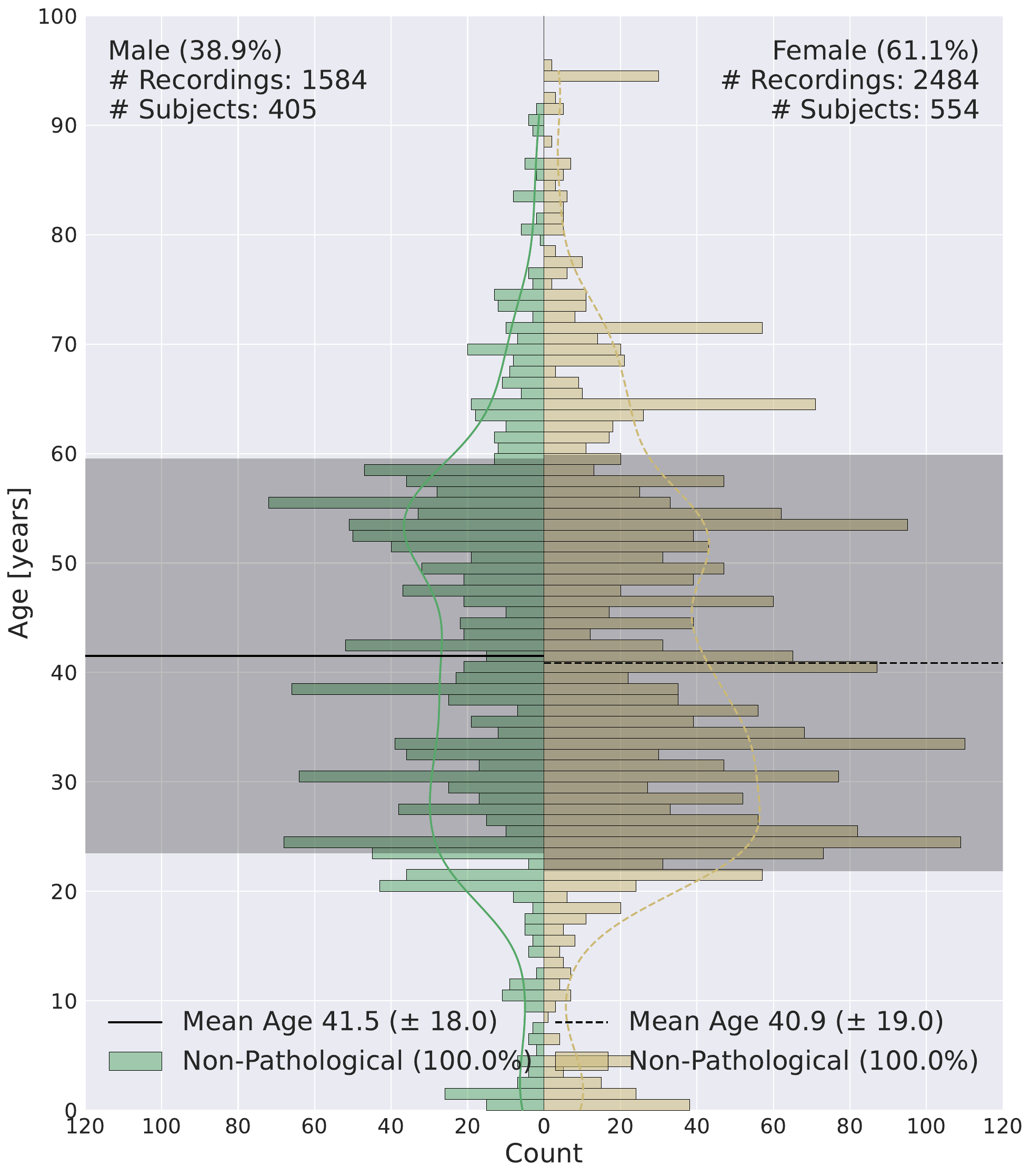}
    \includegraphics[width=.48\textwidth]{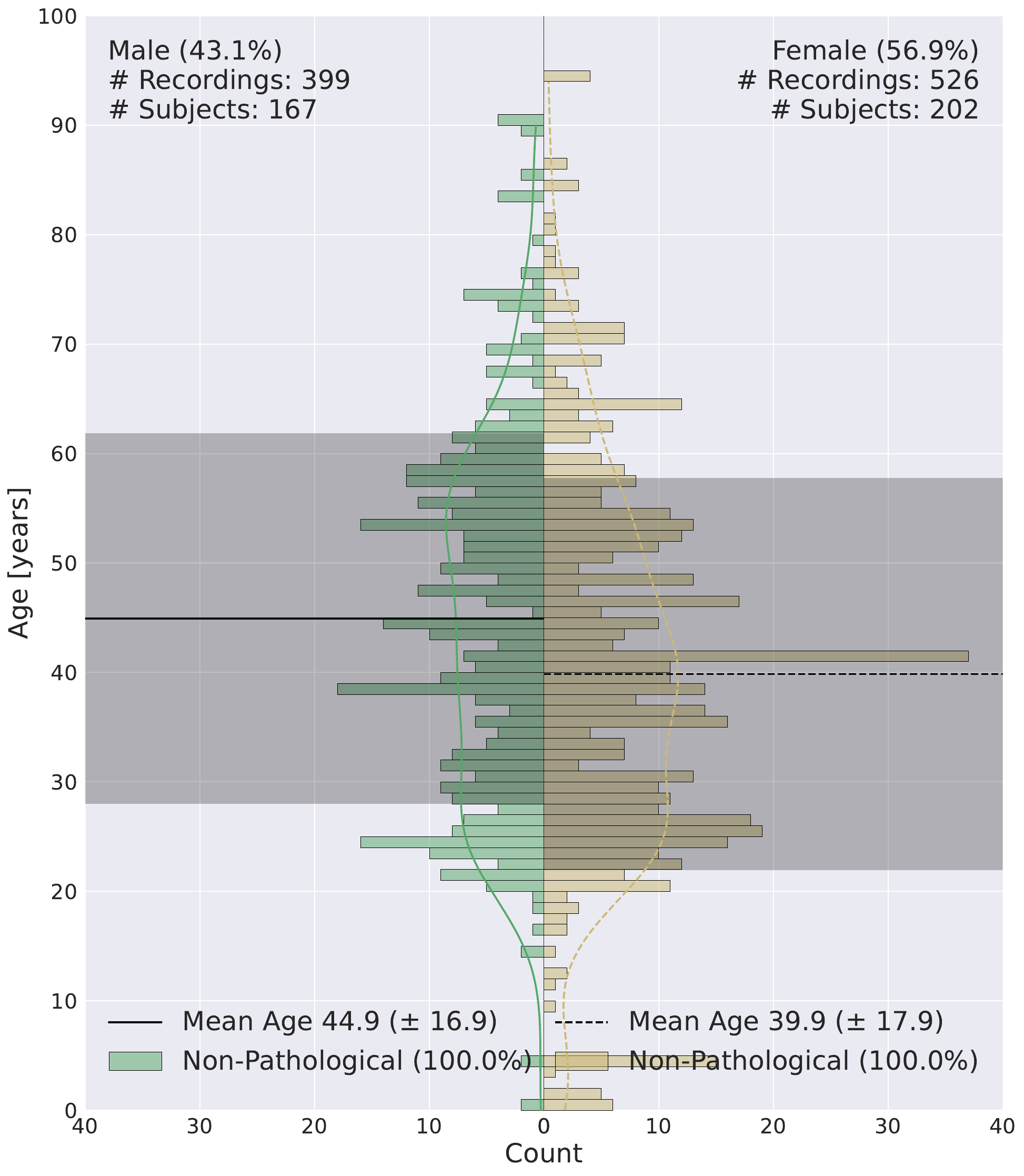}
    \caption{RNP. Left: 2 min \& dirty ages. Right: 15 min \& dirty ages.}
    \label{fig:lnp_pyramids}
\end{figure}

\begin{figure}[htb!]
    \centering
    \includegraphics[width=.48\textwidth]{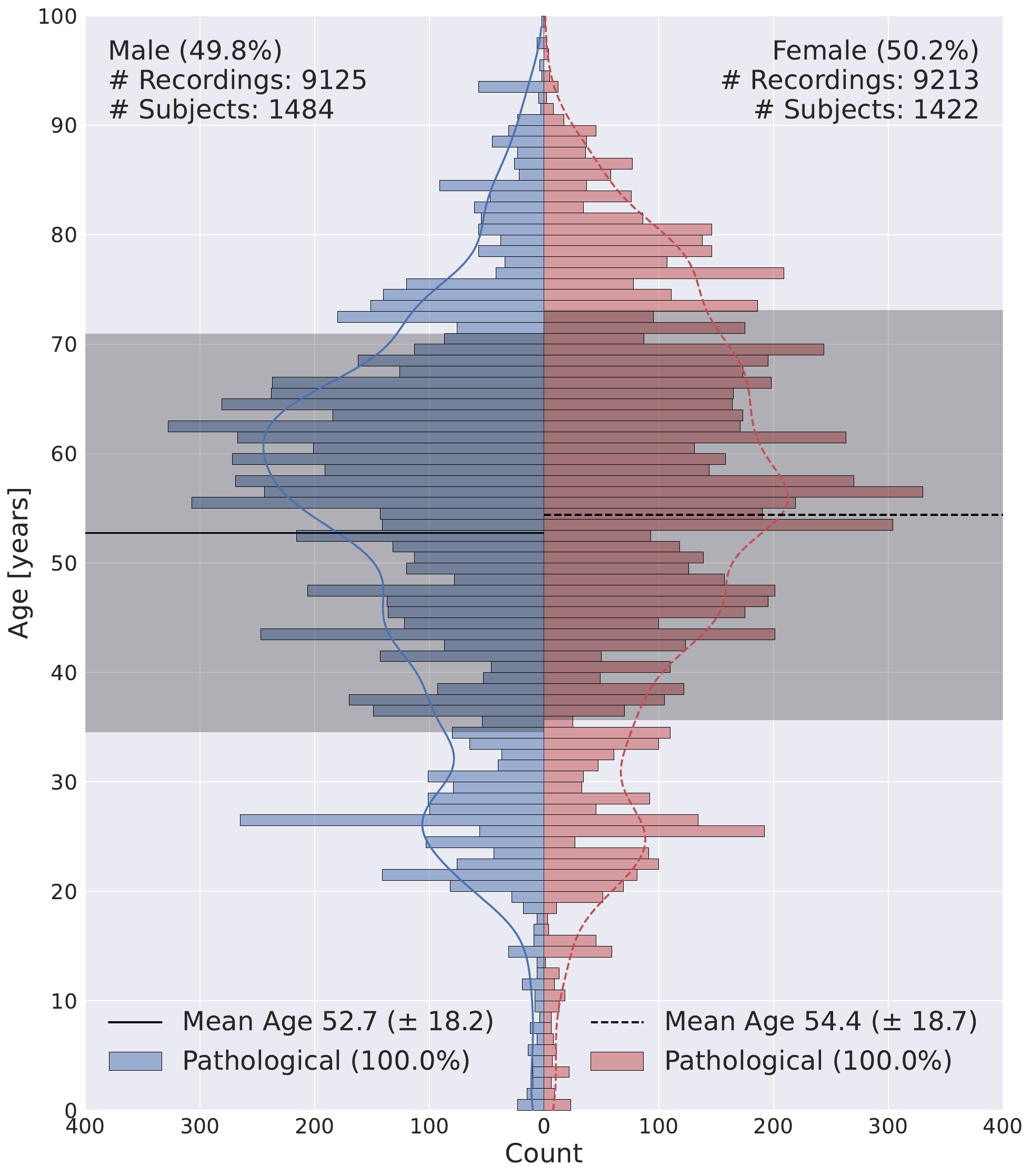}
    \includegraphics[width=.48\textwidth]{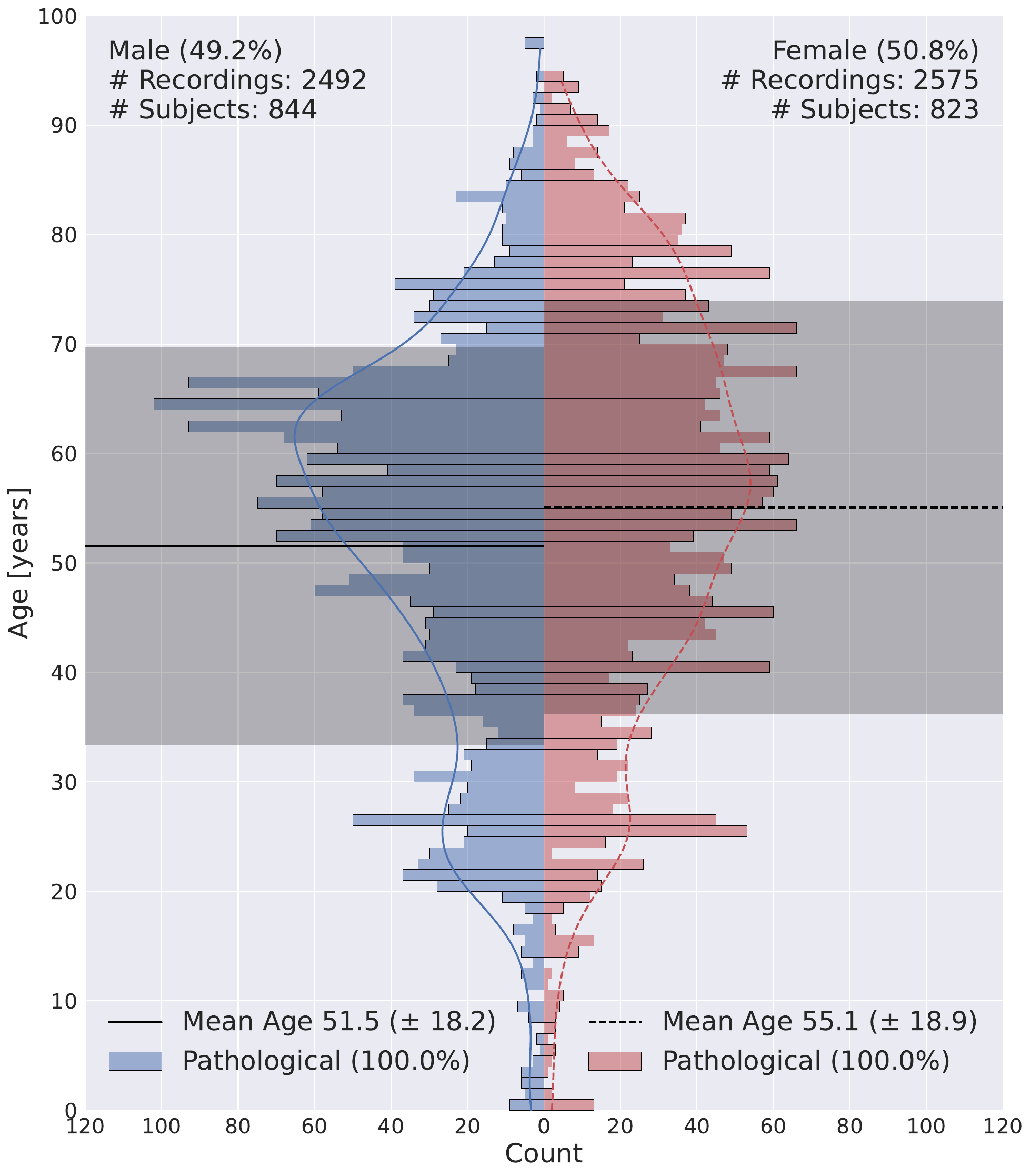}
    \caption{Left: 2 min \& dirty ages. Right: 15 min \& dirty ages.}
    \label{fig:lp_pyramids}
\end{figure}

\begin{figure}[htb!]
    \centering
    \includegraphics[width=.48\textwidth]{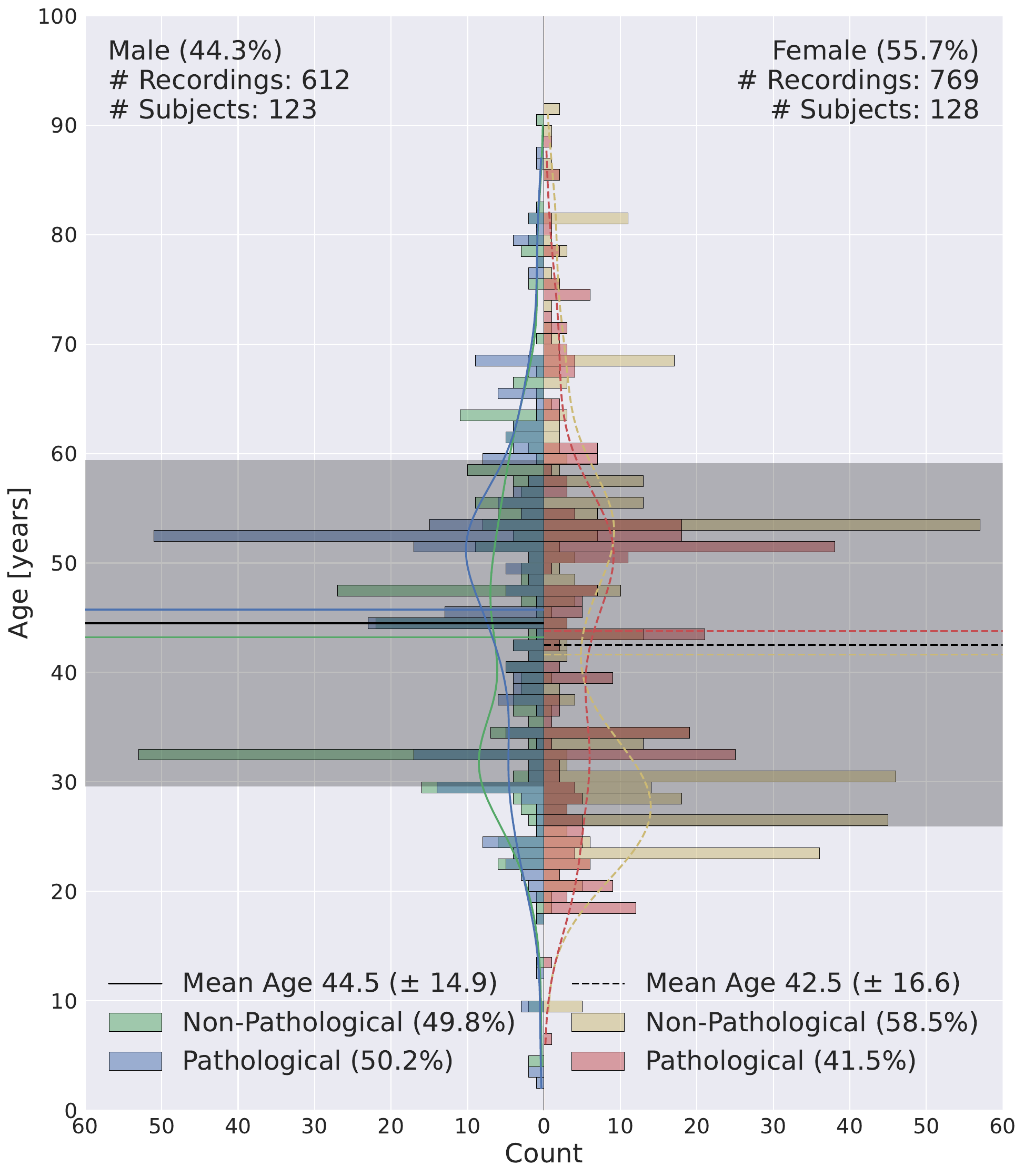}
    \includegraphics[width=.48\textwidth]{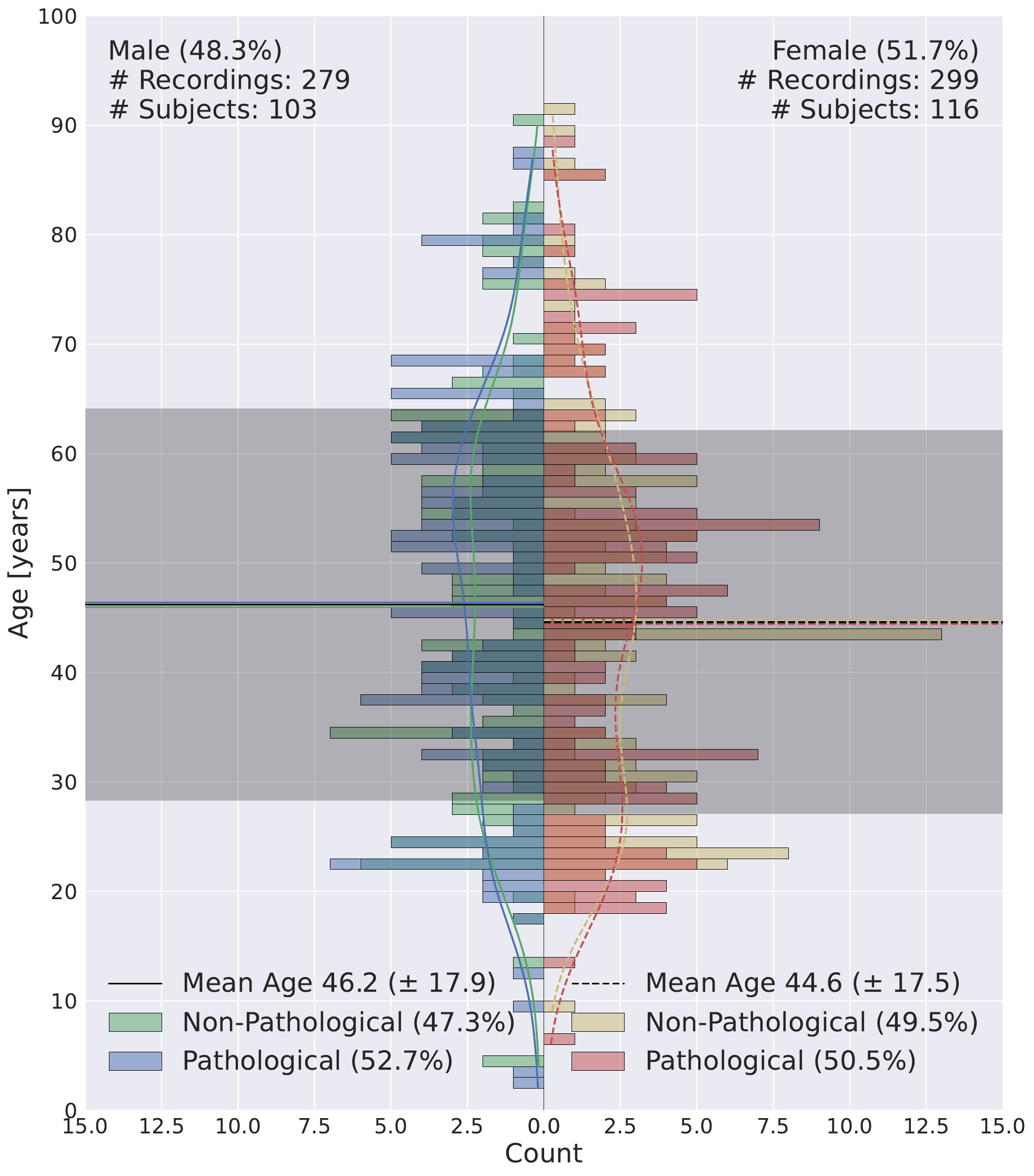}
    \caption{TNPP. Left: 2 min \& dirty ages. Right: 15 min \& dirty ages.}
    \label{fig:lnpp_pyramids}
\end{figure}

\begin{figure}[htb!]
    \centering
    \includegraphics[width=.48\textwidth]{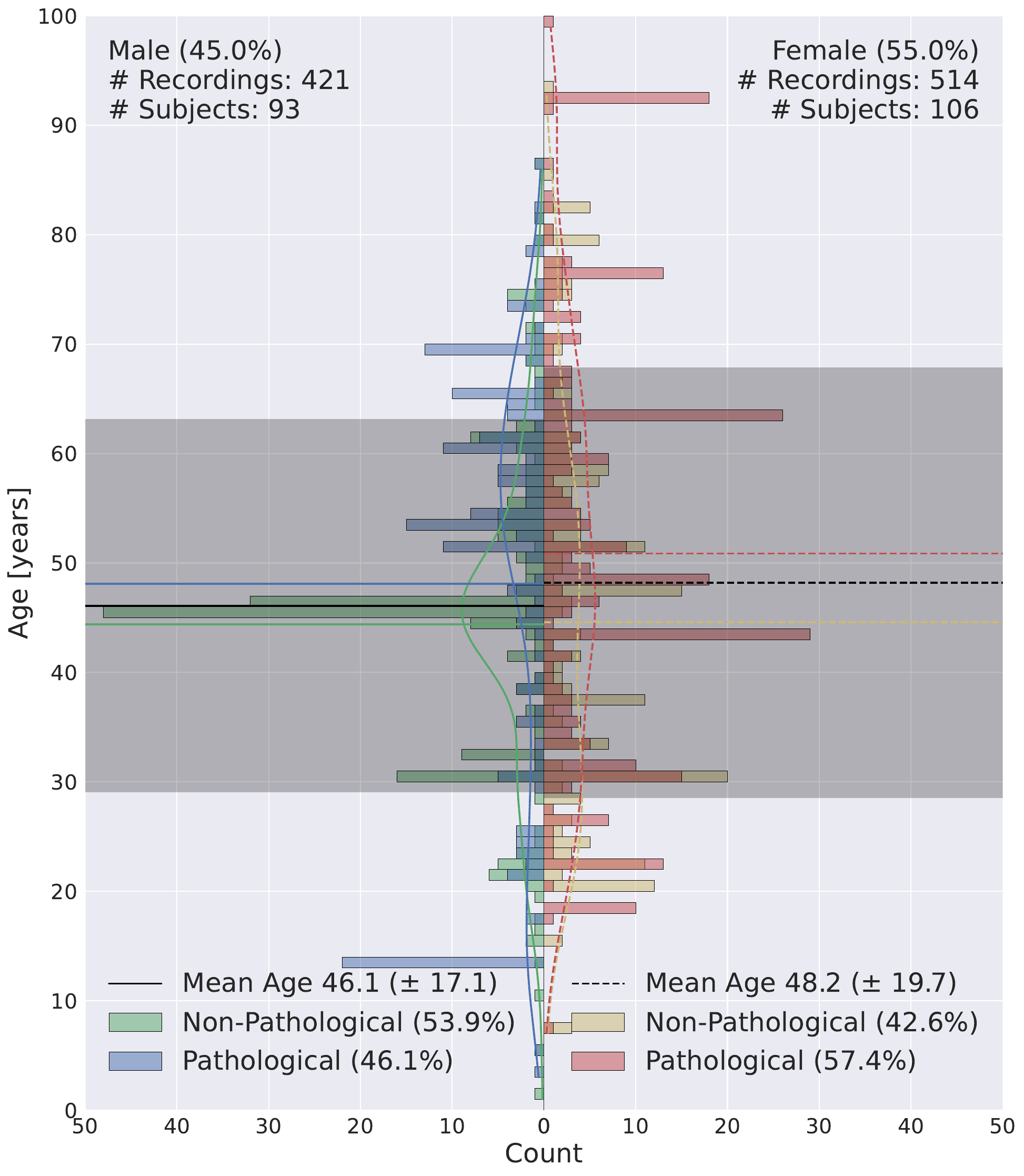}
    \includegraphics[width=.48\textwidth]{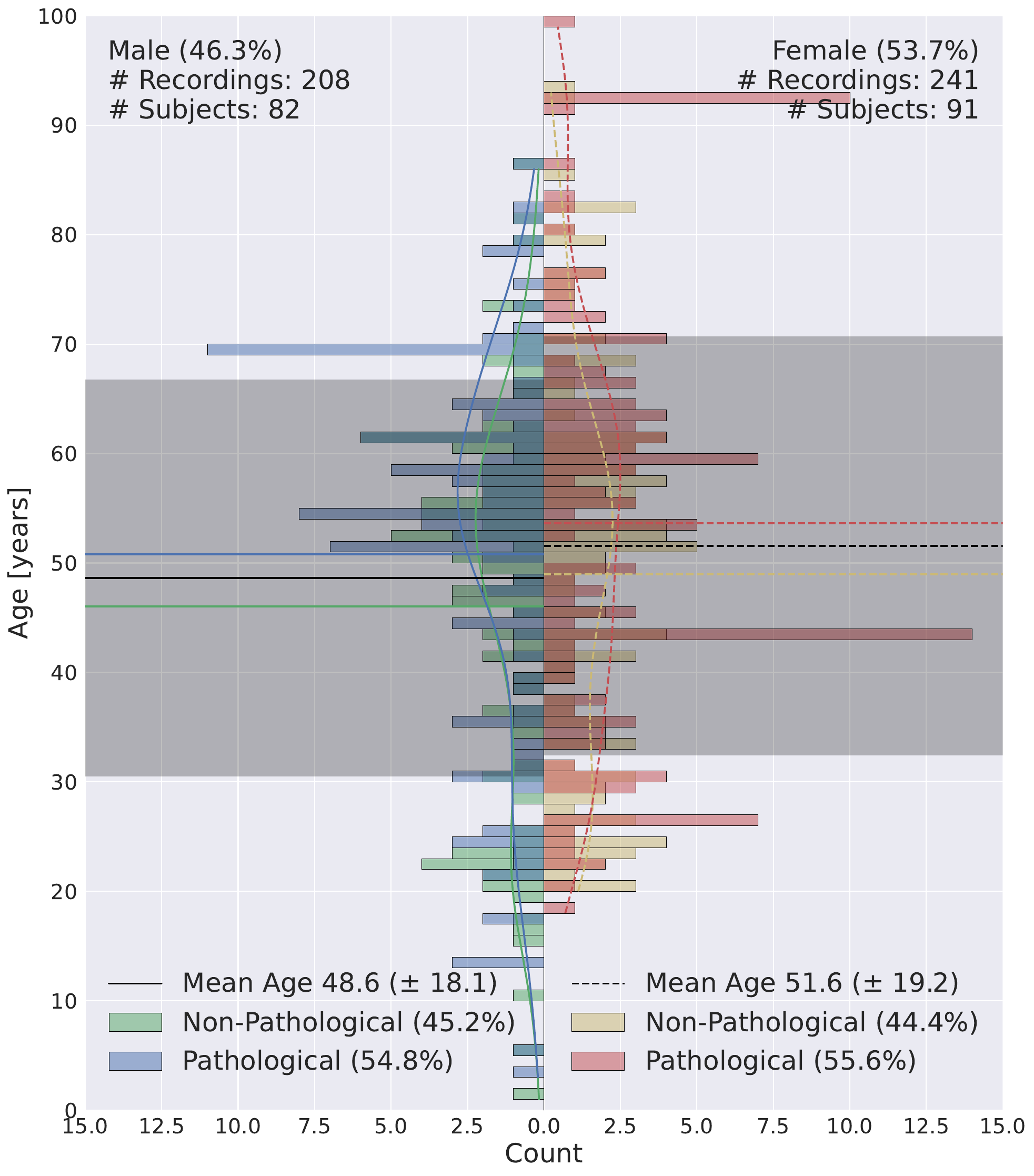}
    \caption{TPNP. Left: 2 min \& dirty ages. Right: 15 min \& dirty ages.}
    \label{fig:lpnp_pyramids}
\end{figure}

We present age pyramids of the datasets with repeated examinations that resulted from different inclusion criteria [Section~\ref{sec:inclusion_exclusion}] in Figures~\ref{fig:lnp_pyramids}–\ref{fig:lpnp_pyramids}.

\subsubsection{Recordings per Subject and Intervals}
\begin{figure}[htb!]
    \centering
    \begin{minipage}[c]{.48\linewidth}
    \centering
    \includegraphics[width=\linewidth]{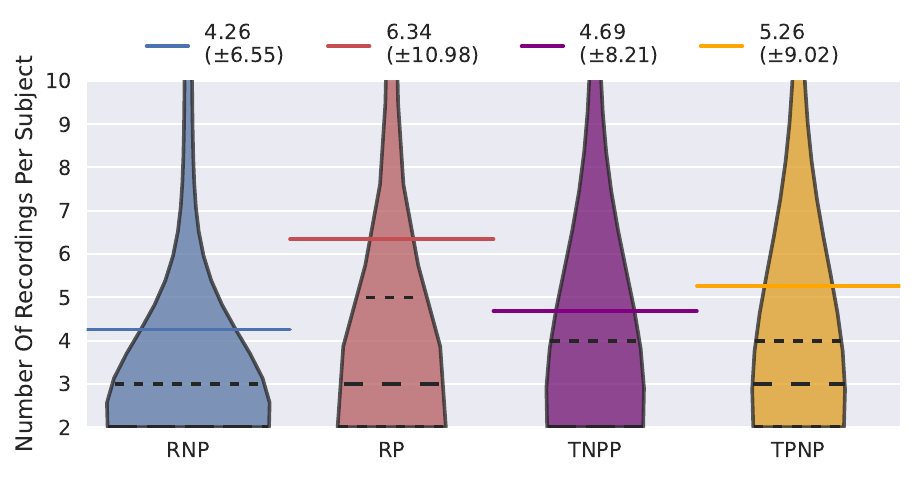}
    \end{minipage}
    \begin{minipage}[c]{.48\linewidth}
    \centering
    \includegraphics[width=\linewidth]{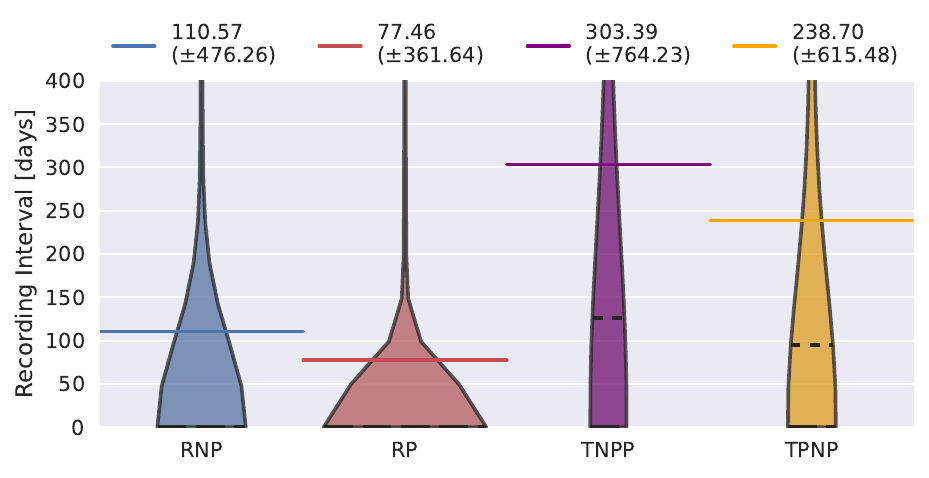}
    \end{minipage}
    \begin{minipage}[c]{.48\linewidth}
    \centering
    \includegraphics[width=\linewidth]{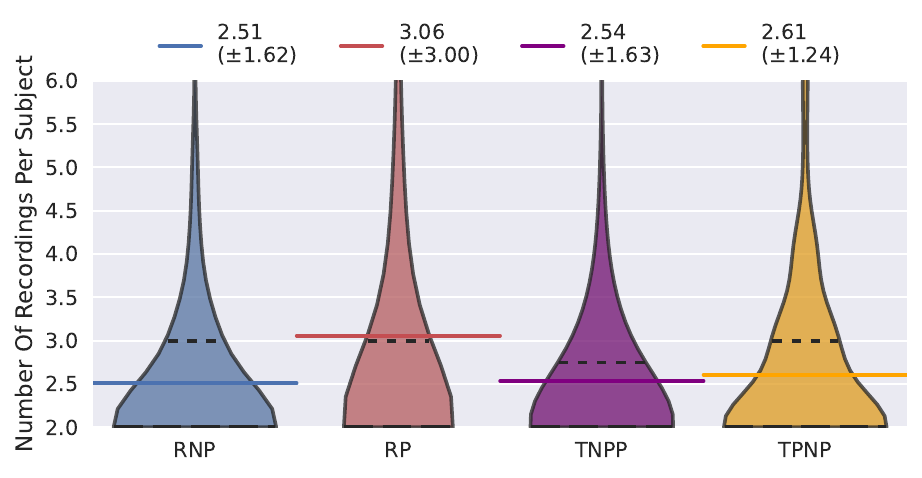}
    a) Number of recordings per subject
    \end{minipage}
    \begin{minipage}[c]{.48\linewidth}
    \centering
    \includegraphics[width=\linewidth]{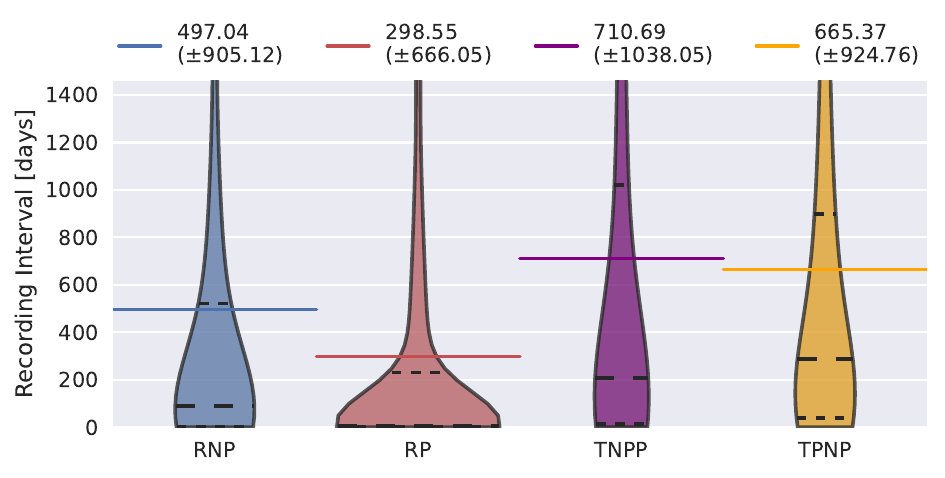}
    b) Recording intervals
    \end{minipage}
    \caption{Top: $\geq$ 2 min \& dirty ages. Bottom: $\geq$ 15 min \& dirty ages. The longer the recordings the less recordings per subject with longer intervals in between. The shorter the recordings the more recordings per subject with shorter intervals in between.}
    \label{fig:longitudinal_rec_intervals_short_unclean}
\end{figure}

We present number of recordings per subject and recording intervals for datasets RNP, RP, TNPP, and TPNP in Figure~\ref{fig:longitudinal_rec_intervals_short_unclean}.

\subsection{Hyperparameters}
In Table~\ref{tab:model_hyperparameters} we present our model hyperparameters. Additional hyperparameter choices are listed in Table~\ref{tab:more_hyperparameters}.
\label{sec:hyperparameters}
\begin{table}[htb!]
    \centering
    \begin{tabular}{c|c}
         batch\_size & 128 \\
         max\_epochs & 35 \\
         n\_chans & 21 \\
         dropout\_prob & 0.0195875974361336 \\
         init\_lr & 0.0004732953501425473 \\
         weigth\_decay & 1.0025958447703478e-07 \\
         n\_filters & 53 \\
         n\_blocks & 5 \\
         kernel\_size & 9 \\
    \end{tabular}
    \caption{Model hyperparameters.}
    \label{tab:model_hyperparameters}
\end{table}

\begin{table}[htb!]
    \centering
    \begin{tabular}{c|c}
         input\_time\_length & 6000 \\
         loss & l1 \\
         channel\_dropout\_prob & .2 \\
    \end{tabular}
    \caption{More hyperparameters.}
    \label{tab:more_hyperparameters}
\end{table}

\subsection{Cross-validation Results}
\label{sec:cv_results}
\begin{figure}[htb!]
    \centering
    \includegraphics[width=\textwidth]{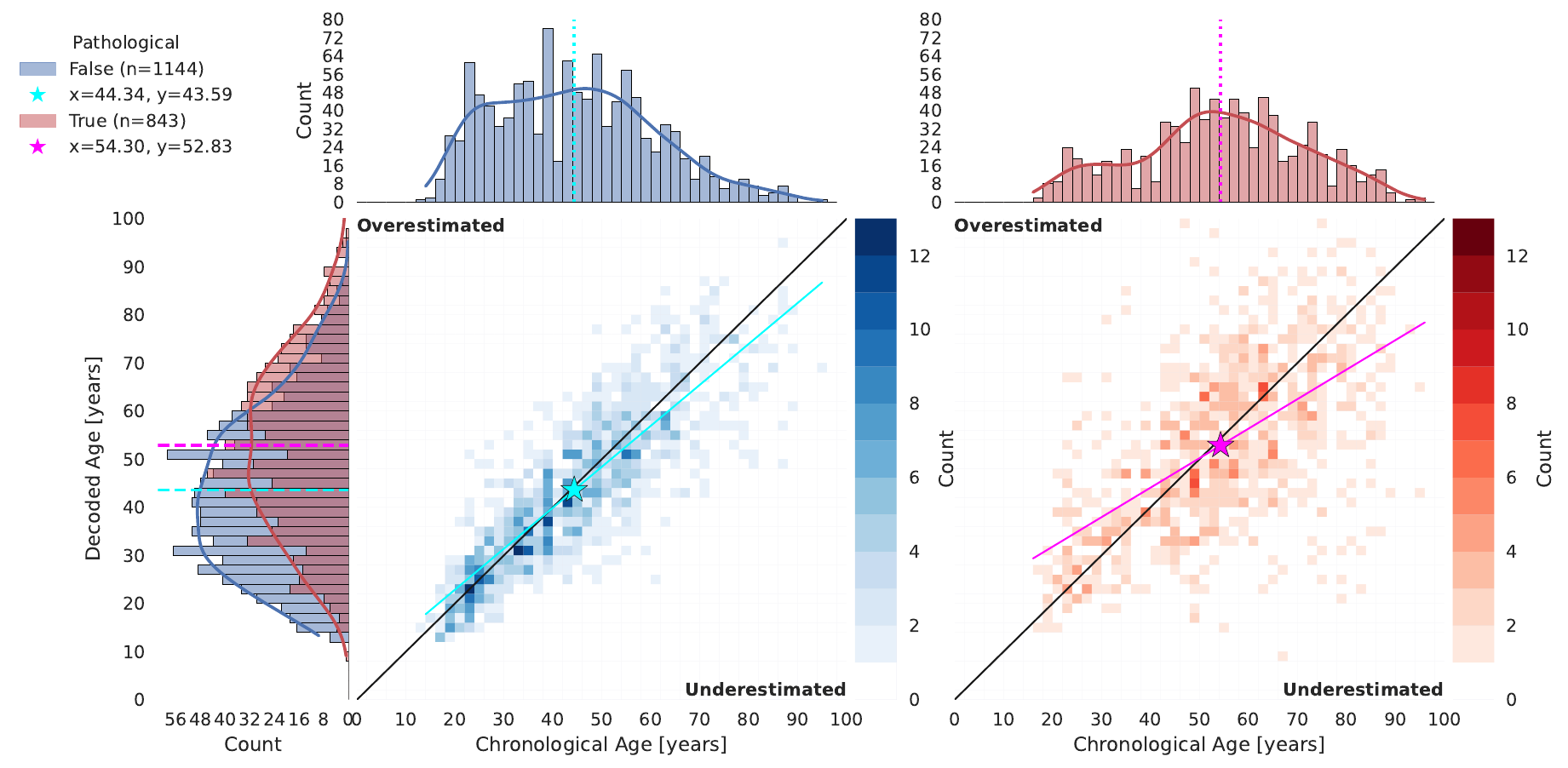}
    \caption{2D histogram of decoded and chronological age with distributions in CV respecting the pathology status. Our models tend to overestimate younger subjects and to underestimate older subjects as very common in these decoding settings. This trend can be consistently observed on non-pathological as well as on pathological subjects through the cyan and magenta line. The effect is stronger for pathological subjects than for non-pathological subjects. The 2D histogram with average markers shown in cyan and magenta reveal a general underestimation of the non-pathological population which is even bigger for the pathological population. }
    \label{fig:cv_age_heatmap}
\end{figure}

In CV our brain age decoders achieved 6.65 years MAE on average. On pathological subjects in contrast, on which the models were not trained, the models did not reach comparable scores with 11.90 years MAE. Figure~\ref{fig:cv_age_heatmap} shows decoded brain ages in relation to chronological ages as well as their distributions.

\begin{figure}[htb!]
    \centering
    \includegraphics[width=\textwidth]{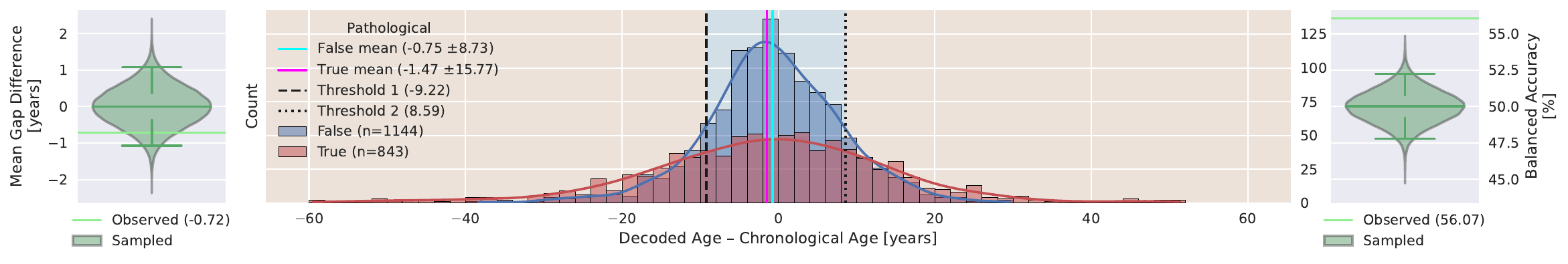}
    \caption{Left: Histogram of brain age gaps respecting the pathology status with average gaps. Right: Permutation test of average gap differences with respect to pathology status observed in CV. The observed average gaps shown in cyan and magenta are negative. The ones for pathological subjects are smaller (-1.47) than the ones for non-pathological subjects (-0.75). There is no statistical evidence that the average gap difference (-0,72) is significant. The gap distribution of brain age gaps of pathological subjects shows roughly twice the amount of variance (+-16) of the distribution of non-pathological subjects (+-8.5).}
    \label{fig:cv_age_gap_histogram}
\end{figure}

We present the distributions of EEG brain age gaps of non-pathological and pathological subjects and the permutation test distributions of average gap differences in Figure~\ref{fig:cv_age_gap_histogram}. 

\begin{figure}[htb!]
    \centering
    \includegraphics[width=\textwidth]{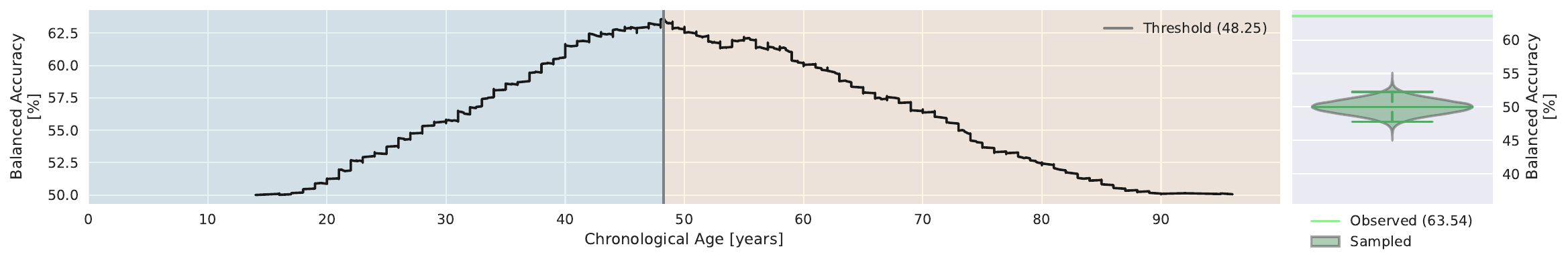}
    \caption{Pathology label assignment based on age threshold with permutation test. A threshold of 48.25 years yielded optimal BACC of 63.54\%.}
    \label{fig:cv_age_thresh}
\end{figure}

\begin{figure}[htb!]
    \centering
    \includegraphics[width=\textwidth]{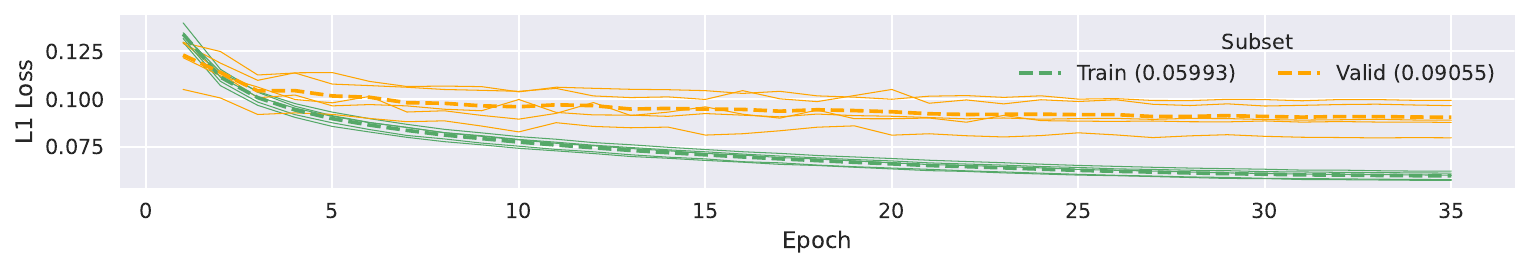}
    \caption{Loss curves are smooth and saturate well before the specified epoch maxium. There is slight variation in final validation score, due to data split.}
    \label{fig:learning_curves}
\end{figure}

Fitting a simple age threshold results in a superior BACC (Figure~\ref{fig:cv_age_thresh}).
The presented learning curves (Figure~\ref{fig:learning_curves}) show neither signs of over- nor underfitting and consistently converge close to the specified epoch maximum. Apart from one run, there is little variance between final loss values of the different runs. Apart from the convincing model fit suggested by Figures~\ref{fig:learning_curves} and \ref{fig:fe_np_age_heatmap}, we observed a prediction bias in our model.

\begin{figure}[htb!]
    \centering
    \includegraphics[width=\textwidth]{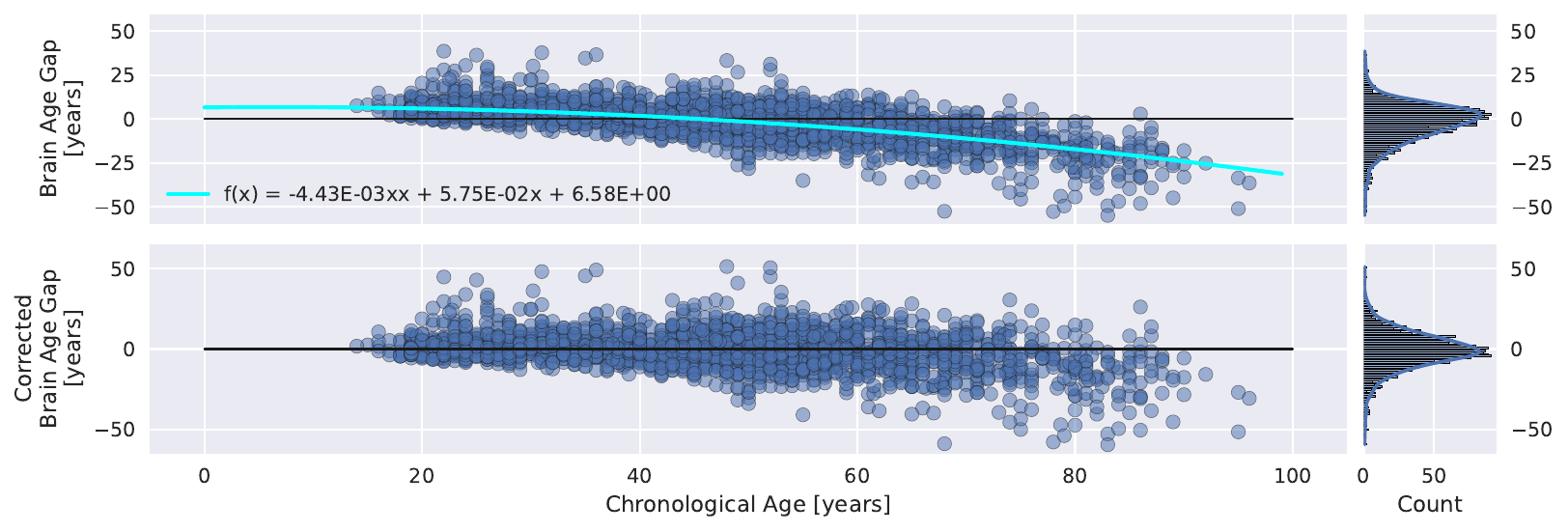}
    \caption{Presentation of fit and application of the quadratic bias model to CV data. Clearly, the gap distribution (top right) is positively skewed, not centered around zero and hence biased. The light blue curve represents the bias model and shows its fit to the data which is not aligned with the baseline. In the lower plot, one can see how the application of the quadratic bias model affects the brain age gaps. As a result, their distribution (bottom right) is symmetrically centered around zero.}
    \label{fig:cv_model_bias}
\end{figure}

We present the quadratic model fit to reduce the bias and the effect of its application in Figure~\ref{fig:cv_model_bias}. 

\subsection{Learning Curves in Final Evaluation}
\begin{figure}[htb!]
    \centering
    \includegraphics[width=\textwidth]{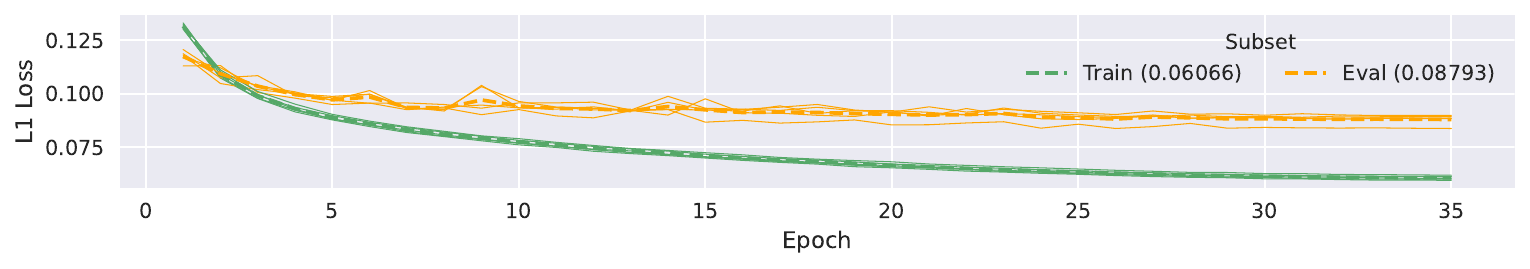}
    \caption{Learning curves of five repetitions of final evaluation. Curves are smooth and saturate well before the specified epoch maximum. Apart from one run, final evaluation loss value is similar for all runs.}
    \label{fig:fe_curves}
\end{figure}

We present the learning curves of our models in FE in Figure~\ref{fig:fe_curves}. There is minor variation in final evaluation loss value, due to varying initizlization seeds.

\subsection{Decoding Scores}
\label{sec:decoding_scores}
As we have also computed the $R^2$ score next to MAE, we report it in Table~\ref{tab:r2_scores}.
\begin{table}[htb!]
    \centering
    \begin{tabular}{c|c|c|c|c|c|c|c|c|c}
        & \multicolumn{2}{|c|}{CV} & \multicolumn{7}{|c}{FE}\\
        \hline
        Dataset & TUAB NP & TUAB P & TUAB NP & TUAB P & RNP & RP &  TNPP & TPNP \\
         \hline
        MAE & 6.65 & 11.90 & 6.60 & 12.82 & 9.27 & 14.87 & 12.71 & 11.02 \\
        $R^2$ & 0.70 & 0.20 & 0.73 & 0.03 & -0.22 & -1.63 & -0.4 & -0.36 \\
    \end{tabular}
    \caption{All decoding scores on all subsets.}
    \label{tab:r2_scores}
\end{table}

\subsection{Analyses on RNP, RP, TNPP, and TPNP}
\label{sec:longitudinal_decoding_results}

\begin{figure}[htb!]
    \centering
    \includegraphics[width=.48\textwidth]{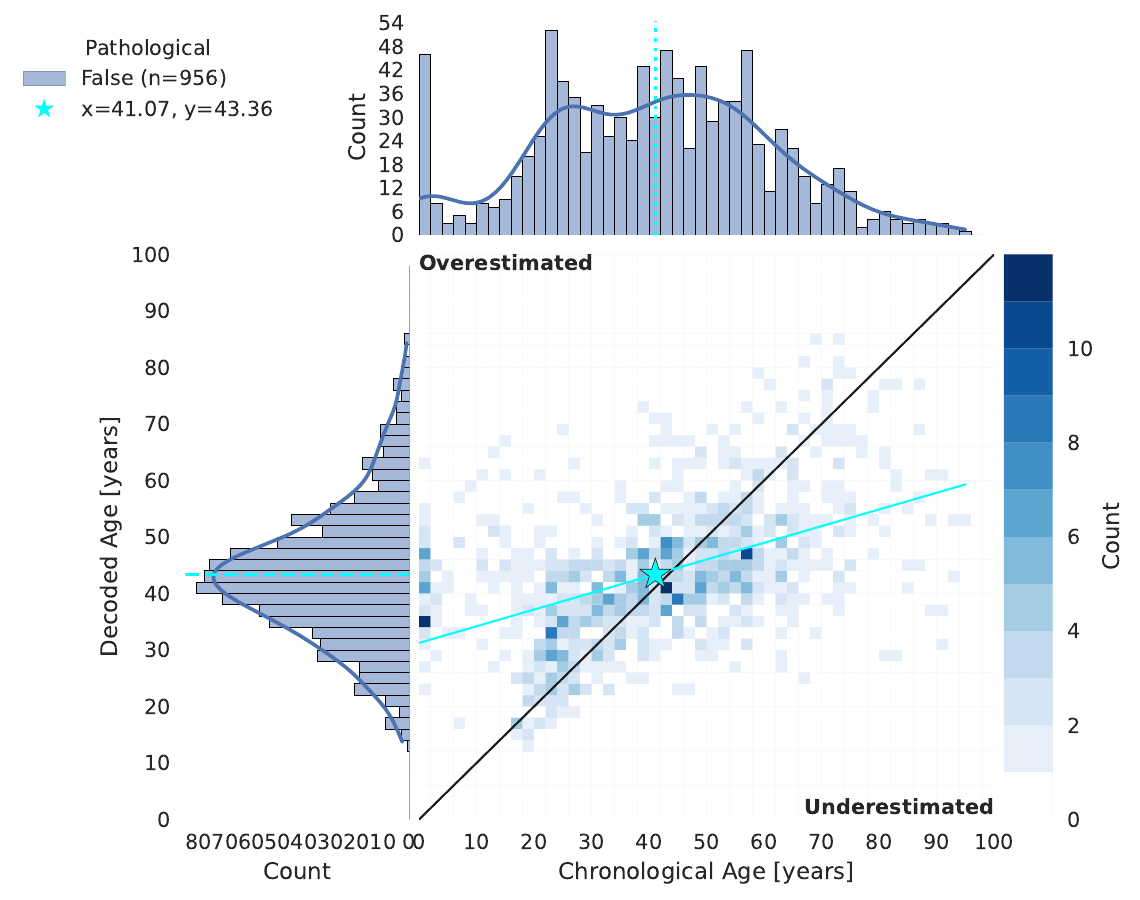}
    \includegraphics[width=.48\textwidth]{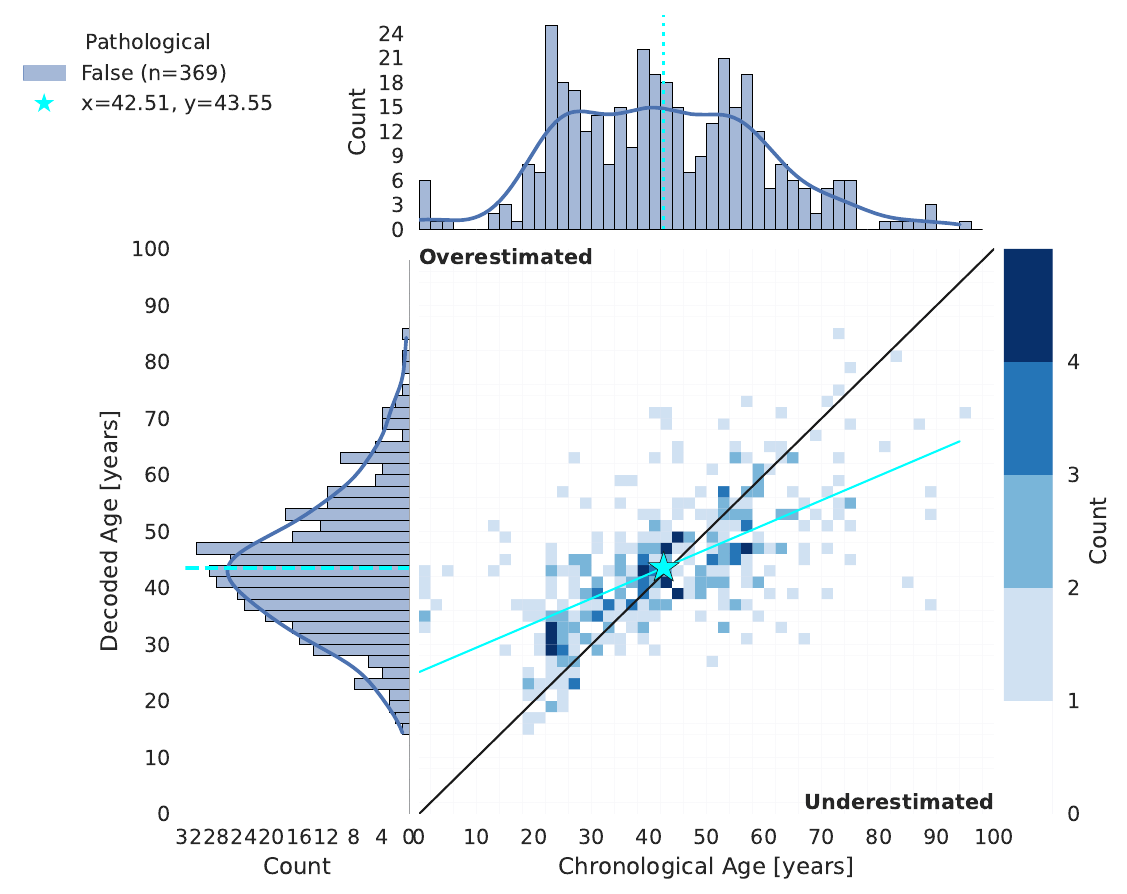}
    \includegraphics[width=.8\textwidth]{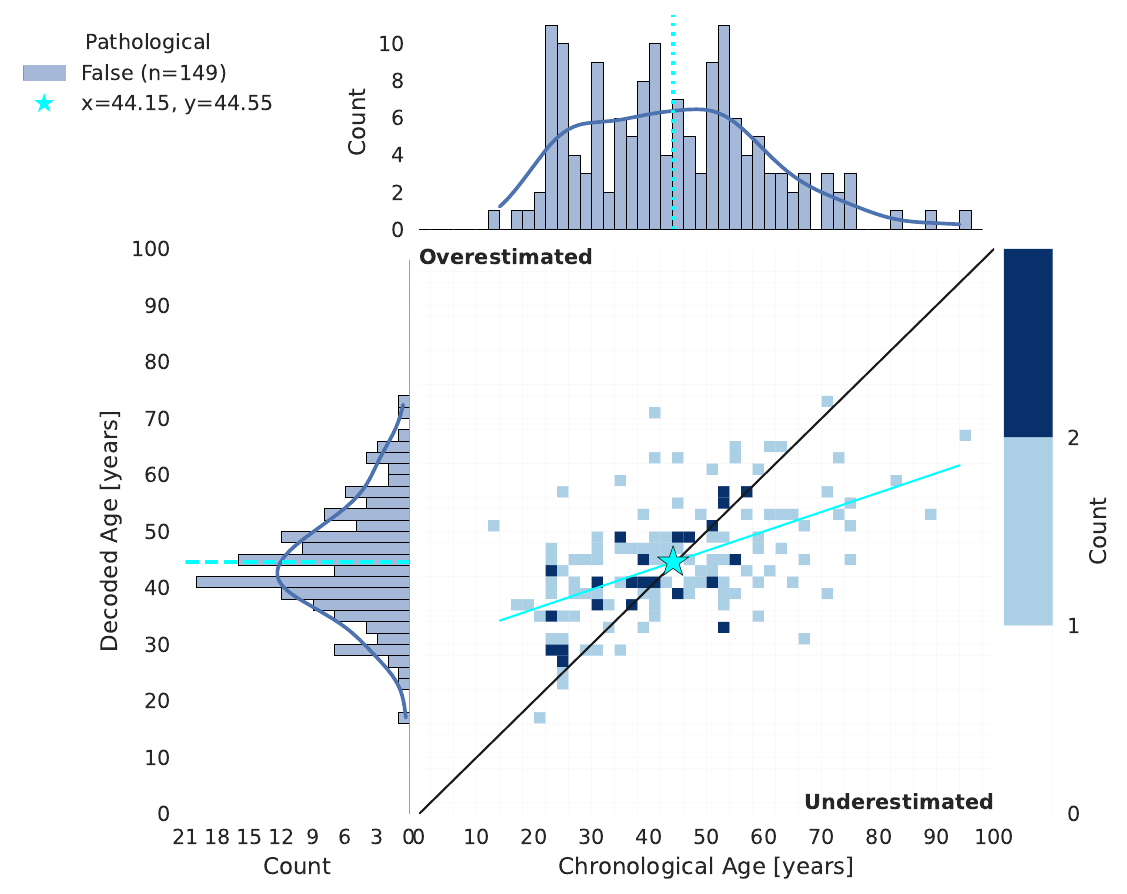}
    \caption{RNP. Left: 2 min \& dirty ages. Right: 15 min \& dirty ages. Bottom: 15 min \& clean ages.}
    \label{fig:lnp_heatmaps}
\end{figure}
\begin{figure}[htb!]
    \centering
    \includegraphics[width=.48\textwidth]{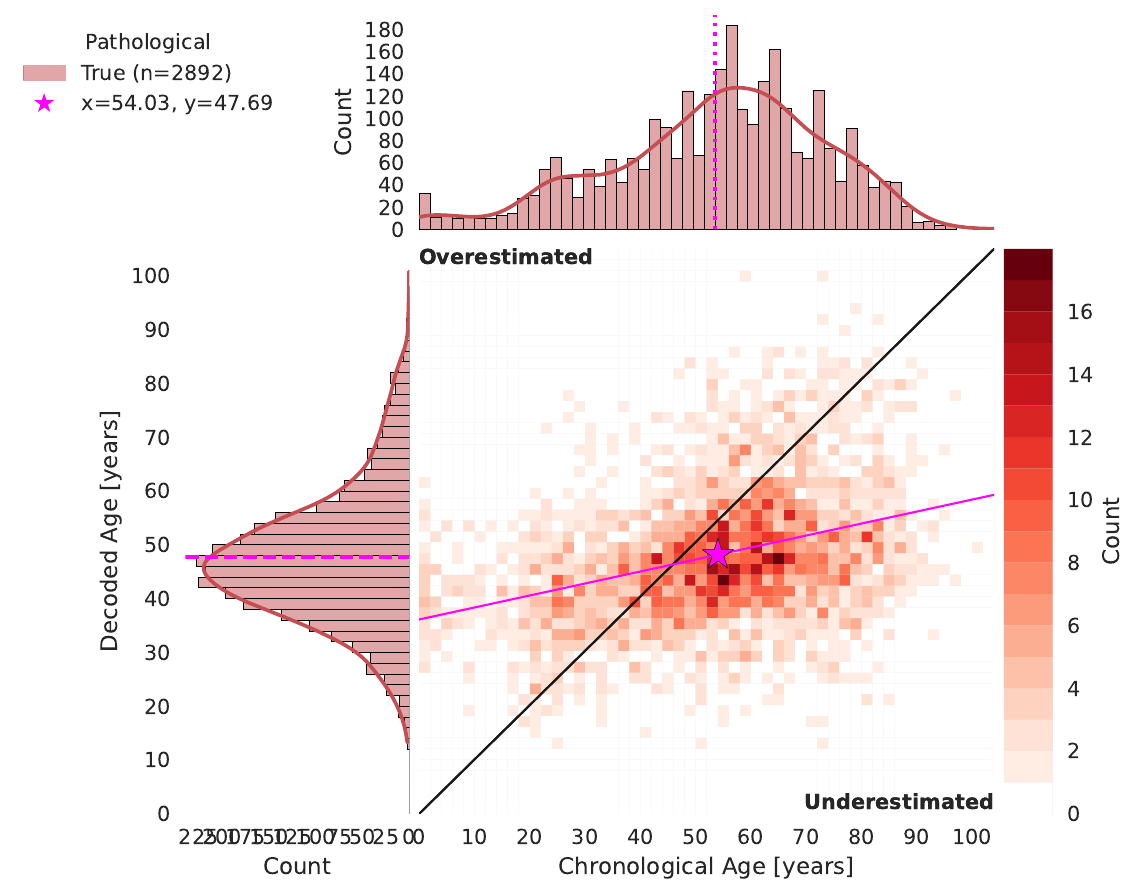}
    \includegraphics[width=.48\textwidth]{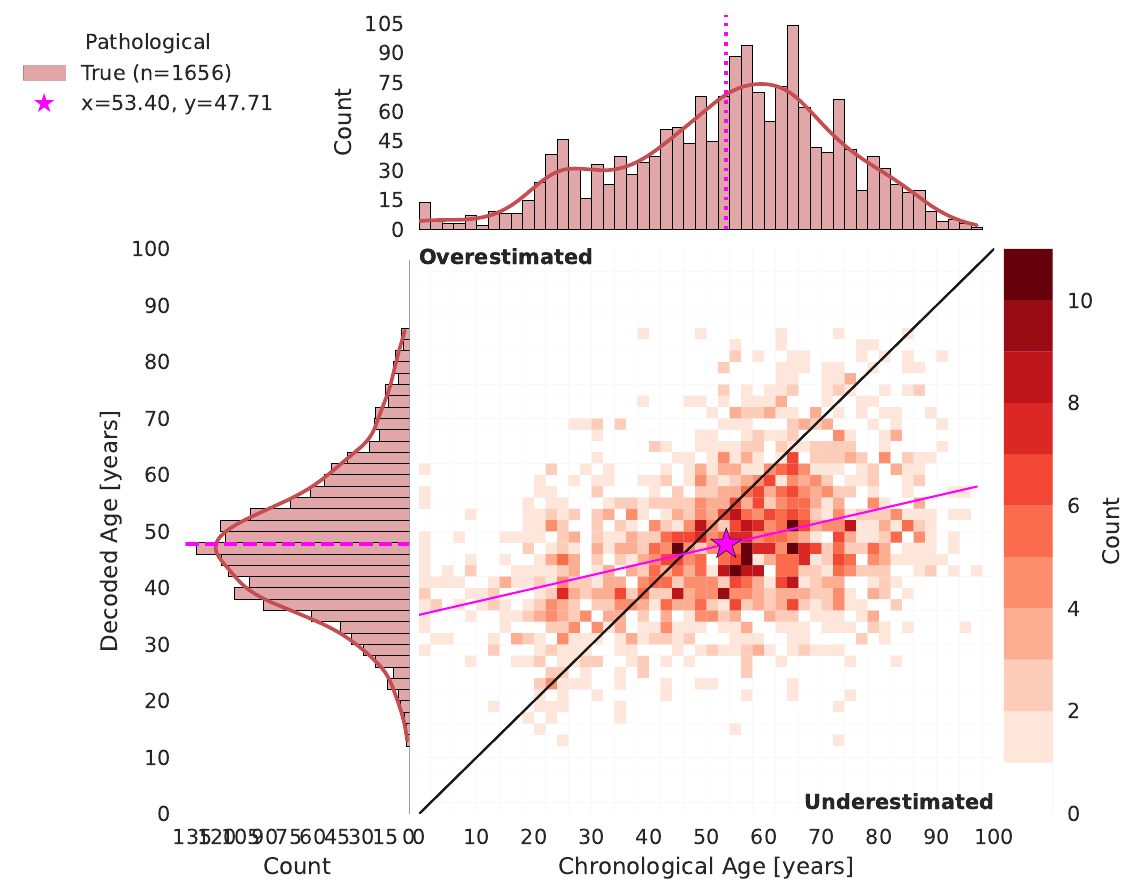}
    \includegraphics[width=.8\textwidth]{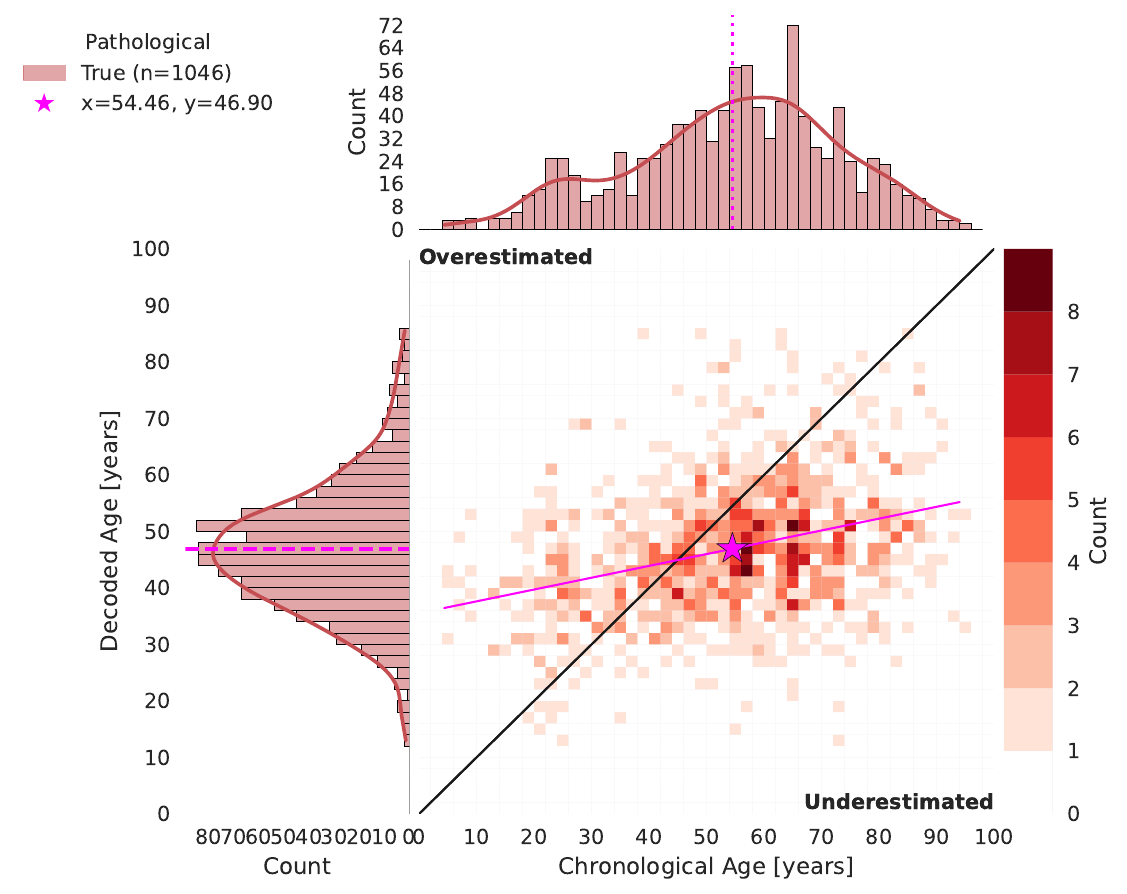}
    \caption{RP. Left: 2 min \& dirty ages. Right: 15 min \& dirty ages. Bottom: 15 min \& clean ages.}
    \label{fig:lp_heatmaps}
\end{figure}
\begin{figure}[htb!]
    \centering
    \includegraphics[width=.75\textwidth]{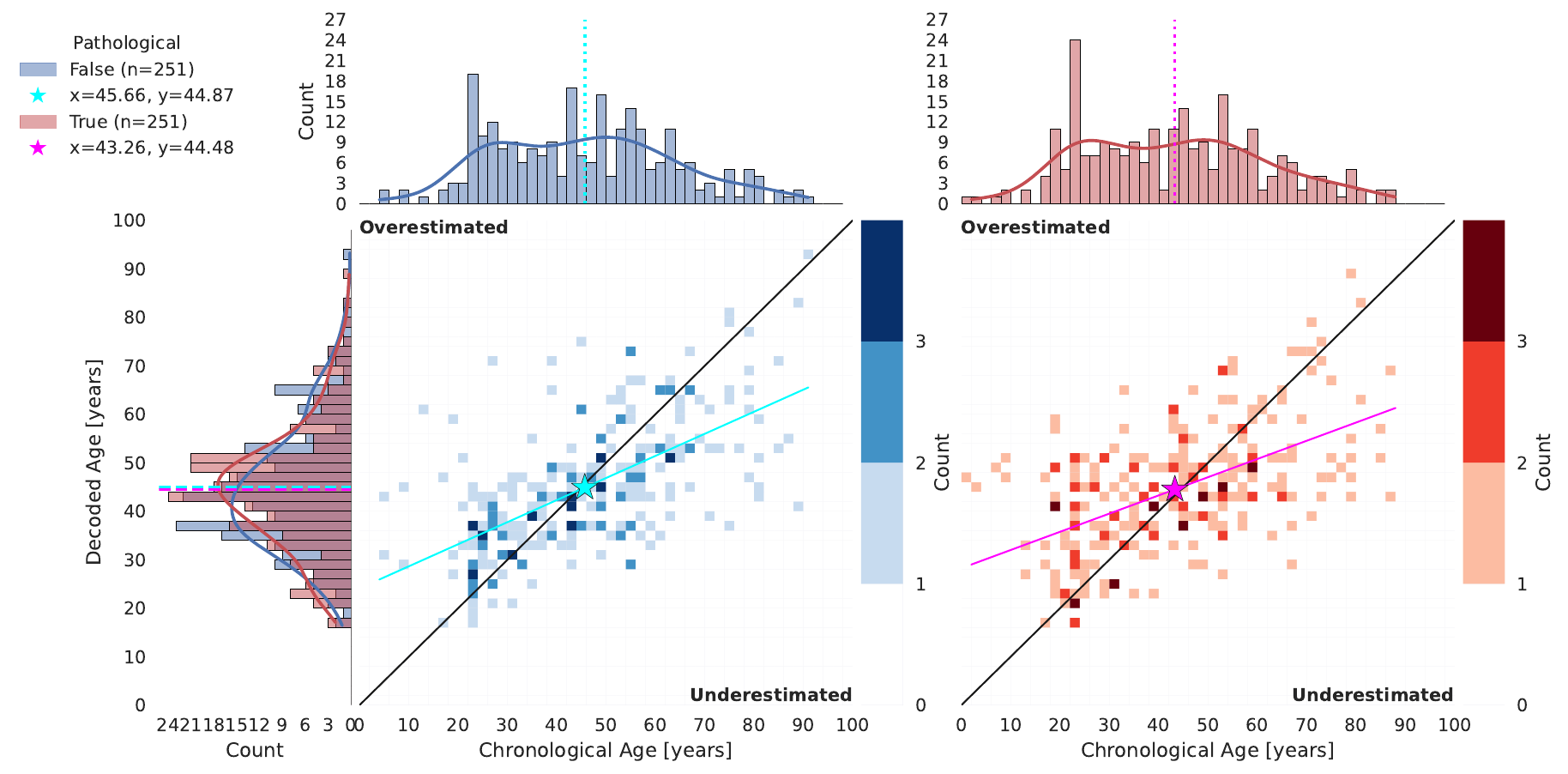}
    \includegraphics[width=.75\textwidth]{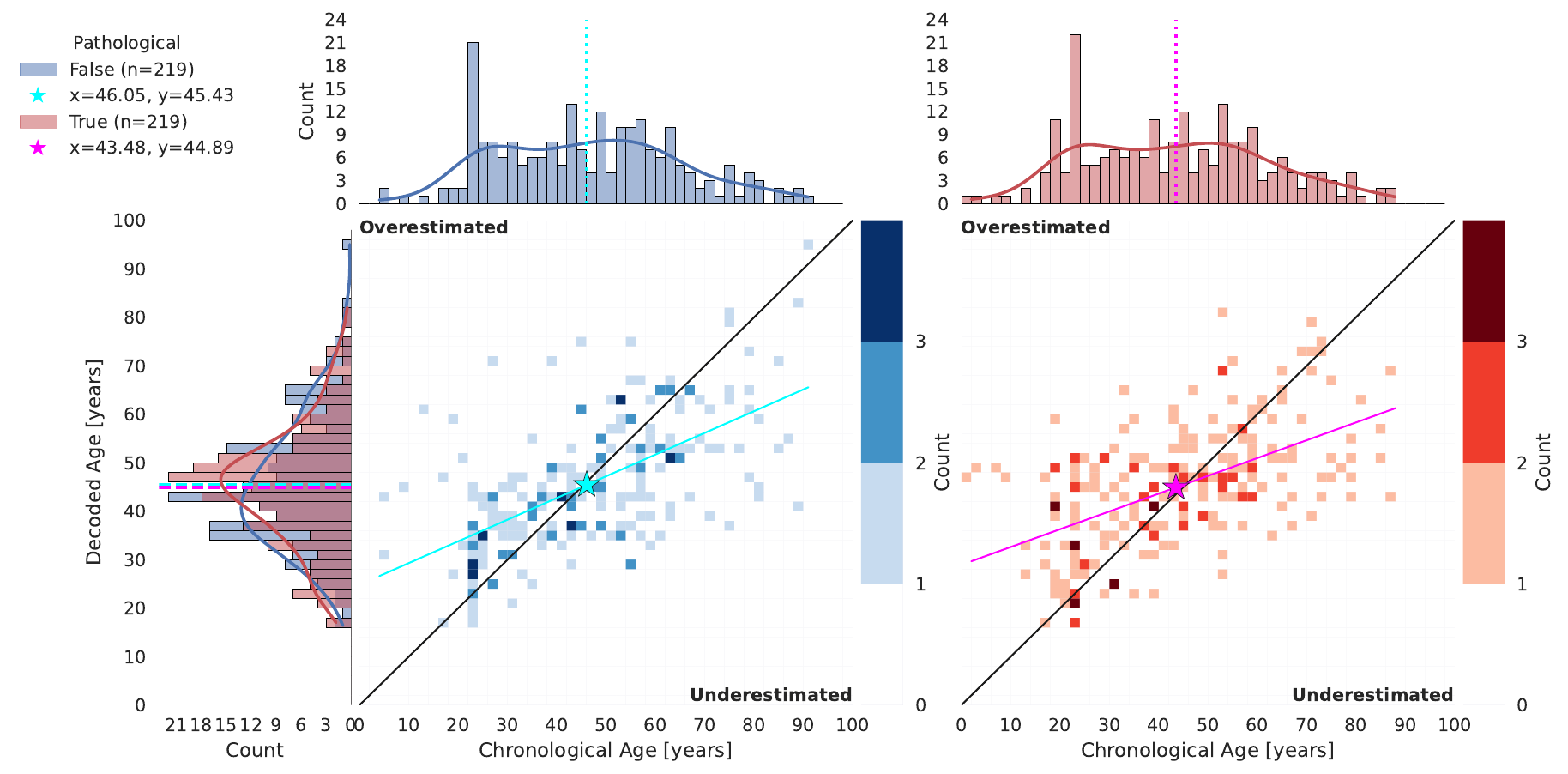}
    \includegraphics[width=.75\textwidth]{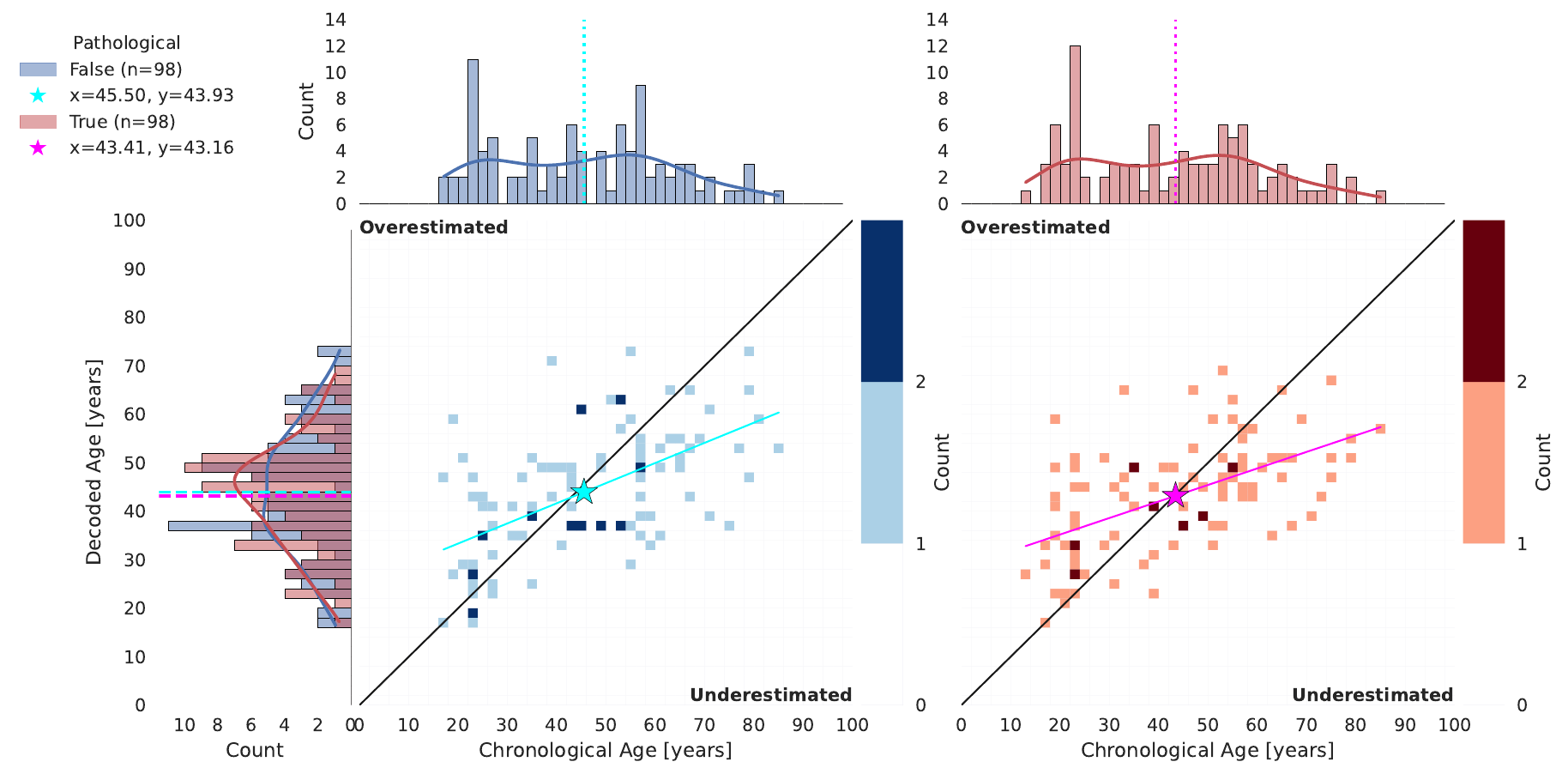}
    \caption{TNPP. Top: 2 min \& dirty ages. Center: 15 min \& dirty ages. Bottomm: 15 min \& clean ages.}
    \label{fig:lnpp_heatmaps}
\end{figure}
\begin{figure}[htb!]
    \centering
    \includegraphics[width=.75\textwidth]{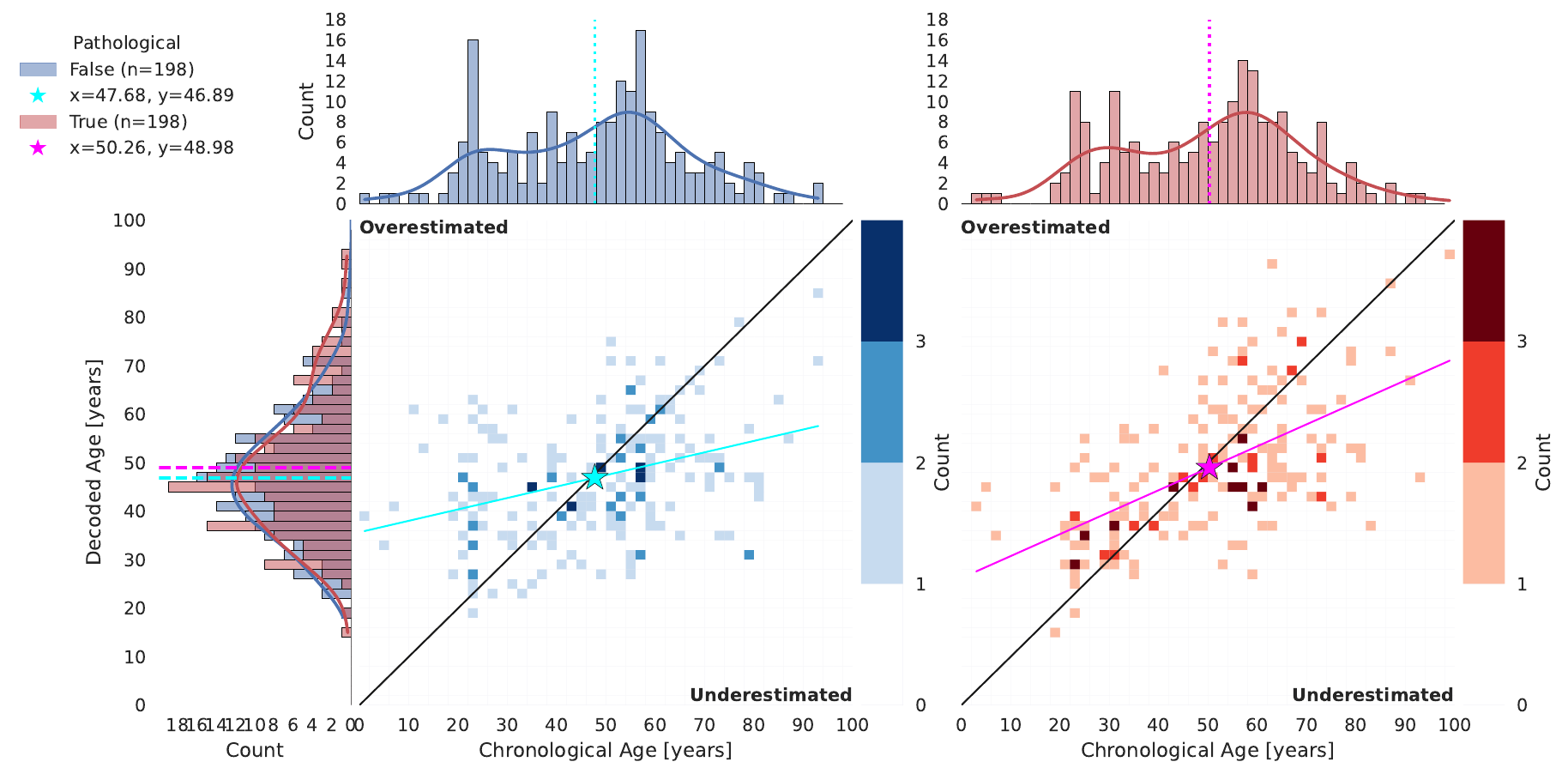}
    \includegraphics[width=.75\textwidth]{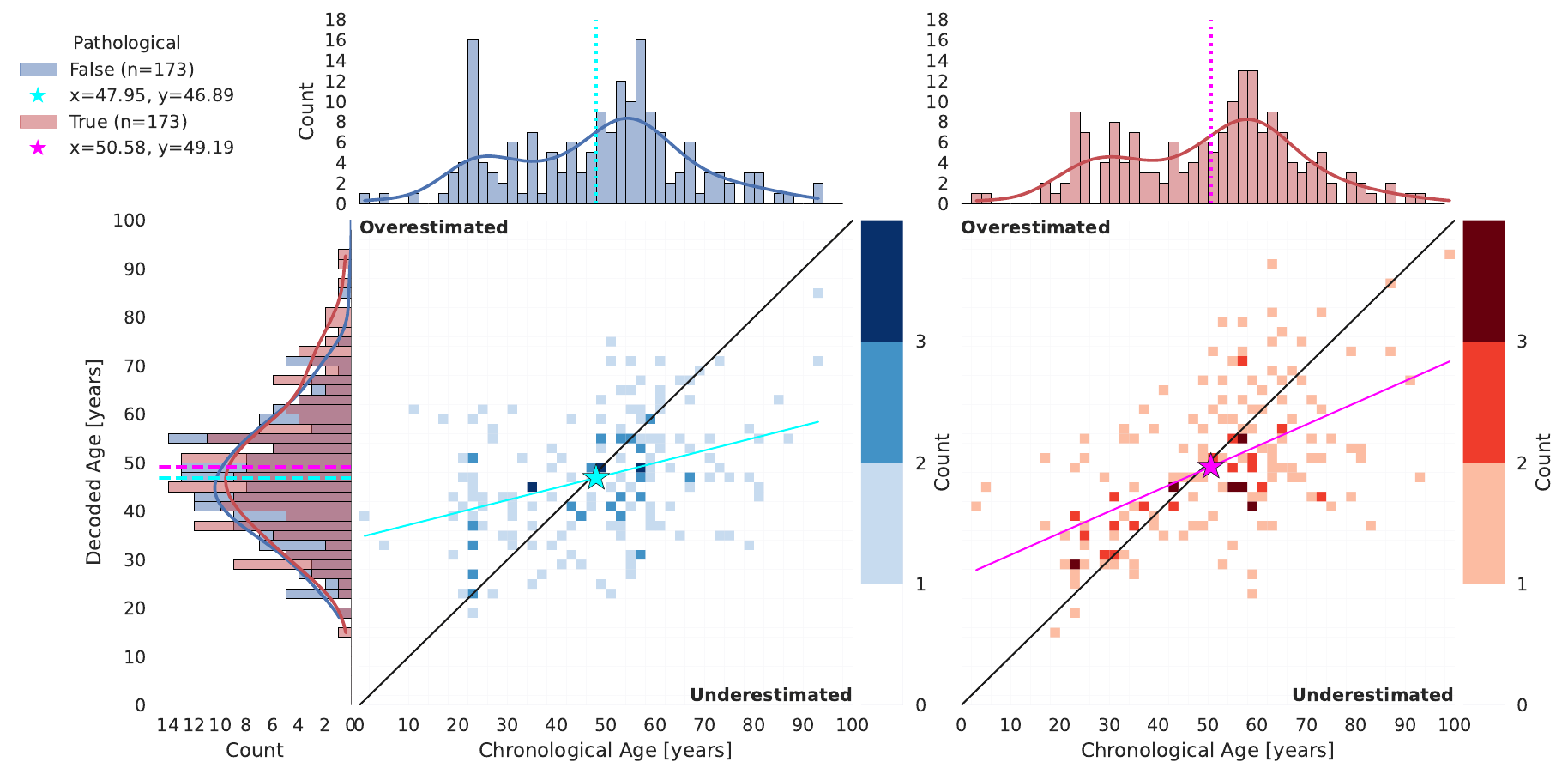}
    \includegraphics[width=.75\textwidth]{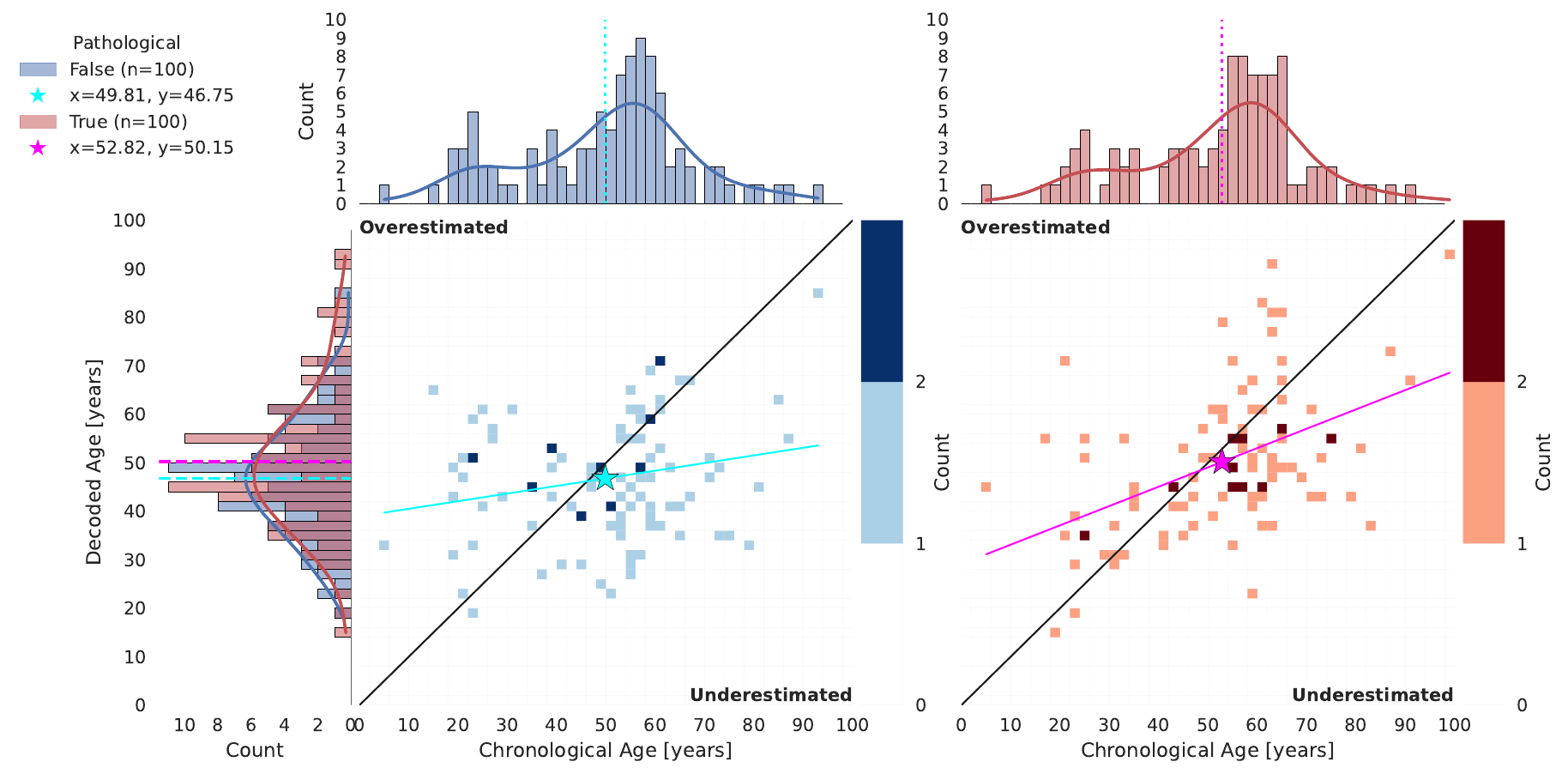}
    \caption{TPNP. Top: 2 min \& dirty ages. Center: 15 min \& dirty ages. Bottom: 15 min \& clean ages.}
    \label{fig:lpnp_heatmaps}
\end{figure}

\begin{figure}[htb!]
    \centering
    \includegraphics[width=\textwidth]{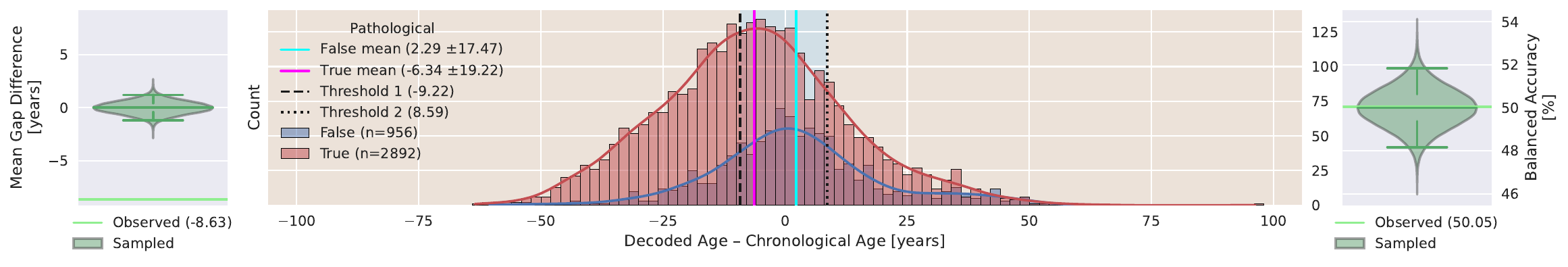}
    \includegraphics[width=\textwidth]{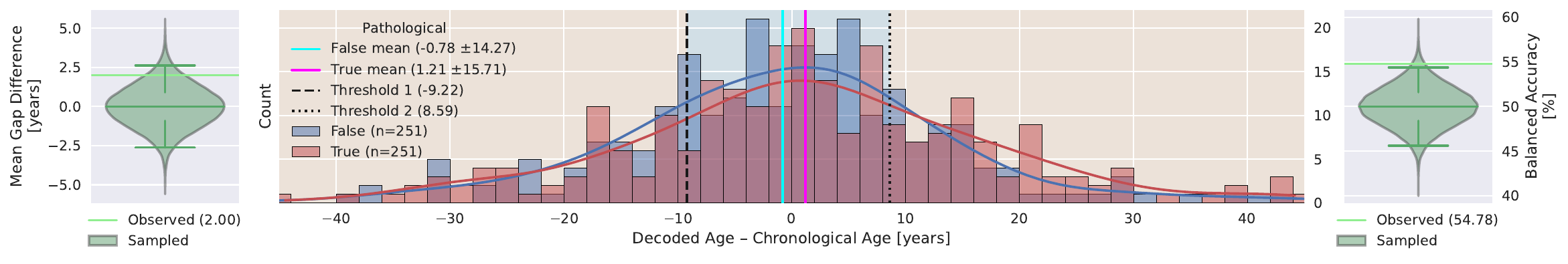}
    \includegraphics[width=\textwidth]{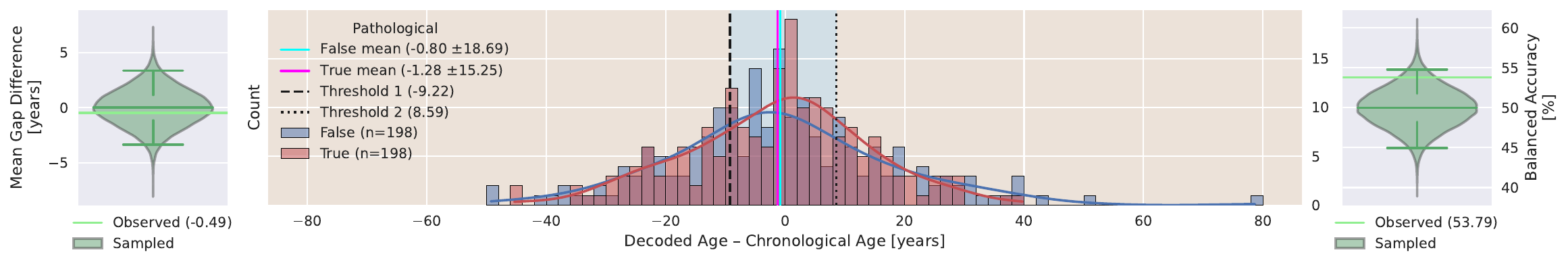}
    \caption{2 min \& dirty ages. Top to bottom: RNP vs. RP, TNPP, TPNP. Gap difference between RNP and RP as well as pathology proxy BACC in LT and TNPP are significant.}
    \label{fig:2_dirty_proxies}
\end{figure}

\begin{figure}[htb!]
    \centering
    \includegraphics[width=\textwidth]{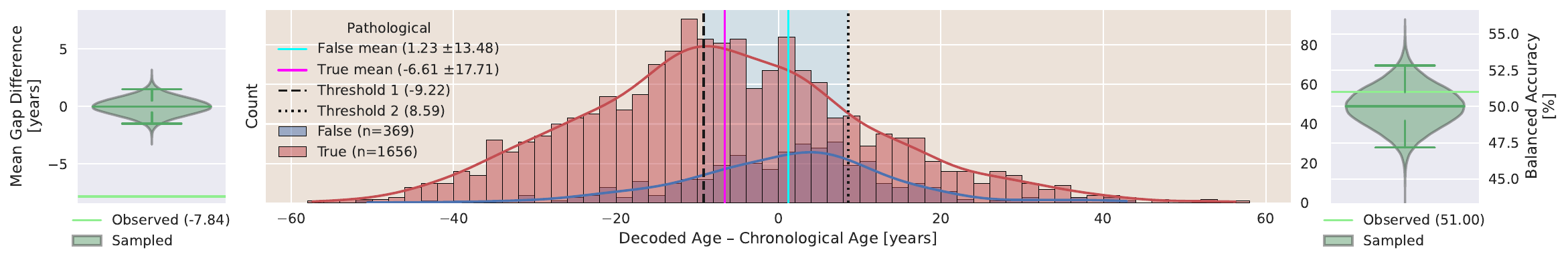}
    \includegraphics[width=\textwidth]{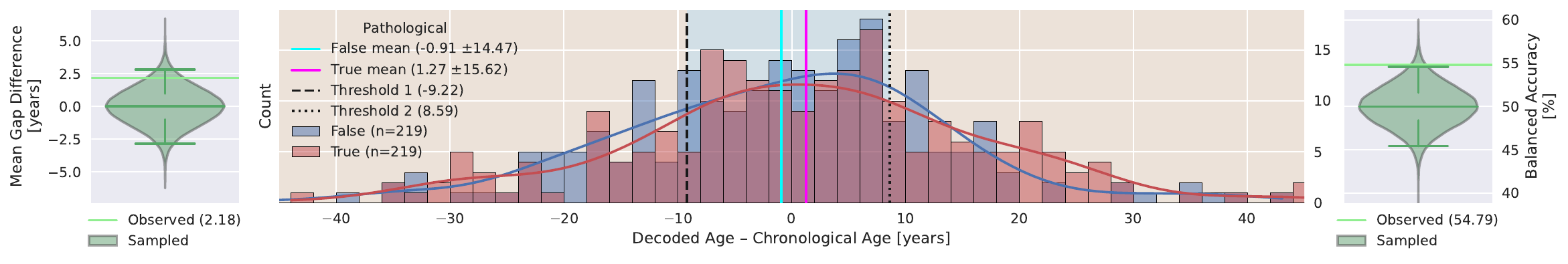}
    \includegraphics[width=\textwidth]{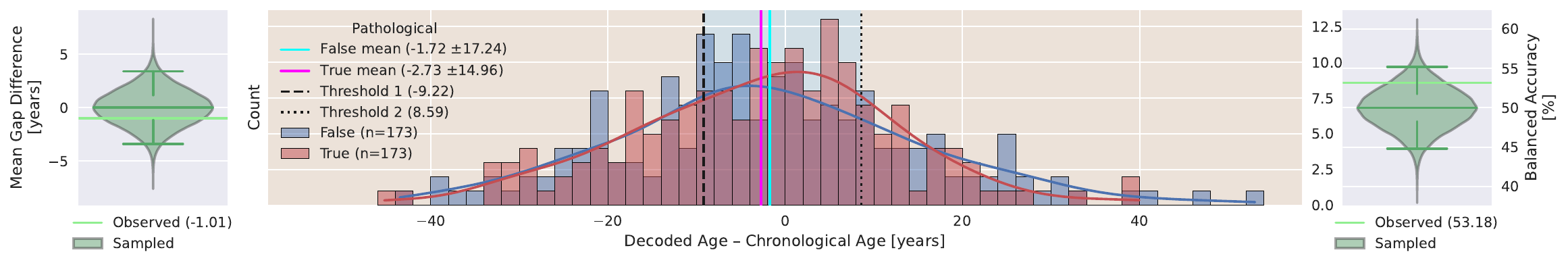}
    \caption{15 min \& dirty ages. Top to bottom: RNP vs. RP, TNPP, TPNP. Gap difference between RNP and RP as well as pathology proxy BACC in TNPP are significant.}
    \label{fig:15_dirty_proxies}
\end{figure}
\end{document}